\documentclass[aps, prd,  showpacs, preprintnumbers, nofootinbib,notitlepage, superscriptaddress]{revtex4-1}

\pdfoutput=1 
\usepackage{amssymb,amsmath,hyperref}
\usepackage{color}
\usepackage{graphicx}
\usepackage{bm}
\usepackage{amsmath,amssymb}
\usepackage{graphicx}
\usepackage{bm}
\usepackage{comment} 
\usepackage{subfigure}
\usepackage{array}
\usepackage{multirow}

\newcommand{\nn}{\nonumber}
\newcommand{\Nstar}{\star}

\newcommand{\Pstar}{{\textbf P}}

\def\fileversion{v1.2a}
\def\filedate{2007/11/23}
\makeatletter

\newbox\swb@xone
\newbox\swb@xtwo
\newbox\swb@xthree
\newbox\swb@xfour
\newdimen\swdimen@ne
\newdimen\swdimentw@

\newcommand{\acontraction}[5][1ex]{%
  \mathchoice
    {\acontraction@\displaystyle{#2}{#3}{#4}{#5}{#1}}%
    {\acontraction@\textstyle{#2}{#3}{#4}{#5}{#1}}%
    {\acontraction@\scriptstyle{#2}{#3}{#4}{#5}{#1}}%
    {\acontraction@\scriptscriptstyle{#2}{#3}{#4}{#5}{#1}}}%
\newcommand{\acontraction@}[6]{%
  \setbox\swb@xone=\hbox{${}#1{}#2{}$}%
  \setbox\swb@xtwo=\hbox{${}#1{}#3{}$}%
  \setbox\swb@xthree=\hbox{${}#1{}#4{}$}%
  \setbox\swb@xfour=\hbox{${}#1{}#5{}$}%
  \swdimen@ne=\wd\swb@xtwo%
  \advance\swdimen@ne by \wd\swb@xfour%
  \divide\swdimen@ne by 2%
  \advance\swdimen@ne by \wd\swb@xthree%
  \vbox{%
    \hbox to 0pt{%
      \kern \wd\swb@xone%
      \kern 0.5\wd\swb@xtwo%
      \acontraction@@{\swdimen@ne}{#6}%
      \hss}%
    \vskip 0.5ex
    \vskip\ht\swb@xtwo}}

\newcommand{\acontraction@@}[3][0.05em]{%
  \hbox{%
    \vrule width #1 height 0pt depth #3%
    \vrule width #2 height 0pt depth #1%
    \vrule width #1 height 0pt depth #3%
    \relax}}

\newcommand{\bcontraction}[5][1ex]{%
  \mathchoice
    {\bcontraction@\displaystyle{#2}{#3}{#4}{#5}{#1}}%
    {\bcontraction@\textstyle{#2}{#3}{#4}{#5}{#1}}%
    {\bcontraction@\scriptstyle{#2}{#3}{#4}{#5}{#1}}%
    {\bcontraction@\scriptscriptstyle{#2}{#3}{#4}{#5}{#1}}}%
\newcommand{\bcontraction@}[6]{%
  \setbox\swb@xone=\hbox{${}#1{}#2{}$}%
  \setbox\swb@xtwo=\hbox{${}#1{}#3{}$}%
  \setbox\swb@xthree=\hbox{${}#1{}#4{}$}%
  \setbox\swb@xfour=\hbox{${}#1{}#5{}$}%
  \swdimen@ne=\wd\swb@xtwo%
  \advance\swdimen@ne by \wd\swb@xfour%
  \divide\swdimen@ne by 2%
  \advance\swdimen@ne by \wd\swb@xthree%
  \lower 0.5ex \vbox{%
    \hbox to 0pt{%
      \kern \wd\swb@xone%
      \kern 0.5\wd\swb@xtwo%
      \bcontraction@@{\swdimen@ne}{#6}%
      \hss}%
    }}

\newcommand{\bcontraction@@}[3][0.05em]{%
  \hbox{%
    \swdimentw@=#3
    \advance\swdimentw@ by -#1
    \vrule width #1 height 0pt depth #3%
    \lower\swdimentw@\hbox{\vrule width #2 height 0pt depth #1}%
    \vrule width #1 height 0pt depth #3%
    \relax}}

\makeatother

\usepackage[T1]{fontenc}
\usepackage[latin9]{inputenc}
\usepackage{graphicx}
\usepackage{esint}
\usepackage{hyperref}
\usepackage{atbegshi}
\usepackage{lipsum}

\usepackage{cancel}
\usepackage[normalem]{ulem}
\usepackage{soul}
\setstcolor{red}
\setul{-.5ex}{2pt}

\def\eqref#1{{(\ref{#1})}}

\begin{document}

 \preprint{\vbox{
\hbox{JLAB-THY-14-1909} 
}}
 \preprint{\vbox{
\hbox{NT@WM-14-04} 
}}

\title{Multichannel $1\rightarrow 2$ transition amplitudes in a finite volume}

\author{Ra\'ul A. Brice\~no{\footnote{\tt rbriceno@jlab.org}}}
\affiliation{Thomas Jefferson National Accelerator Facility, 12000 Jefferson Avenue, Newport
  News, VA 23606, USA}
  \author{Maxwell T. Hansen{\footnote{\tt mth28@uw.edu}}}
\affiliation{Department of Physics, University of Washington,  Box 351560, Seattle, WA 98195, USA}
 \author{Andr\'e Walker-Loud{\footnote{\tt walkloud@wm.edu}}}
\affiliation{Thomas Jefferson National Accelerator Facility, 12000 Jefferson Avenue, Newport
  News, VA 23606, USA}
  \affiliation{Department of Physics, College of William and Mary, Williamsburg, Virginia 23187-8795, USA}

\begin{abstract}
We perform a model-independent, non-perturbative investigation of two-point and three-point finite-volume correlation functions in the energy regime where two-particle states can go on-shell. 
We study three-point functions involving a single incoming particle and an outgoing two-particle state, relevant, for example, for studies of meson decays (e.g., $B^0 \rightarrow K^*\ell^+\ell^-\rightarrow \pi K\ell^+\ell^-$) or meson photo production (e.g., $\pi\gamma^*\rightarrow \pi\pi$).
We observe that, while the spectrum solely depends on the on-shell scattering amplitude, the correlation functions also depend on \emph{off-shell} amplitudes.
The main result of this work is a generalization of the Lellouch-L\"{u}scher formula relating matrix elements of currents in finite and infinite spatial volumes. We extend that work by considering a theory with multiple, strongly-coupled channels and by accommodating external currents which inject arbitrary four-momentum as well as arbitrary angular momentum.
The result is exact up to exponentially suppressed corrections governed by the pion mass times the box size.
We also apply our master equation to various examples, including the two processes mentioned above as well as examples where the final state is an admixture of two open channels.

 \end{abstract}

\maketitle
\section{Introduction\label{sec:intro}}

There are a number of matrix elements involving hadronic two-body initial and/or final states for which a direct calculation with lattice QCD would provide a significant advancement for nuclear and particle physics.
For example, the calculation of proton-proton fusion through the weak interactions, $pp\rightarrow d e^+\nu_e$, would allow for a direct theoretical prediction of this fundamental process which powers the sun.
The MuSun Collaboration will measure a related process, muon capture on deuterium~\cite{Andreev:2010wd}.
At low energies, these two processes are described by the same two-nucleon contact interaction~\cite{Ando:2001es}, providing an opportunity to over-constrain these reactions for which there is currently discrepancy between experimental results~\cite{Bardin,Cargnelli} and theory calculations~\cite{Ando:2001es,Chen:2005ak}.
Another example of particular interest is the heavy meson decay $B^0\rightarrow K^*\ell^+\ell^-\rightarrow \pi K\ell^+\ell^-$. Tentative tension exists between experimental results~\cite{Wei:2009zv,Aaltonen:2011ja,Lees:2012tva,Aaij:2013iag,Aaij:2013qta} and Standard Model predictions~\cite{Bobeth:2012vn,Descotes-Genon:2013wba,Hambrock:2013zya,Beaujean:2013soa} for this process, so that better constraining the latter would clearly be valuable.

The opportunity to test the Standard Model with these decays motivated early quenched lattice QCD calculations~\cite{Bowler:1993rz,Bernard:1993yt,Burford:1995fc,Abada:1995fa,Abada:2002ie,Bowler:2004zb,Becirevic:2006nm}. The newly observed tension between theory and experiment has motivated dynamical lattice QCD calculations to determine the hadronic transition amplitudes. These also find evidence for deviations from the Standard Model~\cite{Horgan:2013hoa,Horgan:2013pva}. There is, however, an important caveat to these calculations known as the \textit{Maiani-Testa no-go theorem}.
This is the observation that there is no simple relation between Euclidean-spacetime correlators and the desired Minkowski-spacetime transition matrix elements, whenever the initial or final states contain multiple hadrons~\cite{Maiani:1990ca}.
As the $K^*(892)$ is a strong resonance of the $K\pi$ scattering system (for $m_\pi \lesssim 400$~MeV), this issue cannot be avoided for lattice QCD calculations of this important quantity. 

The formalism to overcome this challenge was first developed by Lellouch and L\"uscher for the $K\rightarrow \pi\pi$ decay amplitude, and is known as the Lellouch-L\"{u}scher or LL method~\cite{Lellouch:2000pv}. The crucial development was to relate the lattice QCD calculations in a \textit{finite spatial volume} to the infinite-volume matrix element. In a finite volume, the two-pion spectrum is a set of discrete energy levels.  If the size of the finite-volume box is tuned such that one of the two-pion energy levels is degenerate with the kaon, then a simple relation exists between the finite-volume matrix element and the infinite-volume $K \rightarrow \pi \pi$ decay amplitude. In particular, the ratio of the finite- and infinite-volume matrix elements is a known function, depending on the two-pion scattering phase-shift near the kaon mass as well as the finite-volume box size and other kinematic variables. The ratio is commonly referred to as the \textit{LL-factor}. The initial work by Lellouch and L\"uscher was restricted to an S-wave two-pion state in the center-of-mass (c.m.) frame.
This formalism has been extended for all decays below the inelastic threshold~\cite{Lin:2001ek} and for systems with nonzero total momentum~\cite{Kim:2005gf, Christ:2005gi} 
(see Refs.~\cite{Blum:2011pu,Blum:2011ng,Blum:2012uk, Boyle:2012ys} for applications of this formalism to lattice QCD calculations of the $K\rightarrow\pi\pi$ decay amplitude%
\footnote{For important theoretical and numerical developments regarding nonleptonic weak decay on the lattice see Refs.~\cite{Ciuchini:1996mq, Testa:1997ne, Dawson:1997ic, Golterman:1997wb,Rossi:1998kc, Buras:2000kx, Pallante:2000ut, Lin:2001ek,Colangelo:2001uv, Lin:2001fi, Laiho:2002jq, Aoki:1997ev, Bijnens:1998mb, Pekurovsky:1998jd, Pekurovsky:1999pn, Pekurovsky:1999er, Mawhinney:2000hk, Capitani:2000bm, Kim:2002np, Laiho:2003uy, Yamazaki:2006ce, Yamazaki:2008hg}.}).
More recently, the formalism has been extended to accommodate decays into multiple, coupled two-particle channels~\cite{Hansen:2012tf} and to describe the processes $\pi^0\rightarrow \gamma\gamma$~\cite{Meyer:2013dxa}, $N\gamma\rightarrow N\pi$~\cite{Agadjanov:2014kha},%
\footnote{During the preparation of this manuscript a similar and independent work by A. Agadjanov, \emph{et al.}, appeared in the literature~\cite{Agadjanov:2014kha}. In their work, the authors considered pion-photoproduction off a nucleon, $N\gamma\rightarrow N\pi$ in the non-relativistic limit. The authors demonstrated how to study transitions amplitudes for systems with nonzero intrinsic spin. In doing so, they restrict the final two-particle state to be at rest and neglect corrections due to partial wave mixing, but they do allow for the finite volume of the systems to have an asymmetry along one of the cartesian axes.} 
as well as $\textbf{2}\rightarrow \textbf{2}$ processes~\cite{Detmold:2004qn, Meyer:2012wk, Bernard:2012bi, Briceno:2012yi}.

In this work, we extend these studies to unambiguously study $\textbf{1}\rightarrow\textbf{2}$ transition amplitudes involving external currents which insert energy, momentum and angular momentum for systems with any number of two-particle channels which mix with arbitrary strong couplings. Our formalism includes all two-particle angular-momentum states, but is valid only for spin-zero particles and only at energies less than the lowest lying multi-particle inelastic threshold.
In order to determine the $\textbf{1}\rightarrow\textbf{2}$ transition amplitudes, both two- and three-point correlation functions are needed.%
\footnote{It is often customary to label correlation functions by the total number of particles in the initial and final states.  However, in this work, we find it more convenient to label all correlation functions with no insertion of external currents as \textit{two-point} correlation functions and those with an external current as \textit{three-point} correlation functions.} 
From the two-point correlation functions, one extracts the finite-volume spectrum and determines the scattering phase shifts.
From appropriate ratios of three- to two-point correlation functions one determines the finite-volume matrix elements of external currents.
These matrix elements can then be related to the corresponding infinite-volume transition amplitudes.

In Section~\ref{sec:corr2}, we review the two-point correlation functions.
The finite-volume corrections to single particle masses are exponentially suppressed in $m_\pi L$ where $L$ is the spatial extent of the volume and $m_\pi$ is the pion mass~\cite{Luscher:1985dn}.  We assume spatial extents such that these exponential corrections can be safely neglected.
In contrast to the single-particle states, the finite-volume energy spectrum above the two-particle threshold cannot be directly identified with infinite-volume observables.
However, the spectrum does encode information about the infinite-volume {on-shell} scattering amplitude.
The formalism to relate these observables to the finite-volume spectrum is known as the \textit{L\"uscher} method~\cite{Luscher:1986pf, Luscher:1990ux}.
This approach has been investigated and generalized in various contexts~\cite{Rummukainen:1995vs,  Beane:2003yx, Beane:2003da, Li:2003jn, Detmold:2004qn, Bedaque:2004kc, Feng:2004ua, Christ:2005gi, Kim:2005gf, Bernard:2008ax, Bour:2011ef, Davoudi:2011md, Leskovec:2012gb, Gockeler:2012yj, Ishizuka:2009bx, Briceno:2013rwa, Briceno:2013lba, Briceno:2013bda, Briceno:2013hya, Liu:2005kr, Hansen:2012tf, Briceno:2012yi,  Li:2012bi, Guo:2012hv, Bernard:2010fp, Li:2014wga} including most recently a method for describing all $\textbf{2}\rightarrow\textbf{2}$ systems with arbitrary quantum numbers, open channels and boundary conditions~\cite{Briceno:2014oea}.%
\footnote{There have also been attempts to generalize this formalism for three-particle systems~\cite{Roca:2012rx,  Polejaeva:2012ut,Briceno:2012rv, Hansen:2013dla}, but a general solution for the three body system in a finite volume has not been found.} 
We recover the well-known quantization condition for a system with any number of two-scalar channels, with arbitrary angular momentum as well as total linear momenta. The result is
\begin{eqnarray}
\label{eq:QC}
{\det\left[\mathbb{K}(E_n)+ \left(\mathbb{F}^V(E_n)\right)^{-1}\right]}=0,
\end{eqnarray} 
and was first obtained in Refs.~\cite{Hansen:2012tf, Briceno:2012yi}.
$\mathbb{K}$ is the two-particle K-matrix (defined in Eq.~\ref{eq:Kmatrix} with a well known relation to the scattering amplitude, Eq.~\ref{MMtildematrix}) and $\mathbb{F}^V$ is a volume-dependent kinematic matrix (defined in Eq.~\ref{eq:FVsum}).
Both of these are matrices over angular momenta as well as all open two-particle channels, and the determinant is understood to act on this direct-product space.
In the energy regime of elastic scattering, this formalism has been extensively implemented in numerical lattice calculations for single channel processes, e.g.~\cite{Li:2007ey, Durr:2008zz, Beane:2010hg, Beane:2011xf, Beane:2012vq, Beane:2012ey,Yamazaki:2012hi, Beane:2011iw, Beane:2013br, Beane:2011sc,  Pelissier:2011ib, Aoki:2007rd, Lang:2011mn, Pelissier:2012pi, Ozaki:2012ce, Buchoff:2012ja, Dudek:2012xn, Dudek:2012gj, Mohler:2013rwa, Lang:2014tia}.
Until recently, the only numerical implementation of the coupled-channel formalism was by Guo in an exploratory numerical calculation of a two-channel system in 1 + 1 dimensional lattice model~\cite{Guo:2013vsa}.
The first lattice QCD application of this formalism was recently performed by the Hadron Spectrum Collaboration in a benchmark calculation of the $\pi K$-$K\eta$ system~\cite{Dudek:2014qha}.

In Section~\ref{sec:corr3}, we generalize the Lellouch-L\"{u}scher result in several ways.
We allow the the current to insert arbitrary momentum and energy to the system and we include multiple strongly-coupled channels as well as angular momentum mixing in all irreps of the relevant finite-volume symmetry group. We derive a non-perturbative master equation that relates the finite-volume matrix elements of currents with the physically relevant infinite-volume counterpart
\begin{equation}
\left|\langle E_{\Lambda_f,n_f}\textbf{P}_f;L|\tilde{\mathcal{J}}_{\Lambda\mu}^{[J,P,|\lambda|]}(0,\textbf{P}_i-\textbf{P}_f)| E_{\Lambda_i,0}\textbf{P}_i;L\rangle\right| =
 \frac{1}{\sqrt{2E_{\Lambda_i,0}}}
\sqrt{
\left[\mathcal{A}^\dagger_{\Lambda_f,n_f;\Lambda\mu}~\mathcal R_{\Lambda_f,n_f}~\mathcal{A}_{\Lambda_f,n_f;\Lambda\mu}\right]
},
\label{eq:matJaieps}
\end{equation}
where $\tilde{\mathcal{J}}_{\Lambda\mu}^{[J,P,|\lambda|]}(0,\textbf{P}_i-\textbf{P}_f)$ is a current whose quantum numbers and labels are thoroughly defined in Section~\ref{sec:current}. 
$| E_{\Lambda_i,0}\textbf{P}_i;L\rangle$ and $| E_{\Lambda_f,n_f}\textbf{P}_f;L\rangle$ respectively denote the initial and final finite-volume states; the former has the energy and the quantum numbers of a single particle while the latter has that of two particles. 
The subscripts $\Lambda_{i,f}$ indicate that angular-momentum space has been projected onto a particular finite-volume irrep, and $n_f$ is an integer labeling the finite-volume level considered.
Our result relates this finite-volume matrix element to
\begin{eqnarray}
\langle a,{P}_f,J_fm_{J_f};\infty|\tilde{\mathcal{J}}^{[J,P,|\lambda|]}_{\Lambda\mu}(0,\textbf{Q};\infty)| {P}_i;\infty\rangle=
\left[\mathcal{A}_{\Lambda\mu;J_fm_{J_f}}\right]_a~(2\pi)^3~\delta^3(\textbf{P}_i-\textbf{P}_f-\textbf{Q})\,,
\label{eq:JaIV}
\end{eqnarray}
were $a$ is a channel index denoting the two particle flavors in the asymptotic state. In Eq.~\ref{eq:matJaieps}, $\mathcal A$ is understood as a column vector (and $\mathcal A^\dagger$ a row) in the combined angular-momentum/channel space. Finally $\mathcal R_{\Lambda_f,n_f}$, defined in Eq.~\ref{eq:calRdef}, is a matrix in the same space that depends only on the strong-interaction as well as the linear extent of the finite volume.  
This is the coupled-channel and arbitrary-angular-momentum generalization of the LL-factor.

In order to illuminate this result, we apply it to several examples for which the expressions are significantly simplified.
In Section~\ref{sec:Kpipi}, we recover the original $K\rightarrow \pi\pi$ matrix element determined with zero-momentum injection~\cite{Lellouch:2000pv, Kim:2005gf, Christ:2005gi, Lin:2001ek}. In Section~\ref{sec:pigammapipi}, we consider the slightly more complex example of $\pi\gamma^*\rightarrow \pi\pi\rightarrow \rho$.  The degeneracy of the two final-state particles prevents even and odd partial wave mixing, even in boosted systems.
Angular momentum conservation and parity requires the final state to be in a $P$-wave with the leading finite-volume contamination from an $F$-wave.
By neglecting this contamination, we obtain an explicit expression for the P-wave LL-factor for such a system, and find large volume deviation from the well known S-wave result.
For final states with non-degenerate particles, even and odd partial waves will generally mix.
In Section~\ref{sec:2Dcase}, we apply our master equation to systems with coupled channels, whether the mixing is physical or induced by the finite volume.
Finally, in Section~\ref{sec:DpipiKK}, we recover the known result for $D\rightarrow\{\pi\pi,K\bar{K}\}$~\cite{Hansen:2012tf}.

In this work we also include two appendices.  In Appendix~\ref{sec:freepoles} we discuss a technical detail of our derivation, the cancellation of free-poles in integrands of correlation functions. For complete generality, in Appendix~\ref{sec:SpinFVTBCs}, we extend the formalism to include effects of twisted boundary conditions (TBCs)~\cite{PhysRevLett.7.46, Bedaque:2004kc} and volumes that are arbitrary rectangular prisms, using the compact notation of Ref.~\cite{Briceno:2014oea}.

\bigskip
\noindent

\section{Two-point correlation functions\label{sec:corr2}}

In this section we derive expressions for the one-particle and two-particle two-point correlation functions in a finite volume. To achieve this we must first define appropriate interpolating operators. These are most conveniently classified according to the irreducible representations (irreps) of the relevant symmetry group. For a system at rest in a finite cubic volume, the symmetry group is the octahedral group, ${O}_h$. In order to accommodate systems with half-integer spin, one has to consider the double cover of the octahedral group, denoted by ${O}^D_h$~\cite{Johnson:1982yq}. For systems in flight with total momentum $\textbf{P}$, the symmetry is reduced to a subgroup of $O_h$ or $O_h^D$, defined by the subset of octahedral transformations which leave~$\textbf{P}$ invariant. This is referred to as a little group and will be labeled LG(\textbf{P}). 

Let $\varphi_{\Lambda\mu}(x_0,\textbf{P})$ denote a single particle interpolating operator at Euclidean time $x_0$ with momentum $\textbf{P}$ and in row $\mu$ of the $\Lambda$ irrep of LG(\textbf{P}).\footnote{For details regarding the construction of these operators from quark and gluonic degrees of freedom we direct the readers to Refs.~\cite{Moore:2005dw, Basak:2005ir, Moore:2006ng, Bernard:2008ax,  Dudek:2009qf, Dudek:2010wm, Luu:2011ep, Edwards:2011jj, Thomas:2011rh, Gockeler:2012yj} and references therein.} Because $\Lambda\mu$ are good quantum numbers in finite volume, the one-particle two-point functions will not mix states in different rows or irreps
\begin{equation}
C^{(1)}_{\Lambda' \mu', \Lambda \mu}(x_0-y_0,\textbf{k}) \equiv \langle 0|\varphi_{\Lambda'\mu'}(x_0, \textbf{k})
\varphi_{\Lambda\mu}^{\dag}(y_0,- \textbf{k})|0\rangle \propto \delta_{\Lambda', \Lambda} \delta_{\mu',\mu} \,.
\end{equation}

In this study, we will focus on the scenario where the single-particle states are either pseudoscalars or scalars.\footnote{For QCD near the physical point there are no stable scalar particles, only pseudoscalar mesons. At unphysical quark masses, by contrast, one finds stable scalar particles as well. Additionally, although LQCD is the motivation for this work, this formalism is model independent and is relevant for studying hadronic physics as well as atomic physics in a finite volume. (See Refs.~\cite{Drut:2012md, Endres:2012cw} and references within for examples of atomic physics calculations performed in a finite volume.) See Ref.~\cite{Agadjanov:2014kha} for insight into how to deal with states with nonzero spin in the non-relativistic limit.} In such cases there is a single one-dimensional irrep that has overlap with the particle of interest, and the irrep is exclusively specified by its momentum. For example, as shown explicitly in Table~\ref{table:irreps}, the pseudoscalar mesons are in the $\mathbb{A}_1^-$ irrep of \(O_h\) when at rest and in the $\mathbb{A}_2$ irrep of LG(\textbf{k}) when in flight. Therefore, it is sufficient to define the single particle interpolating operators in terms of their momenta and we will drop the $\Lambda\mu$ subscript. We thus introduce 
\begin{eqnarray}
C^{(1)}(x_0-y_0,\textbf{k}) \equiv \langle 0|\varphi(x_0,\textbf{k})
\varphi^{\dag}(y_0,-\textbf{k})|0\rangle
=
e^{-E^{(1)}_{k}(x_0-y_0)}
|\langle 0|\varphi(0,\textbf{k})|E^{(1)} \textbf{k};L\rangle|^2
+\mathcal{O}\left( L^3  \frac{e^{-E^{(1)}_{3,th}(x_0-y_0)}}{E^{(1)}_{3,th}}\right),
\label{eq:corr11}
\end{eqnarray}
where $L$ is the linear extent of the finite cubic spatial volume and $E_{k}^{(1)}, E^{(1)}_{3,th}$ denote the lowest two eigenvalues of the moving-frame Hamiltonian, in the subspace that has overlap with \(\langle 0 \vert \varphi_{\Lambda \mu}(0,\textbf{k})\). We have assumed $x_0>y_0$ to order the operators before inserting a complete set of states. As the subscript suggests, in QCD the first excited energy $E^{(1)}_{3,th}$ corresponds to a state in the vicinity of the three-particle threshold for spinless particles. 

\begin{table}
\begin{center}
\subtable[]{ \label{table:subduca}
\begin{tabular}{c|c|c|c} 
\hspace{.1cm}Group\hspace{.1cm}&
$J^P$& 
 \hspace{.1cm}$\Lambda(\mu)$\hspace{.1cm}& 
\hspace{.1cm}$[{C}^{J}_{\Lambda}]_{\mu,\lambda}~$\hspace{.1cm}\\\hline 
$O_h$&$0^\pm$&$\mathbb{A}_1^\pm(1)$&1\\
$\frac{\textbf{Q}L}{2\pi}=(0,0,0)$
&$1^\pm$&$\mathbb{T}_1^\pm(1)$&$\delta_{1,\lambda}$\\
&$1^\pm$&$\mathbb{T}_1^\pm(2)$&$\delta_{0,\lambda}$\\
&$1^\pm$&$\mathbb{T}_1^\pm(3)$&$\delta_{-1,\lambda}$\\
&$2^\pm$&$\mathbb{T}_2^\pm(1)$&$\delta_{1,\lambda}$\\
&$2^\pm$&$\mathbb{T}_2^\pm(2)$&$(\delta_{2,\lambda}-\delta_{-2,\lambda})/\sqrt{2}$\\
&$2^\pm$&$\mathbb{T}_2^\pm(3)$&$\delta_{-1,\lambda}$\\
&$2^\pm$&$\mathbb{E}^\pm(1)$&$\delta_{0,\lambda}$\\
&$2^\pm$&$\mathbb{E}^\pm(2)$&$(\delta_{2,\lambda}+\delta_{-2,\lambda})/\sqrt{2}$\\ 
\hline 
\end{tabular}

} \subtable[]{ \label{table:subducb}
\begin{tabular}{c|c|c|c} 
\hspace{.1cm}LG(\textbf{Q})\hspace{.1cm}&
$|\lambda|^{\tilde{\eta}}$& 
 \hspace{.1cm}$\Lambda(\mu)$\hspace{.1cm}& 
\hspace{.1cm}$\mathcal{S}^{\tilde{\eta},{\lambda}}_{\Lambda\mu}~$\hspace{.1cm}\\\hline 
Dic$_{4}$&$0^+$&$\mathbb{A}_1(1)$&1\\
$\frac{\textbf{Q}L}{2\pi}=(0,0,n)$&$0^-$&$\mathbb{A}_2(1)$&1\\
  		&$1$&$\mathbb{E}(1)$&$(\delta_{s,+}+\tilde{\eta}\delta_{s,-})/\sqrt{2}$\\
  		&$1$&$\mathbb{E}(2)$&$(\delta_{s,+}-\tilde{\eta}\delta_{s,-})/\sqrt{2}$\\ 
		 &$2$&$\mathbb{B}_1(1)$&$(\delta_{s,+}+\tilde{\eta}\delta_{s,-})/\sqrt{2}$\\
  		&$2$&$\mathbb{B}_2(1)$&$(\delta_{s,+}-\tilde{\eta}\delta_{s,-})/\sqrt{2}$\\ 
		\hline 
Dic$_{2}$&$0^+$&$\mathbb{A}_1(1)$&1\\
$\frac{\textbf{Q}L}{2\pi}=(n,n,0)$&$0^-$&$\mathbb{A}_2(1)$&1\\
  		&$1$&$\mathbb{B}_1(1)$&$(\delta_{s,+}+\tilde{\eta}\delta_{s,-})/\sqrt{2}$\\
  		&$1$&$\mathbb{B}_2(1)$&$(\delta_{s,+}-\tilde{\eta}\delta_{s,-})/\sqrt{2}$\\ 
  		&$2$&$\mathbb{A}_1(1)$&$(\delta_{s,+}+\tilde{\eta}\delta_{s,-})/\sqrt{2}$\\
  		&$2$&$\mathbb{A}_2(1)$&$(\delta_{s,+}-\tilde{\eta}\delta_{s,-})/\sqrt{2}$\\ 
		\hline 
Dic$_{3}$&$0^+$&$\mathbb{A}_1(1)$&1\\
$\frac{\textbf{Q}L}{2\pi}=(n,n,n)$&$0^-$&$\mathbb{A}_2(1)$&1\\
  		&$1$&$\mathbb{E}(1)$&$(\delta_{s,+}+\tilde{\eta}\delta_{s,-})/\sqrt{2}$\\
  		&$1$&$\mathbb{E}(2)$&$(\delta_{s,+}-\tilde{\eta}\delta_{s,-})/\sqrt{2}$\\ 
  		&$2$&$\mathbb{E}(1)$&$(\delta_{s,+}-\tilde{\eta}\delta_{s,-})/\sqrt{2}$\\
  		&$2$&$\mathbb{E}(1)$&$-(\delta_{s,+}-\tilde{\eta}\delta_{s,-})/\sqrt{2}$\\ 
		\hline 
\end{tabular}
}
\caption{ \label{table:irreps} (a)~Shown are the subduction coefficients, $[{C}^{J}_{\Lambda}]_{\mu,\lambda}$ used to project states onto the irreps of $O_h$. (b) Shown are the subduction coefficients determined in Ref.~\cite{Thomas:2011rh}, for $|\lambda| \leq 2$, where $s=sign(\lambda)$ and $\tilde{\eta}=(-1)^{l+J}$ used to project operators onto the irreps of the Dic$_{4}$, Dic$_{2}$, and Dic$_{3}$ groups as shown in Eq.~\ref{eq:projboost}. }
\end{center}
\end{table}

\begin{figure*}[t]
\begin{center}
\subfigure[]{
\label{fig:FVcorr}
\includegraphics[scale=0.45]{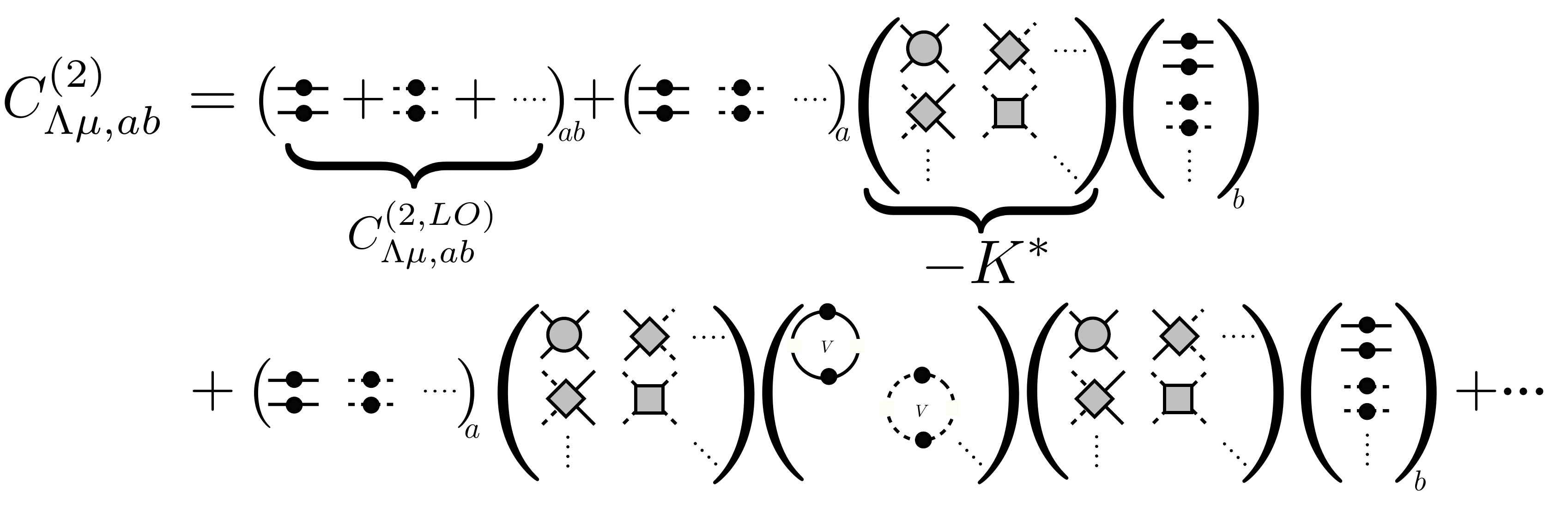}}\\
\subfigure[]{
\label{fig:kernel}
\includegraphics[scale=0.25]{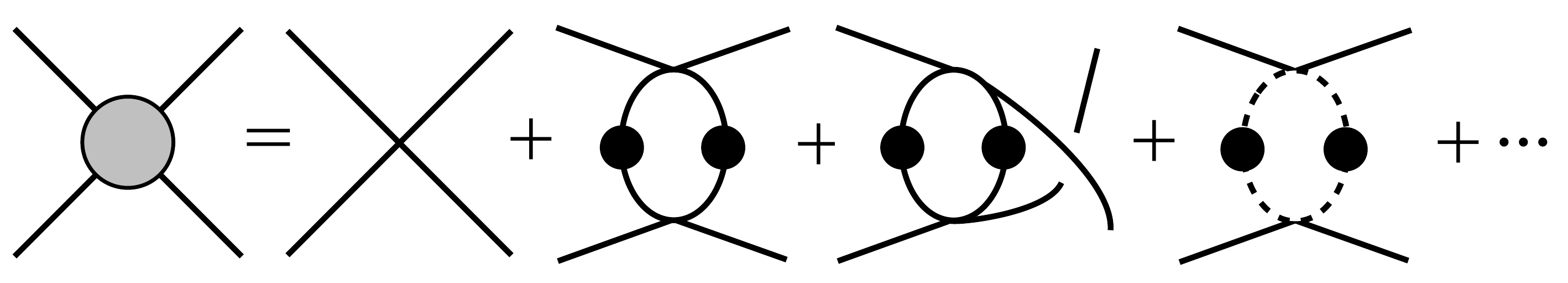}}
\subfigure[]{
\label{fig:1bodyprop}
\includegraphics[scale=0.25]{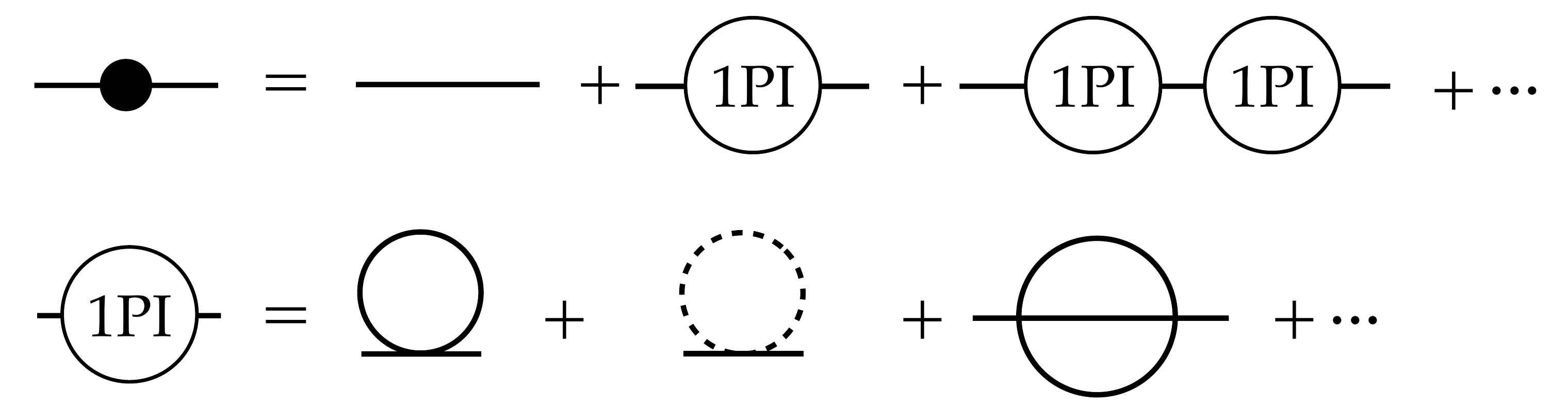}}

\caption{a) Shown is the definition of the finite volume two-particle correlation function. The solid lines denote two-particles in the $``1"$ channel, dashed lines denote particle in the $``2"$ channel. The correlation function is written in terms of the c.m.~kernel, ${K}^*$, and the fully dressed single particle propagators. b) Shown is ${K}^*$ for the first channel, which is the sum of all two-particle irreducible s-channel diagrams. Explicitly shown are examples of diagrams that are included in the kernel: contact interactions, t- and u-channel diagrams. In general, all diagrams allowed by the underlying theory where the intermediate particles cannot all simultaneously go on-shell are absorbed into the kernel. As described in the text, in this study we are restricted to energies where only two-particle states are allowed to go on-shell. c) Shown is the definition of the fully dressed one particle propagator in terms of the one particle irreducible (1PI) diagrams.   }\label{fig:corr2}
\end{center}
\end{figure*}

One can also calculate the correlation function's leading time dependence directly from the fully dressed single particle propagator (see Fig.~\ref{fig:1bodyprop}) 
\begin{eqnarray}
C^{(1)}(x_0-y_0,\textbf{k})&=&L^3
\int \frac{d P_{0}}{2\pi}
~\left(\frac{1}{2\omega_{\textbf{k}}(iP_0+\omega_{\textbf{k}})}+\cdots\right)e^{iP_{0}(x_0-y_0)}\nn\\
&=&
L^3\frac{e^{-\omega_{\textbf{k}}(x_0-y_0)}}{2\omega_{\textbf{k}}}
+\mathcal{O}\left(L^3~\frac{e^{-E^{(1)}_{3,th}(x_0-y_0)}}{E^{(1)}_{3,th}}\right) ,
\label{eq:corr12}
\end{eqnarray}
where $\omega_{\textbf{k}}=\sqrt{m^2+\textbf{k}^2}$, with \(m\) equal to the physical infinite-volume pole mass. In the first line, the ellipses denote corrections that are finite at the single particle pole. This includes terms with poles at higher values of imaginary \(P_0\) which correspond to higher energy states. We emphasize that, in arriving at this identity, we have used the {on-shell} renormalization convention in which the residue of the single particle propagator is set to 1. This convension is equivalently expressed as 
\begin{eqnarray}
\langle 0|\phi(0,\textbf{0})|E^{(1)} \textbf{k};\infty\rangle=1 \,,
\label{eq:infiniteVopnorm}
\end{eqnarray}
 where $\phi(x_0,\textbf{x})$ is the Fourier transform of $\varphi(x_0,\textbf{k})$ and $|E^{(1)} \textbf{k};\infty\rangle$ is the infinite-volume one-particle state with relativistic normalization
\begin{equation}
\langle E^{(1)} \textbf{k}';\infty |E^{(1)} \textbf{k};\infty\rangle = 2 \omega_{\textbf{k}} (2 \pi)^3 \delta^3(\textbf{k}' - \textbf{k}) \,.
\label{eq:infiniteVnorm}
\end{equation}

By comparing Eqs.~\ref{eq:corr11} and \ref{eq:corr12}, we deduce $E_k^{(1)}=\omega_{\textbf{k}}$ and
\begin{eqnarray}
|\langle 0|\varphi(0,\textbf{k})|E^{(1)} \textbf k; L\rangle|=\sqrt{\frac{L^3}{2\omega_{\textbf k}}}.
\label{eq:phiLambda}
\end{eqnarray}
These relations hold up to exponentially suppressed corrections of the form \(e^{-mL}\), which we discuss in more detail below. We stress that Eq.~\ref{eq:phiLambda} is only a statement of renormalization convention on \(\varphi\) together with the normalization convention for finite-volume states
\begin{equation}
\langle E^{(1)} \textbf k; L | E^{(1)} \textbf k; L \rangle = 1\,. 
\end{equation}
As will become evident in Section~\ref{sec:corr3}, the wave-function renormalization does not impact the final result, Eq.~\ref{eq:matJaieps}. Any other choice for the residue would exactly cancel in the ratio used to access finite-volume matrix elements. The motivation for deriving Eq.~\ref{eq:phiLambda} in the manner just presented is that it provides a straightforward warm-up for our analysis of the two-particle two-point correlation function, to which we now turn.

The two-particle correlation function can be determined by considering an alternative energy range and using two- instead of one-particle interpolating fields. For the sake of generality, we consider a system with N coupled two-particle channels. We label the masses in the \(j\)th channel $m_{j,1}$ and $m_{j,2}$, with $m_{j,1}\leq m_{j,2}$. We continue to restrict our attention to spin zero particles. The particles in the $jth$ channel can go on-shell if the c.m.~energy $E^*$ satisfies $m_{j,1}+m_{j,2} \leq E^* < E^*_{th}$. Here $E^*_{th}$ is the energy of the first allowed multi-particle threshold, boosted to the c.m.~frame.\footnote{For a system with a $\mathbb Z_2$ symmetry, such as G-parity for $\pi\pi$ in the isospin limit of QCD, this corresponds to the lowest four-particle threshold. For systems without such symmetries it corresponds to the lowest three-particle threshold.} In practice we must require $ E^* \ll E^*_{th}$, because if $E^*$ is too close to the multi-particle threshold then the neglected exponentially suppressed corrections become enhanced.

 The {on-shell} c.m.~relative momentum for the $jth$ channel satisfies
\begin{eqnarray}
\label{momentumcc}
k^{*2}_{j,on}=\frac{E^{*2}}{4}-\frac{(m_{j,1}^2+m_{j,2}^2)}{2}+\frac{
(m_{j,1}^2-m_{j,2}^2)^2}{4E^{*2}}.~~
\end{eqnarray}
Functions and coordinates evaluated in the c.m.~frame will always have a superscript $``*"$, and it is important to remember that a function $f$ in a moving frame that depends on $k$ can always be related to the c.m.~frame function $f^*$ via $f^*(k^*) \equiv f(k)$. This just defines a coordinate change and does not imply anything about the Lorentz representation of \(f\). Coordinates in the moving frame and c.m.~frame are related by standard Lorentz transformations. For example, if we consider a particle with mass $m$, momenta $\textbf{k}$ and $\textbf{k}^*$ in the moving and c.m.~frames, then
\begin{eqnarray}
\sqrt{m^2+k^{*2}}=\gamma(\sqrt{m^2+k^{2}}-\beta {k}_{||}), \hspace{1cm}
{k}^*_{||}&=&\gamma({k}_{||}-\beta \sqrt{m^2+k^{2}}), \hspace{1cm}
{k}^*_{\perp}={k}_{\perp}, 
\end{eqnarray}
where $\gamma = \frac{E}{E^*}$ and $\beta=\frac{|\textbf{P}|}{E}$.

Two-particle interpolating operators in a given irrep can be written as a linear combination of products of single particle interpolating operators with appropriate Clebsch-Gordan coefficients~\cite{Thomas:2011rh, Dudek:2012gj, Moore:2005dw, Basak:2005ir, Moore:2006ng, Bernard:2008ax,  Dudek:2009qf, Dudek:2010wm,  Edwards:2011jj}. By first considering an energy range where only a single channel is present, one can readily write down the relevant two-body operator
\begin{eqnarray}
\label{eq:twobody_oper}
\mathcal{O}_{\Lambda\mu}(x_0,\textbf{P},|\textbf{P}-\textbf{k}|,|\textbf{k}|) = 
\sum_{R \in {\rm LG(\textbf{P})}} \mathcal{C}(\textbf{P}\Lambda\mu; R \textbf{k}; R (\textbf{P}-\textbf{k}))  
\varphi({x_0,  R \textbf{k}}) {\tilde \varphi}({x_0,   R (\textbf{P}-\textbf{k})} ),
\end{eqnarray}
where in general $\varphi$ and $\tilde \varphi$ may be identical or non-identical operators and $R$ is understood as an element of the representation of LG(\textbf{P}) defined by action on three-dimensional spatial vectors. In order to minimize unnecessary notation, we will suppress the dependence of \(\mathcal O\) on $|\textbf{P}-\textbf{k}|$ and $|\textbf{k}|$ from now on.\footnote{Throughout this work $\mathcal{O}$ will denote an operator that has overlap with a two-particle state and $\varphi$ will refer to a single particle operator. Of course in general these must couple to all states with the appropriate quantum numbers.} 

To completely specify the Clebsch-Gordan coefficients, we now introduce \(\{\textbf k\}_\Pstar\) as the set of all momenta that are reached by applying a rotation in LG(\textbf P) to \(\textbf k\). We then denote the irreps of particles one and two by $\Lambda_1(\{\textbf{P}-\textbf{k}\}_\Pstar)$ and $\Lambda_2(\{\textbf{k}\}_\Pstar)$ respectively, and define the Clebsch-Gordan coefficient, $\mathcal{C}(\textbf{P}\Lambda\mu; R \textbf{k}; R (\textbf{P}-\textbf{k}))  $, to project the two particles in $\Lambda_1(\{\textbf{P}-\textbf{k}\}_\Pstar)\otimes\Lambda_2(\{\textbf{k}\}_\Pstar)$ onto $\Lambda(\textbf{P}),\mu$. This may also be expressed as an innerproduct of states
\begin{equation}
\mathcal{C}(\textbf{P}\Lambda\mu; R \textbf{k}; R (\textbf{P}-\textbf{k})) \equiv \big \langle \Lambda(\textbf P), \mu \big \vert \Lambda_1(\{\textbf{P}-\textbf{k}\}_\Pstar), R (\textbf P - \textbf k) ; \Lambda_2(\{\textbf{k}\}_\Pstar), R \textbf k \big \rangle\,,
\end{equation}
from which follows
\begin{equation}
\sum_{R \in {\rm LG(\textbf{P})}} \vert \mathcal{C}(\textbf{P}\Lambda\mu; R \textbf{k}; R (\textbf{P}-\textbf{k})) \vert ^2 = 1 \,.
\end{equation}

The simplest nontrivial example of this operator construction is reached by setting the total momentum to zero, setting $\textbf{k}=\frac{2\pi}{L}\hat{\textbf{k}}\equiv q_{(1)}\hat{\textbf{k}}$, and taking the two-particle operator to be in the $\mathbb{A}_1^+$ irrep
\begin{eqnarray}
\label{eq:twobody_operA1p}
\mathcal{O}_{\mathbb{A}_1^+}(x_0,\textbf{0}) &=&
\frac{\sigma}{\sqrt{6}}\left[
\varphi(x_0,  q_{(1)}\hat{\textbf{z}} ) \tilde{\varphi}(x_0,   -q_{(1)}\hat{\textbf{z}} ) 
+\varphi(x_0,  -q_{(1)}\hat{\textbf{z}} ) \tilde{\varphi}(x_0,   q_{(1)}\hat{\textbf{z}} ) 
+\varphi(x_0,  q_{(1)}\hat{\textbf{x}} ) \tilde{\varphi}(x_0,   -q_{(1)}\hat{\textbf{x}} ) \right.\nn\\
&+&\left.\varphi(x_0,  -q_{(1)}\hat{\textbf{x}} ) \tilde{\varphi}(x_0,   q_{(1)}\hat{\textbf{x}} ) 
+\varphi(x_0,  q_{(1)}\hat{\textbf{y}} ) \tilde{\varphi}(x_0,   -q_{(1)}\hat{\textbf{y}} ) 
+\varphi(x_0,  -q_{(1)}  \hat{\textbf{y}}) \tilde{\varphi}(x_0,   q_{(1)}\hat{\textbf{y}} ) \right],
\end{eqnarray}
where $\sigma = \sqrt{1/2}$ if ${\varphi}$ and $\tilde{\varphi}$ are the same operators and $\sigma=1$ otherwise. If we give the system a nonzero boost along $\hat{\textbf{z}}$, then the symmetry group is reduced to LG($\hat{\textbf{z}}$). Consider the scenario where the momentum of the $\varphi$ field has magnitude $q_{(1)}$ and that of $\tilde{\varphi}$ has magnitude $\sqrt{2}q_{(1)}$. With these single-particle operators, we can construct a two-particle operator that transforms in the $\mathbb{A}_1$ irrep~\cite{Dudek:2012gj}
\begin{eqnarray}
\label{eq:twobody_operA1}
\mathcal{O}_{\mathbb{A}_1}(x_0,q_{(1)}\hat{\textbf{z}}) &=&
\frac{1}{2}\left[
\varphi(x_0,  q_{(1)}\hat{\textbf{x}} ) 
\tilde{\varphi}(x_0,   
q_{(1)}
( \hat{\textbf{z}}- \hat{\textbf{x}}))
+\varphi(x_0,  q_{(1)}\hat{\textbf{y}} ) 
\tilde{\varphi}(x_0,   q_{(1)}
( \hat{\textbf{z}}- \hat{\textbf{y}}))
 \right.\nn\\
&+&\left.\varphi(x_0,  -q_{(1)}\hat{\textbf{x}} )
 \tilde{\varphi}(x_0,   q_{(1)}
( \hat{\textbf{z}}+ \hat{\textbf{x}}))
+\varphi(x_0, - q_{(1)}\hat{\textbf{y}} ) 
\tilde{\varphi}(x_0,   q_{(1)}
( \hat{\textbf{z}}+ \hat{\textbf{y}}))
\right].
\end{eqnarray} 

In general, there might be $N$ open channels contributing to a given state. For example, an infinite volume $\pi\pi$ state can mix with a $K\bar{K}$ state, and both must thus have nonzero overlap with the corresponding finite-volume state.\footnote{We note that for physical particle masses the $K \bar K$ threshold exceeds the four pion threshold. Since coupling to this state is ignored in the present formalism, application to $\pi \pi$, $K \bar K$ is only valid for unphysical heavy pions and will otherwise introduce systematic uncertainties on extracted quantities.} It is convenient to introduce an index, e.g. ``$a$'', to the interpolating operator in Eq.~\ref{eq:twobody_oper} to indicate the infinite-volume channel that it interpolates
\begin{eqnarray}
\label{eq:twobody_oper_N}
\mathcal{O}_{\Lambda\mu}({x_0,\textbf{P}})
\longrightarrow
\mathcal{O}_{\Lambda\mu,a}({x_0,\textbf{P}}).
\end{eqnarray}
For example, $\mathcal{O}_{\Lambda\mu,a}$ could refer to a $\pi\pi$-like or a $K\bar{K}$-like operator. 
With this, we can write a generic correlation function for a two-particle system that has been projected onto a given irrep as
\begin{eqnarray}
C^{(2)}_{\Lambda\mu,ab}(x_0-y_0,\textbf{P})&=&\langle 0|\mathcal{O}_{\Lambda'\mu',a}(x_0,\textbf{P})
\mathcal{O}^{\dag}_{\Lambda\mu,b}(y_0,-\textbf{P})|0\rangle\nn\\
&&\hspace{-2cm}=\delta_{\Lambda,\Lambda'}\delta_{\mu,\mu'}\sum_{n}e^{-E_{\Lambda,n}(x_0-y_0)}
\langle 0|\mathcal{O}_{\Lambda\mu,a}(0,\textbf{P})|E_{\Lambda,n}\textbf{P};L\rangle
\langle E_{\Lambda,n}\textbf{P};L|\mathcal{O}^\dag_{\Lambda\mu,b}(0,-\textbf{P})|0\rangle
+\mathcal{O}\left(L^6\frac{e^{-E_{th}(x_0-y_0)}}{E^2_{th}}\right) ,~~~~
\label{eq:corr21}
\end{eqnarray}
where $E_{\Lambda,n}$ is the $nth$ two-particle eigenenergy of the $\Lambda$-irrep of LG(\textbf{P}). This is the two-body analog of Eq.~\ref{eq:corr11}. In general we expect multiple two-body states below the first multi-particle threshold, $E_{th}$, and hence include a sum over $n$.  

The correlation function can also be written in terms of the interactions of the two-particle system. The leading order (LO) contribution to the correlation function (first diagram in Fig.~\ref{fig:FVcorr}) is determined by considering the limit in which the interactions vanish, and as a result the different channels cannot mix. We find
\begin{equation}
\label{eq:corr2LO}
C^{(2,LO)}_{\Lambda\mu,ab}(x_0-y_0,\textbf{P}) = L^6 \int \frac{d P_{0}}{2\pi}e^{iP_{0}(x_0-y_0)} \widetilde C^{(2,LO)}_{\Lambda\mu,ab}(P_0,\textbf P) \,,
\end{equation}
where
\begin{equation}
\widetilde C^{(2,LO)}_{\Lambda\mu,ab}(P_0,\textbf P) \equiv \delta_{ab}\frac{1}{\eta}    \int \frac{d k_{0}}{2\pi}  \sum_{R \in LG(\textbf{P})}  \mathcal{C}(\textbf{P}\Lambda\mu; R \textbf{k}; R (\textbf{P}-\textbf{k})) G(k) G(P-k) \mathcal{C^*}(\textbf{P}\Lambda\mu; R \textbf{k}; R (\textbf{P}-\textbf{k})) \,.
\end{equation}
Here we have introduced the fully dressed propagator
\begin{equation}
G(k) \equiv \int d^4 x e^{-ikx} \langle 0 \vert T \phi(x) \phi^\dagger (0) \vert 0 \rangle \,,
\end{equation}
with on-shell renormalization \(\lim_{k_0 \rightarrow i \omega_k} (k^2 + m^2) G(k) = 1 \). We have also introduced the symmetry factor $\eta$ which is equal to 1/2 if the particles identical and have momenta that are related by LG(\textbf P) rotations, and equal to 1 otherwise.

Observe here that \(G(k)\) is the infinite-volume fully dressed propagator. Really \(\widetilde C^{(2,LO)}_{\Lambda\mu,ab}(P_0,\textbf P)\) should be constructed from the finite-volume analog of \(G(k)\). However, as long as \([P_0^2 + \textbf P^2]^{1/2}\) has an imaginary part with magnitude below \(E^*_{th}\), then using the infinite-volume propagator only incurs exponentially suppressed corrections of the form \(e^{- m_\pi L}\), with \(m_\pi\) the lightest mass in the spectrum. This is discussed in more detail in the context of the Bethe-Salpeter kernel below. We deduce that our expression for \(\widetilde C^{(2,LO)}_{\Lambda\mu,ab}(P_0,\textbf P)\) is only valid in a strip of the complex \(P_0\) plane which runs along the real axis and is bounded by \([P_0^2 + \textbf P^2] = - E^{*2}_{th}\).

We now complete the analysis of \(C^{(2,LO)}_{\Lambda\mu,ab}(x_0-y_0,\textbf{P})\), by evaluating the \(k_0\) and \(k_0'\) integrals. First we define
\begin{equation}
\omega_{j,1}\equiv\sqrt{m_{j,1}^2+(\textbf{P}-\textbf{k})^2}\,, \ \  \omega_{j,2}\equiv\sqrt{m_{j,2}^2+\textbf{k}^2} \,.
\end{equation}
In performing the \(k_0\) and \(k_0'\) integrals we encircle the pole at \(i\omega_{j,2}\) and this fixes the $``2"$ particle in the $jth$ channel to be on-shell with free energy $\omega_{j,2}$. By energy conservation, the $``1"$ particle will have energy $-iP_0-\omega_{j,2}$. Specifically we find
\begin{eqnarray}
\label{eq:corr2LO}
C^{(2,LO)}_{\Lambda\mu,ab}(x_0-y_0,\textbf{P})&=& \delta_{ab}\frac{L^6}{\eta}\int \frac{d P_{0}}{2\pi}e^{iP_{0}(x_0-y_0)}
\sum_{R \in LG(\textbf{P})} 
~\frac{
|\mathcal{C}(\textbf{P}\Lambda\mu; R \textbf{k}; R (\textbf{P}-\textbf{k}))  |^2}{4~\omega_{a,1}~\omega_{a,2}(iP_0+(\omega_{a,1}+\omega_{a,2}))}+\mathcal{O}\left(L^6\frac{e^{-E_{th}(x_0-y_0)}}{E^2_{th}}\right) \,.~~~~~~
\end{eqnarray}
Note here that the first term gives a pole in the \(P_0\) plane that sits in the region where our expression for \(\widetilde C^{(2,LO)}_{\Lambda\mu,ab}(P_0,\textbf P)\) is valid. We do not control the exact form of the second term, which is an exponential that decays according to some above-threshold energy. The precise form of the above threshold term is not needed for our final result.

To include higher orders we need only use the fact that the correlation function, defined in Eq.~\ref{eq:corr21}, is correctly reproduced by the all-orders summation of a skeleton expansion built from Bethe-Salpeter kernels and fully dressed propagators. In particular we define the next-to-leading-order (NLO) correlator as the contribution built from a single insertion of the Bethe-Salpeter kernel, $K$. The kernel is depicted in Fig.~\ref{fig:kernel} and is defined as the sum of all amputated four-point diagrams that are two-particle irreducible in the s-channel. We find
\begin{equation}
C^{(2,NLO)}_{\Lambda\mu,ab}(x_0-y_0,\textbf{P})  = L^6  \int \frac{d P_0}{2 \pi} e^{i P_0(x_0-y_0)} \widetilde C^{(2,NLO)}_{\Lambda\mu,ab}(P_0,\textbf{P})  \,, 
\end{equation}
where
\begin{equation}
\begin{split}
\label{eq:NLOdef}
\widetilde C^{(2,NLO)}_{\Lambda\mu,ab}(P_0,\textbf{P}) & = - \frac{1}{L^3} 
\sum_{R,R' \in {\rm LG(\textbf{P})}}
  \mathcal{C}(\textbf{P}\Lambda\mu; R' \textbf{k}; R' (\textbf{P}-\textbf{k})) \\
& \hspace{-30pt} \times  \int \frac{d k_0'}{2 \pi}  \int \frac{d k_0}{2 \pi}   G(k') G(P-k') K(P,k,k')  G(k) G(P-k)\mathcal{C^*}(\textbf{P}\Lambda\mu; R \textbf{k}; R (\textbf{P}-\textbf{k})) \,. 
\end{split}
\end{equation}
In general, the kernel is a function of volume, but since the c.m.~energy is restricted to satisfy $m_{1}+m_{2}~\leq E^*\ll  E^*_{th}$ the intermediate particles appearing in the kernel cannot all simultaneously go on-shell. This implies that the summands appearing in diagrams are smooth functions of summed momenta. Therefore one can show using Poisson's resummation formula  
\begin{eqnarray}
\label{poisson}
\left[~\frac{1}{L^3}\sum_{\mathbf{q}}\hspace{-.5cm}\int~\right] f(\textbf{q})=  
\sum_{\textbf{n}\neq 0}    \int \frac{d\mathbf{q}}{(2\pi)^3} f(\textbf{q})~{e^{iL \mathbf{n}\cdot \mathbf{q}}}, \nn
\end{eqnarray}
that the difference between finite- and infinite-volume kernels is exponentially small in \(m_\pi L\). In writing the Poisson resummation formula the following notation has been introduced
\begin{eqnarray}
\left[~\frac{1}{L^3}\sum_{\mathbf{q}}\hspace{-.5cm}\int~\right]\equiv\left(\frac{1}{L^3}\sum_{\mathbf{q}}-\int\frac{d\textbf{q}}{(2\pi)^3}\right).
\end{eqnarray}
Since we neglect these corrections, the result discussed here holds only for $m_\pi L\gg 1$. We will neglect any terms in the correlation function that are exponentially suppressed with the mass of any particle in any coupled channel since $\mathcal{O}(e^{-m_{i} L})\leq \mathcal{O}(e^{-m_\pi L})$. These corrections have been previously determine for $\pi\pi$~\cite{Bedaque:2006yi} and $NN$ systems~\cite{Sato:2007ms} in an S-wave, as well as the $\pi\pi$ system in a P-wave in Ref.~\cite{Chen:2012rp, Albaladejo:2013bra}.

 Higher order contributions to the correlation function can be readily evaluated by  making the following replacement 
\begin{equation}
- [K(P,k,k')]_{a,b} \longrightarrow - [\mathbb T_L(P,k,k')]_{a,b} \,,
\end{equation}
where
\begin{equation}
- [\mathbb T_L(P,k,k')]_{a,b} = - [K(P,k,k')]_{a,b} +  \int\frac{d l_{0}}{2\pi}\frac{\xi_j}{L^3}\sum_{\textbf{l}} [K(P,k,l)]_{a,j} G_j(l) G_j(P-l) [\mathbb T_L (P,l,k')]_{j,b} \,,
\label{eq:TLk}
\end{equation} 
and the summation over the intermediate channel $j$ is implicit. 
 
A convenient expression for \(\mathbb T_L\) can be found utilizing the machinery developed by Kim, Sachrajda, and Sharpe~\cite{Kim:2005gf}. In order to determine the finite-volume corrections to the correlation function, it is sufficient to know the difference between the finite-volume momentum sum and the infinite-volume momentum integral acting on the two-particle poles. Using a principal-value prescription to define the integral at the pole, we observe
 \begin{eqnarray}
\xi_j\left[~\frac{1}{L^3}\sum_{\mathbf{l}}\hspace{-.5cm}\int~\right]~
\text{P.V.}\frac{  [K(P,k,l)]_{a,j}[K(P,l,k')]_{j,b}}{4~\omega_{1,\textbf{P}-\textbf{l}}~\omega_{2,\textbf{l}}(\omega_{1,\textbf{P}-\textbf{l}}+\omega_{2,\textbf{l}}-P_{0,M})}
 &\equiv&- [K^*_{off,on}\mathbb{F}^V K^*_{on,off}]_{a,b}+\mathcal{O}(e^{-m_\pi L}),
 \label{eq:FVsum}
 \end{eqnarray}
where the c.m.~kernel, $K^*_{off,on}$, is the kernel for a system where the two incoming particles are evaluated {on-shell}, while the outgoing particles may in general be \emph{off-shell}. Here we have also introduced the Minkowski energy $P_{0,M}\equiv-iP_{0}$. Note, if one chooses to use an $i\epsilon$ prescription for the propagator, this would lead to a second contribution to the right-hand side of Eq.~\ref{eq:FVsum}, due to the residue of the infinite-volume integral on the left hand side.

In writing the right-hand side of Eq.~\ref{eq:FVsum}, the kernels and the finite volume function have been written as matrices over angular momentum. The matrix elements of $\mathbb{F}^V$ in the spherical harmonic basis are found to be~\cite{Kim:2005gf,Hansen:2012tf, Briceno:2012yi}         
\begin{eqnarray}
&& \left[\mathbb{F}^V_j\right]_{lm_l;l'm_{l'}}=-\frac{\xi_j}{8\pi P_{0,M}^*}\left[\sum_{l'',m''}\frac{(4\pi)^{3/2}}{k^{*{l''}}_{j,on}}c_{l''m''}^{\mathbf{d}}(k^{*2}_{j,on};{L}) 
 \int d\Omega~Y^*_{ lm_l}Y^*_{l''m''}Y_{l'm_{l'}}\right].
\label{eq:deltaGPBCs}
\end{eqnarray}
The function $c^{\textbf{d}}_{lm}$ is defined as
\begin{eqnarray}
\label{eq:clm}
c^\mathbf{d}_{lm}(k^{*2}_j; {L})
=\frac{\sqrt{4\pi}}{\gamma L^3}\left(\frac{2\pi}{L}\right)^{l-2}\mathcal{Z}^\mathbf{d}_{lm}[1;(k^*_j {L}/2\pi)^2],
\hspace{1cm}
\mathcal{Z}^\mathbf{d}_{lm}[s;x^2]
= \sum_{\mathbf r \in \mathcal{P}_{\mathbf{d}}}\frac{|\mathbf{r}|^lY_{lm}(\mathbf{r})}{(r^2-x^2)^s},\label{eq:clm}
\end{eqnarray} 
where $\gamma=P_{0,M}/P_{0,M}^*$, the sum is performed over $\mathcal{P}_{\mathbf{d}}=\left\{\mathbf{r}\in \textbf{R}^3\hspace{.1cm} | \hspace{.1cm}\mathbf{r}={\hat{\gamma}}^{-1}(\mathbf m-\alpha_j \mathbf d) \right\}$, $\textbf{m}$ is a triplet integer, $\mathbf d$ is the normalized boost vector $\mathbf d=\mathbf{P}L/2\pi$, $\alpha_j=\frac{1}{2}\left[1+\frac{m_{j,1}^2-m_{j,2}^2}{P_{0,M}^{*2}}\right]$~\cite{Davoudi:2011md, Fu:2011xz, Leskovec:2012gb}, and $\hat{\gamma}^{-1}\textbf{x}\equiv{\gamma}^{-1}\textbf{x}_{||}+\textbf{x}_{\perp}$,  with $\textbf{x}_{||}(\textbf{x}_{\perp})$ denoting the $\textbf{x}$ component that is parallel(perpendicular) to the total momentum, $\textbf{P}$. In Appendix~\ref{sec:SpinFVTBCs} we give the generalization of this for asymmetric volumes with twisted boundary conditions. 

We mention a subtlety here with the definition of $c_{lm}^{\textbf d}(k^{*2}_j;L)$ for $k^{*2}_j<0$. The definitions given above continue to hold for subthreshold momenta, but only if the appropriate analytic continuation is implemented. To understand this in detail we first observe that the sum defining $\mathcal Z^{\textbf d}_{lm}$ diverges for $s<3/2+l/2$ and in particular diverges for $s=1$. The function $\mathcal Z^{\textbf d}_{lm}$ is thus understood to be defined via analytic continuation from $s>3/2+l/2$. To make this definition more apparent in the present context we give the equivalent form from Kim, Sachrajda and Sharpe\footnote{Our definition of $c_{lm}$ differs from Ref.~\cite{Kim:2005gf} by an overall sign.}
\begin{equation}
c^\mathbf{d}_{lm}(k^{*2}_{j,on}; {L}) = - \frac{1}{\gamma L^3} \sum_{\textbf k^*} \frac{\exp[\alpha(k^{*2}_{j,on}-k^{*2})]}{k^{*2}_{j,on}-k^{*2}} k^{*l} \sqrt{4 \pi} Y_{lm}\big(\hat{\textbf k}^* \big) + \delta_{l0}~\mathrm{P.V.}\! \int \frac{d \textbf k^*}{(2 \pi)^3}\frac{\exp[\alpha(k^{*2}_{j,on}-k^{*2})]}{k^{*2}_{j,on}-k^{*2}} \,,
\end{equation}
where the sum is over all $\textbf k^* \in (2 \pi/L) \mathcal P_{\textbf d}$ and the limit $\alpha \rightarrow 0^+$ is understood. This definition makes the ultraviolet regularization, which is implicit in the analytic continuation in $s$, more explicit. For continuation to $k^{*2}_{j,on}<0$ it is convenient to rewrite the integral as an $i \epsilon$ prescription and a remainder
\begin{align}
\mathrm{P.V.}\! \int \frac{d \textbf k^*}{(2 \pi)^3}\frac{\exp[\alpha(k^{*2}_{j,on}-k^{*2})]}{k^{*2}_{j,on}-k^{*2} } &=  \int \frac{d \textbf k^*}{(2 \pi)^3}\frac{\exp[\alpha(k^{*2}_{j,on}-k^{*2})]}{k^{*2}_{j,on}-k^{*2} + i \epsilon } + \frac{i k^*_{j,on}}{4 \pi} \,.
\end{align}
The subthreshold continuation of the left hand ride is defined as the following limit of the right-hand side
\begin{equation}
\lim_{k^*_{j,on} \rightarrow i \kappa_j}\left[ \int \frac{d \textbf k^*}{(2 \pi)^3}\frac{\exp[\alpha(k^{*2}_{j,on}-k^{*2})]}{k^{*2}_{j,on}-k^{*2} + i \epsilon } + \frac{i k^*_{j,on}}{4 \pi}\right]
=-\left[ \int \frac{d \textbf k^*}{(2 \pi)^3}\frac{\exp[-\alpha(\kappa_j^2+k^{*2})]}{\kappa_j^2+k^{*2}  } + \frac{\kappa_j}{4 \pi}\right]\,,
\end{equation}
where $\kappa_j$ is the binding momentum of the $jth$ channel.

We next turn to the Bethe-Salpeter kernel which, like $\mathbb F^V$, can be expressed as a matrix in angular momentum
\begin{equation}
K_{off,off}^*(P_{0}^*, \textbf k^*_i, \textbf k^*_f)=4\pi
\sum_{{l,m_{l},l',m_{l'}}}
 Y_{lm_{l}}(\hat{\textbf{k}}^*_f)
Y^*_{l'm_{l'}}(\hat{\textbf{k}}^*_i) ~[K_{off,off}^*(P_0^*,k^*_i,k^*_j)]_{lm_l,l'm_{l'}}.
\label{MJMlsbasis}
\end{equation} 
Here we consider a kernel in which both the initial and final states are off-shell. More precisely, we assume $k_{i,0} = i \omega_{\textbf k_i}$ and $k_{f,0} = i \omega_{\textbf k_f}$, but no additional constraints. These relations, which arise from contour integration as discussed, do not give on-shell two-particle states since \(P_0-k_{0,i} \neq i \omega_{\textbf P - \textbf k_{i}}\) and \(P_0-k_{0,f} \neq i \omega_{\textbf P - \textbf k_{f}}\). Nevertheless, it is still possible to change to the c.m.~frame, expressing the kernel in terms of \((P_{0}^*, \textbf k^*_i, \textbf k^*_f)\) as indicated above. Note that the matrix defined in Eq.~\ref{MJMlsbasis} is diagonal,
\begin{equation}
[K_{off,off}^*(P_0^*,k^*_i,k^*_j)]_{lm_l,l'm_{l'}} \propto \delta_{l,l'} \delta_{m_l,m_l'}\,.
\end{equation}
 This follows from the rotational invariance of the infinite-volume theory, equivalently from the fact that the only angular dependence in the c.m.~frame is $\hat{\textbf{k}}_i^*\cdot\hat{\textbf{k}}^*_f$. Finally, we comment that the on-shell kernel is accessed by constraining the three momenta magnitudes to \(k^*_i=k^*_f=k^*_{on}\). We return to this discussion in the context of the quantization condition below.

\begin{figure*}[t]
\begin{center}\subfigure[]{ 
\includegraphics[scale=0.45]{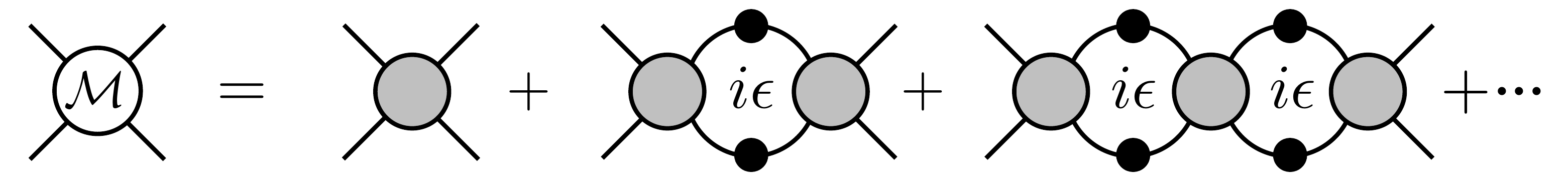}
\label{fig:M_mat}}
\subfigure[]{ 
\includegraphics[scale=0.45]{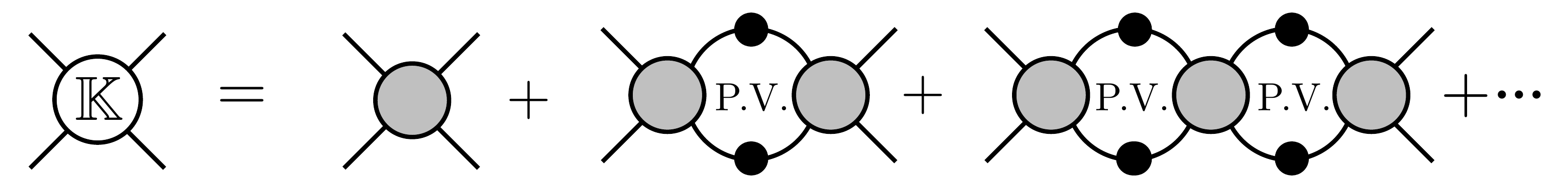}
\label{fig:K_mat}}\\
\caption{ 
 In order to illustrate the differences and similarities between a) the scattering amplitude, $\mathcal{M}$, and b) the K-matrix, $\mathbb{K}$, we show their diagrammatic representation for the single-channel case in terms of the kernel (defined in Fig.~\ref{fig:kernel}) and infinite volume loops. The infinite volume loops of the scattering amplitude are evaluated using the $i\epsilon$ prescription, while those of the K-matrix are evaluated using the principal value, as shown. For multichannel scenarios one simply upgrades the kernels and two-particle loops to be matrices in the number of open channels as depicted in Fig.~\ref{fig:corr2}. Note that the single particle propagators are fully dressed as defined in Fig.~\ref{fig:1bodyprop}.
}\label{fig:MK_mat}
\end{center}
\end{figure*}

Directly following Kim, Sachrajda and Sharpe by summing over all possible insertions of the Bethe-Salpeter kernel, we find 
\begin{equation}
\label{eq:TLdef}
- \mathbb T_L =  \mathbb K_{off,off} - \mathbb K_{off,on} \frac{1}{1+ \mathbb{F}^V \mathbb K} \mathbb{F}^V \mathbb K_{on,off} \,.
\end{equation}
Here we have introduced the two-to-two K-matrix, which is defined as the sum of all infinite-volume, amputated $\textbf{2}\rightarrow \textbf{2}$ diagrams with loop integrals defined via principal-value prescription\footnote{The use of pole prescription here is somewhat subtle. If we restrict the Euclidean-signature time component \(P_0\) to be real, then no pole prescription is needed. However if \(P_0\) is imaginary and thus \(P_{0,M}\) is real, then poles appear in the region of integration. Our definition requires always performing time component integrals with \(P_{0,M}\) off the real axis, as in the standard \(i \epsilon\) prescription. This produces integrals over spatial components of the form of Eq.~\ref{eq:FVsum}. These are always to be evaluated with real \(P_{0,M}\) and with the principal value pole prescription. Alternatively one may use the \(i \epsilon\) prescription for both the time-component and spatial-component parts of each two-particle loop integral, but then one must take only the real part.}
\begin{equation}
 [\mathbb K(P,k,k')]_{a,b} \equiv - [K(P,k,k')]_{a,b} - \xi_j \,  \text{P.V.} \! \int \! \! \frac{d \textbf l}{(2 \pi)^3} \int \frac{d l_{0}}{2\pi}    [K(P,k,l)]_{a,j} G_j(l) G_j(P-l) [\mathbb K (P,l,k')]_{j,b} \,.
 \label{eq:Kmatrix}
\end{equation}
This object is explicitly shown in Fig.~\ref{fig:K_mat} for a single channel scenario. Observe that in Eq.~\ref{eq:TLdef} we have given subscripts on \(\mathbb K\) to indicate whether the incoming and outgoing states are on or off shell. $ \mathbb K$ with no subscript is reserved for the on-shell K-matrix.

We contrast the K-matrix to the scattering amplitude, $\mathcal{M}$, which is defined as the sum of all infinite-volume, amputated $\textbf{2}\rightarrow \textbf{2}$ diagrams with integration defined via $i\epsilon$ prescription (as shown in Fig.~\ref{fig:M_mat} for a single channel)
\begin{equation}
[\mathcal M(P,k,k')]_{a,b} \equiv - [K(P,k,k')]_{a,b} - \xi_j    \! \int \! \! \frac{d^4 l}{(2 \pi)^4}     [K(P,k,l)]_{a,j} G_j(l) G_j(P-l) [\mathcal M (P,l,k')]_{j,b} \,.
\end{equation}
The on-shell K-matrix can be directly related to the on-shell scattering amplitude by introducing a kinematic matrix that is diagonal over the N open channels $\mathbb{P}=\text{diag}(\sqrt{\xi_1q^{*}_1},\sqrt{\xi_2q^{*}_2},\ldots,\sqrt{\xi_Nq^{*}_N})/\sqrt{4\pi E^*}$. For a system with angular momentum $J=l=l'$, the amplitudes $\mathcal{M}_{J}$ and $\mathbb{K}_{J}$ are related via~\cite{Hansen:2012tf},
\begin{eqnarray}
\label{eq:MKrel}
\mathcal{M}_{J}^{-1}=\mathbb{K}_{J}^{-1}-i\mathbb{P}^{2}/2,\label{MMtildematrix}
\end{eqnarray}
and the scattering amplitude and the $S$-matrix via
\begin{eqnarray}
i\mathcal{M}_{J}=\mathbb{P}^{-1}~{(S_{J}-\mathbb{I})}~\mathbb{P}^{-1}.\label{Smatrix}
\end{eqnarray}

Substituting \(\mathbb T_L\) for \(K\) in Eq.~\ref{eq:NLOdef} gives the full correlation function
\begin{eqnarray}
C^{(2)}_{\Lambda\mu,ab}(x_0-y_0,\textbf{P})&=&
L^6\int \frac{d P_{0}}{2\pi}e^{iP_{0}(x_0-y_0)} \left \{ \widetilde C^{[2, \mathcal O(\mathbb K)]}_{\Lambda\mu,ab}(P_0,\textbf{P})  -
\mathbb{Y}_{\Lambda \mu} \left[
\frac{1}{L^3}~\frac{1}{\mathbb{K}+\left(\mathbb{F}^{V}\right)^{-1}}~\right]_{ab} \mathbb{Y}^\dag_{\Lambda \mu}\nn \right \}\\
\label{eq:corrP0}&&\hspace{9cm}+\mathcal{O}\left(L^6{e^{-E_{th}(x_0-y_0)}}/{E^2_{th}}\right).
\end{eqnarray}
The first term of the integrand is defined as
\begin{align}
 \widetilde C^{[2, \mathcal O(\mathbb K)]}_{\Lambda\mu,ab}(P_0,\textbf{P}) & \equiv \widetilde C^{(2,LO)}_{\Lambda\mu,ab}(P_0,\textbf{P}) +  \frac{1}{L^3} 
 \sum_{R,R' \in {\rm LG(\textbf{P})}}  \mathcal{C}(\textbf{P}\Lambda\mu; R' \textbf{k}; R' (\textbf{P}-\textbf{k})) \\
& \hspace{-20pt} \times  \int \frac{d k_0'}{2 \pi}  \int \frac{d k_0}{2 \pi}   G(k') G(P-k') \mathbb K(P,k,k')  G(k) G(P-k)\mathcal{C^*}(\textbf{P}\Lambda\mu; R \textbf{k}; R (\textbf{P}-\textbf{k})) \,.
\end{align}
We have also introduced new notation for the second term
\begin{multline}
\label{eq:Ymatrix}
[\mathbb Y_{\Lambda \mu}]_{l,m} \sqrt{4 \pi} Y_{lm}(\hat k^*) \equiv \sum_{R \in \mathrm{LG}(\textbf P)} C(\textbf P \Lambda \mu, R(\textbf P - \textbf k'), R \textbf k') \int \frac{d k_0'}{2 \pi} G(k_0',R \textbf k') G(P_0 - k_0', R(\textbf P - \textbf k')) \mathbb K(P,k, k' ) \,.
\end{multline}

We stress that $\mathbb Y$ depends on {\em off-shell} K-matrices. This dependence is unavoidable in the two-particle correlation function and will persist in our final result. However, we will see that the off-shell contributions cancel when we consider the ratio of correlation functions that is needed to isolate the matrix-element of an external current between finite-volume energy eigenstates.


In order to evaluate the integral over $P_0$ we first note that the free poles of the integrand exactly cancel. This is a nontrivial observation that cannot be reached unless one formally keeps all partial wave contributions that have overlap with the irrep of interest. In particular, in Appendix~\ref{sec:freepoles}, along with showing an explicit proof of the cancelation of the free poles, we show that by truncating the scattering amplitude to be in an S-wave the free poles in general do not cancel. The cancellation of free poles assures that the only contribution to Eq.~\ref{eq:corrP0} is from integration around poles of the interacting system. To evaluate these, we introduce 
\begin{eqnarray}
\label{eq:mathbbM}
\mathbb{M}(P_{0,M})\equiv\mathbb{K}(P_{0,M})+ \left(\mathbb{F}^V(P_{0,M})\right)^{-1}.
\end{eqnarray}

Now note that the finite-volume two-particle spectrum is given by energies for which \(\mathbb{M}(P_{0,M})\) has a vanishing eigenvalue. This is L\"uscher's quantization condition, given in Eq.~\ref{eq:QC} above. At this stage we think it useful to discuss the connection of this result to previous work. We first observe that, although the condition in terms of \(\mathbb{M}(P_{0,M})\) is most convenient for the bulk of our analysis, here it is useful to reexpress it as
\begin{equation}
\label{eq:altQcond}
\det[(\mathbb K(P_{0,M}))^{-1} + \mathbb F^V(P_{0,M})] = 0\,.
\end{equation}
This is reached by multiplying $\mathbb M$ by $\mathbb K^{-1}$ on the left and by $\mathbb F^V$ on the right before taking the determinant. It gives an equivalent condition since multiplying by these matrices does not change the locations of zero eigenvalues. Substituting Eq.~\ref{eq:MKrel} into this form then gives
\begin{equation}
\det[(\mathcal M(P_{0,M}))^{-1} + \mathbb F_{i \epsilon}^V(P_{0,M})] = 0 \,,
\end{equation}
where $\mathbb F_{i \epsilon}^V(P_{0,M}) \equiv i \mathbb P^2/2  + \mathbb F^V(P_{0,M})$. This shows the equivalence of the present result to those appearing in Refs.~\cite{Luscher:1986pf, Luscher:1990ux, Rummukainen:1995vs, Kim:2005gf, Hansen:2012tf, Briceno:2012yi}. 

Next we consider Eq.~\ref{eq:altQcond} for energies in the vicinity of the lowest two-particle threshold. In this case we need only consider the S-wave scattering for the lowest two-particle channel. The quantization condition becomes
\begin{equation}
\frac{\xi}{8 \pi P_{0,M}^*} \left [ k^* \cot \delta(k^*) -  4 \pi c_{00}^{\textbf  d}(k^{*2};L) \right] = 0 \,.
\end{equation}
We may analytically continue this result below threshold by replacing $k^* = i \vert k^* \vert = i \kappa$. In this continuation $4 \pi c_{00}^{\textbf  d}(k^{*2};L)= - \kappa$ plus exponentially suppressed corrections.\footnote{We stress that the exponentially suppressed corrections may be large near threshold so that keeping such terms may be important.} We deduce
\begin{equation}
 k^* \cot \delta(k^*) \bigg \vert_{k^* = i \kappa} +  \kappa = 0\,,
\end{equation}
which is the standard, infinite-volume condition for a bound state.

Returning to the $P_0$ integral in Eq.~\ref{eq:corrP0}, we write the inverse of $\mathbb M(P_{0,M})$ in terms of a determinant and adjugate, 
\begin{eqnarray}
\frac{1}{\mathbb{M}(P_{0,M})}\equiv\frac{1}{\det[ \mathbb{M}(P_{0,M})]}\text{adj}[\mathbb{M}(P_{0,M})].
\label{eq:mathbbMadj}
\end{eqnarray}
This equation defines the adjugate which is also equal to the transpose of the cofactor matrix.\footnote{{This is also commonly known as the adjoint of a matrix, but given the context of this work, this nomenclature could be confused with the Hermitian conjugate. Therefore, we refer to this matrix as the adjugate to avoid confusion.}}
 It implies that, as \(P_{0,M}\) approaches a two-particle energy, \(\mathbb{M}(P_{0,M})^{-1}\) will diverge in proportion to \(\det[ \mathbb{M}(P_{0,M})]^{-1}\) such that \(\text{adj}[\mathbb{M}(P_{0,M})]\) remains finite. This separation, into diverging prefactor times finite matrix, makes Eq.~\ref{eq:mathbbMadj} useful for evaluating the residue of the two-particle poles. 
Looking at the variation of the quantization condition about the energy eigenvalues, we find
\begin{eqnarray}
{\det[ \mathbb{M}(P_{0,M})]}|&=&
{\det[ \mathbb{M}(E_n)]}
+(P_{0,M}-E_n)\left.\text{tr}\left[ \text{adj}[\mathbb{M}(P_{0,M})]\frac{\partial \mathbb{M}(P_{0,M})}{\partial P_{0,M}}\right]\right|_{ P_{0,M}=E_n}+\mathcal{O}((P_{0}-iE_n)^2)\\
&=&
-i(P_{0}-iE_n)\left.\text{tr}\left[ \text{adj}[\mathbb{M}(P_{0,M})]\frac{\partial \mathbb{M}(P_{0,M})}{\partial P_{0,M}}\right]\right|_{  P_{0,M}=E_n}+\mathcal{O}((P_{0}-iE_n)^2) \,.
\end{eqnarray}
With this in hand, one can perform the integral in Eq.~\ref{eq:corrP0} to find 
\begin{eqnarray}
\label{eq:corrP1}
C^{(2)}_{\Lambda\mu,ab}(x_0-y_0,\textbf{P})&=&
L^3\sum_{n}
e^{-E_{\Lambda,n}(x_0-y_0)}~
\left[ ~{\mathbb{Y}}_{\Lambda\mu,n}~R_{\Lambda,n}~\mathbb{Y}^\dag_{\Lambda\mu,n}~\right]_{ab},\\
R_{\Lambda,n}&=&\left.\text{adj}[\mathbb{M}(P_{0,M})]~{\text{tr}\left[ \text{adj}[\mathbb{M}(P_{0,M})]\frac{\partial \mathbb{M}(P_{0,M})}{\partial P_{0,M}}\right]}^{-1}\right|_{P_{0,M}=E_{\Lambda,n}},
\label{eq:RLambdan}
\end{eqnarray}
where $\mathbb{Y}_{\Lambda,n}$ is the value of $\mathbb{Y}$ (defined in Eq.~\ref{eq:Ymatrix}) evaluated at the $nth$ interactive two-particle pole. Here the sum over \(n\) runs over a finite set of energies that lie below the next multi-particle threshold. We are constrained to this region because our expression for the integrand of the \(P_0\) integral was only valid for $\mathrm{Im} \, P_0^* < 4m$, as already discussed above. 

By comparing this result to Eq.~\ref{eq:corr21}, we find that the matrix elements of the interpolating operators in general satisfy
\begin{eqnarray}
\label{eq:Onab}
\langle 0|\mathcal{O}_{\Lambda\mu,a}(0,\textbf{P})|E_{\Lambda,n}\textbf{P};L\rangle
\langle E_{\Lambda,n}\textbf{P};L|\mathcal{O}^\dagger_{\Lambda\mu,b}(0, - \textbf{P})|0\rangle
&=&
L^{3}~\left[~{\mathbb{Y}}_{\Lambda \mu,n}~R_{\Lambda,n}~\mathbb{Y}^\dag_{\Lambda \mu,n}~\right]_{ab} 
\end{eqnarray}
and in the case that $a=b$ this implies
\begin{eqnarray}
\label{eq:On}
|\langle 0|\mathcal{O}_{\Lambda\mu,a}(0,\textbf{P})|E_{\Lambda,n}\textbf{P};L\rangle|
&=&
L^{3/2}~\sqrt{\left[
~{\mathbb{Y}}_{\Lambda\mu,n}~R_{\Lambda,n}~\mathbb{Y}^\dag_{\Lambda\mu,n}~\right]_{aa}},
\end{eqnarray}
where the repeated indices on the right-hand side are \emph{not} summed. Equations~\ref{eq:Onab} and \ref{eq:On} are the main results of this section.

\subsection{Single channel S-wave result\label{sec:swavecheck}}
Here we consider the case where the orbital angular momentum is restricted to the S-wave. For this scenario to be relevant, the irrep of interest should have strong overlap with the S-wave and all higher contributions should be small. These conditions hold, for example, for the $\pi\pi$ system near threshold. At rest the LO contamination to the S-wave is due to $l=4$ and in the moving frame the LO contamination is due to $l=2$, both of which are suppressed at low energies. In this scenario $\mathbb{M}$ is a one by one matrix and its adjugate is one. Using Eq.~\ref{eq:RLambdan} one obtains that the residue at the $nth$ pole is
\begin{eqnarray}
R_{S,n} = \big [ \partial \mathbb M/\partial P_{0,M} \big ]^{-1} \vert_{P_{0,M}=E_{\Lambda,n}} = \left [  \frac{8 \pi E^*_n}{\xi q^*_n} \frac{1}{\cos^2 \delta_S} \big[\frac{\partial (\delta_S+\phi_{00}^\textbf{d})}{\partial P_{0,M}} \big] \vert_{P_{0,M}=E_{\Lambda,n}} \right ]^{-1},
\label{eq:RnSwave}
\end{eqnarray}
where we have introduced the pseudophase $\phi^\textbf{d}_{lm}$ with $(lm)$ angular momentum in the moving frame
\begin{eqnarray}
q^*_{\Lambda,n}\cot\phi^\textbf{d}_{lm}=-\frac{4\pi}{q^{*l}_{\Lambda,n}} c^\mathbf{d}_{lm}(q^{*2}_{\Lambda,n}; {L}).
\label{eq:philm}
\end{eqnarray} 

Combining this with the S-wave projection of $\mathbb Y$, defined in Eq.~\ref{eq:Ymatrix} above, we conclude
{\begin{eqnarray}
|\langle 0|\mathcal{O}_{\Lambda\mu}(0,\textbf{P})|E_{\Lambda,n}\textbf{P};L\rangle|
&=&
L^{3/2}~\left(
\frac{\xi q^*_n}{8 \pi E^*_n} \frac{|\mathbb{Y}_{S}|^2~\cos^2 \delta_S}{\big[{\partial (\delta_S+\phi_{00}^\textbf{d})}/{\partial P_{0,M}} \big] \vert_{P_{0,M}=E_{\Lambda,n}}}\right)^{1/2}.
\label{eq:SwaveLL}
 \end{eqnarray}
We stress here that $\mathbb Y_S$ contains the dependence of this matrix element of the specific operator used. We also recall that, in the case where the operator is built from single-particle interpolating fields as in Eq.~\ref{eq:twobody_oper}, then $\mathbb Y_S$ can be expressed in terms of an off-shell K-matrix, Eq.~\ref{eq:Ymatrix}. This off-shellness is unavoidable, since the single particle interpolating fields are evaluated at finite-volume momenta, which are different from the momenta that are determined by the finite-volume spectrum of the interacting theory. In Section~\ref{sec:corr3} we show how the dependence on $\mathbb Y$ cancels when one constructs ratios to access finite-volume matrix elements. 

Also it is worth mentioning that Eq.~\ref{eq:SwaveLL} clearly explains why, if one constructs an operator with a particular set of discrete momenta, the resulting correlation function will have largest overlap with the nearest eigenstate. This is because the amplitude of each exponential scales as $\sim\sqrt{|\mathbb{Y}_S|^2}$. From the definition in Eq.~\ref{eq:Ymatrix}, one observes that this diverges in the limit that the energy of two free particles with the quantum numbers of the two-particle operator, $E_{free}$, coincides with the finite-volume interacting energy, $E_{\Lambda,n}$. In fact, one can shown that near this pole, the overlap factor scales as $\sim {|E_{\Lambda,n}-E_{free}|}^{-1}$. In Section~\ref{sec:Kpipi} we show that this result reproduces the well known LL-factor in a moving frame.}
\subsection{$\pi\pi$ in a P-wave \label{sec:pipiPwave}}
In the case that the two particles of interest are degenerate, parity is still a good quantum number, even when the total momentum is nonzero. As a result, odd and even partial waves in the $\pi\pi$ systems do not mix. Therefore, when interested in studying scattering in the P-wave $\pi\pi$ channel, the LO partial wave contamination to consider is due to the F-wave. By neglecting this contribution, $\mathbb{M}$ can be written as a one by one matrix, and the quantization condition can be in general be written as
\begin{align}
\cot\delta_{P}+\cot\phi^\textbf{d}_{P}&=0,
\end{align}
where the pseudo phase $\phi^\textbf{d}_{P}$ can be written in terms of the pseudophases defined in Eq.~\ref{eq:philm}
\begin{align}
\cot\phi^\textbf{d}_{P}&\equiv\left(\cot\phi^\textbf{d}_{00}
+\alpha_{20,\Lambda}^\mathbf{d}\cot\phi^\textbf{d}_{20}
+\alpha_{22,\Lambda}^\mathbf{d}\cot\phi^\textbf{d}_{22} 
\right).
\label{eq:Ppseudophase}
\end{align}
For systems with $\textbf{d}=\textbf{0}$ and cubic volumes, the $c^\textbf{d}_{2m}$ exactly vanish. For systems with non-zero total momenta or for asymmetric volumes, $c^\textbf{d}_{2m}$ do not necessarily vanish and the values of $\alpha_{20,\Lambda}^\mathbf{d}$ and $\alpha_{22,\Lambda}^\mathbf{d}$ for ${\mathbf{d}^2}\leq3$ are shown in Table~\ref{table:alphad}.

\begin{center}
\begin{table} 
\begin{tabular}{c|c|c|c} 
\hspace{.1cm}\textbf{d}\hspace{.1cm}& 
 \hspace{.1cm}$(00n)$\hspace{.1cm}& 
 \hspace{.1cm}$(nn0)$\hspace{.1cm}& 
 \hspace{.1cm}$(nnn)$\hspace{.1cm}
 \\\hline
& $\alpha_{20,\mathbb{A}_1}^\mathbf{d} =\frac{2}{\sqrt{5}}$
& $\alpha_{20,\mathbb{A}_1}^\mathbf{d} =-\frac{1}{\sqrt{5}},\hspace{.5cm}
\alpha_{22,\mathbb{A}_1}^\mathbf{d} =-i\sqrt{\frac{6}{5}}$
& $\alpha_{22,\mathbb{A}_1}^\mathbf{d} =-2i\sqrt{\frac{6}{5}}$
 \\ 
 & $\alpha_{20,\mathbb{E}}^\mathbf{d} =-\frac{1}{\sqrt{5}}$
  & 
 $\alpha_{20,\mathbb{B}_1}^\mathbf{d} =-\frac{1}{\sqrt{5}},\hspace{.5cm}
  \alpha_{22,\mathbb{B}_1}^\mathbf{d} =i\sqrt{\frac{6}{5}}$
 & $\alpha_{22,\mathbb{E}}^\mathbf{d} =i\sqrt{\frac{6}{5}}$\\
 && 
 $\alpha_{20,\mathbb{B}_2}^\mathbf{d} =\frac{2}{\sqrt{5}}$
\vspace{.05cm}\\ \hline
\end{tabular}
\caption{Nonzero values of $\alpha_{20,\Lambda}^\mathbf{d}$ and $\alpha_{22,\Lambda}^\mathbf{d}$ for ${\mathbf{d}^2}\leq3$. For the $\mathbb{T}_1^-$ irrep of ${O}^D_h$, the $c^\mathbf{d}_{2m}$ vanish, therefore there is no need to define $\alpha_{2m,\Lambda}^\mathbf{d}$ for this irrep. 
}
\label{table:alphad}
\end{table}
\end{center} 

The overlap factor of the two-particle interpolating operator with the $nth$ finite volume eigenstate for a two-particle systems in a P-wave follows form Eq.~\ref{eq:SwaveLL}, after substituting for the definition of the pseudo phase and using the P-wave phase shift 
\begin{eqnarray}
|\langle 0|\mathcal{O}_{\Lambda\mu}(0,\textbf{P})|E_{\Lambda,n}\textbf{P};L\rangle|
&=&
L^{3/2}~\left(
\frac{ \xi q^*_n}{8 \pi E^*_n} \frac{|\mathbb{Y}_{P}|^2~\cos^2 \delta_P}{\big[{\partial (\delta_P+\phi_{P}^\textbf{d})}/{\partial P_{0,M}} \big] \vert_{P_{0,M}=E_{\Lambda,n}}}\right)^{1/2}.
\label{eq:PwaveLL}
\end{eqnarray} 
In Section~\ref{sec:pigammapipi} we show this leads to the needed LL-factor for $\pi\gamma^*\rightarrow\pi\pi$ when the final state is in a P-wave.  

Note that we have left the symmetry factor $\xi$ unspecified here. A P-wave state is antisymmetric under exchange of individual particle momenta. Thus, for bosons, overall exchange symmetry implies that the P-wave states must also be antisymmetric under exchange of particle labels. This in turn implies that only non-identical pions can participate in P-wave scattering. However, if we use two-pion isospin states to define scattering quantities, then $\xi=1/2$ must nonetheless be used. To make sense of this consider, for example, the two pion state with $I=1, M_I=1$
\begin{equation}
\vert I=1, M_I = 1 \rangle = \frac{1}{\sqrt{2}} \left(\vert \pi^+ \pi^0\rangle - \vert \pi^0 \pi^- \rangle \right ) \,.
\end{equation}
This can be used to construct and compare P-wave scattering amplitudes, and K-matrices, defined using isospin states and using unsymmetrized states. One finds
\begin{equation}
[\mathbb K_{I=1, M_I=1}]_{P} = 2 [ \mathbb K_{ \pi^+ \pi^0 \rightarrow \pi^+ \pi^0}]_{P} \,,
\end{equation}
where the superscript $P$ indicates that we are only considering $l=l'=1$ entries. This discrepancy in K-matrices implies a difference in the values for $[\mathbb Y_{I=1, M_I=1}]_{P}$ and $[\mathbb Y_{\pi^+ \pi^0}]_{P}$, as defined in Eq.~(\ref{eq:Ymatrix}). However, the finite-volume matrix element of a given operator must independent of this choice. Eq.~(\ref{eq:PwaveLL}) gives consistent predictions as long as one uses $\xi=1$ with $[\mathbb Y_{\pi^+ \pi^0}]_{P}$ and $\xi=1/2$ with $[\mathbb Y_{I=1, M_I=1}]_{P}$. See also the discussion after Eq.~(\ref{eq:LLfactorP}) below.


\subsection{$\pi K$ for $J\leq1$ \label{sec:piKSPwave}}

As a slightly more complicated example, we consider the $\pi K$ operator. For such a system with zero total momentum, parity is a good quantum number and as a result odd and even partial waves do not mix. If we restrict the angular momentum to satisfy $J\leq 1$, the system could be in a S- or P-wave. The corresponding cubic irreps would be the $\mathbb{A}_1^+$ and $\mathbb{T}_1^-$, and the matrix elements of their respective operators are described by Eqs.~\ref{eq:SwaveLL} and \ref{eq:PwaveLL}, respectively. For boosted system, parity is no longer a good quantum number. As a result odd and even partial waves will mix. Neglecting D-wave contamination, one finds that for boosted systems at least one irrep will have large overlap with P-wave states and no overlap with the S-wave. One can readily identify such irreps as $\mathbb{E}$ for $\textbf{d}=(00n)$, $\mathbb{B}_1$ and $\mathbb{B}_2$ for $\textbf{d}=(nn0)$, and $\mathbb{E}$ for $\textbf{d}=(nnn)$. For these irreps, the overlap factor is again shown in Eq.~\ref{eq:PwaveLL}. The non-vanishing values for $\alpha_{20,\Lambda}^\mathbf{d}$ and $\alpha_{22,\Lambda}^\mathbf{d}$ for ${\mathbf{d}^2}\leq3$ are given in Table~\ref{table:alphad}. The $\mathbb{A}_1$ irrep for these boost vectors is an admixture of S- and P-wave. As an example, consider the $\mathbb{A}_1$ in the  $\text{Dic}_4$ group, which is the symmetry group for $\textbf{d}=(00n)$. This irrep mixes the  $(l,m)=\{(0,0),(1,0),\ldots\}$ partial waves. In this space one can write down the finite volume function $\mathbb{F}^V$ and the K-matrix
\begin{eqnarray}
\label{eq:KmatpiK}
 \text{Dic}_4 ~\mathbb{A}_1:~~~~~~
\mathbb{F}^{V}_{\mathbb{A}_1}&=&
\frac{q^*_{\mathbb{A}_1,on}}{8\pi E^*_{\mathbb{A}_1}}\left(
\begin{array}{cc}
\cot\phi^\textbf{d}_{00}  & 
  \cot\phi^\textbf{d}_{10} \\
  \cot\phi^\textbf{d}_{10}&
\cot\phi^\textbf{d}_{00} + {2}/\sqrt{{5}} \cot\phi^\textbf{d}_{20}   \\
\end{array}
\right),\\
 \mathbb{K}_{on,on;\mathbb{A}_1}&=&
\frac{8\pi E^*_{\mathbb{A}_1}}{q^*_{\mathbb{A}_1,on}}
\left(
\begin{array}{cc}
[{\cot\delta_S}]^{-1}  & 
  0 \\
0&
[{\cot\delta_P}]^{-1}  \\
\end{array}
\right).
\label{eq:FVA1Dic4}
\end{eqnarray}
{

The quantization condition can be written as 
\begin{eqnarray}
\text{Dic}_4 ~\mathbb{A}_1&:& 
{\det[\mathbb{M}_{\mathbb{A}_1}]}={\det\left[\mathbb{K}_{on,on;\mathbb{A}_1}+ \left(\mathbb{F}^V_{\mathbb{A}_1}\right)^{-1}\right]}=0.
\label{eq:A1Dic4}
\end{eqnarray}
In order to evaluate $|\langle 0|\mathcal{O}_{\mathbb{A}_1,0,\textbf{P}}|E_{\mathbb{A}_1,n}\textbf{P};L\rangle|$, we first need to evaluate the adjugate of ${\mathbb{M}}_{\mathbb{A}_1}$,
\begin{eqnarray}
\rm{adj}[{\mathbb{M}}_{\mathbb{A}_1}]=
\left(
\begin{array}{cc}
\left[{\mathbb{M}}_{\mathbb{A}_1}\right]_{22}
 & 
-\left[{\mathbb{M}}_{\mathbb{A}_1}\right]_{12} \\
-\left[{\mathbb{M}}_{\mathbb{A}_1}\right]_{21}&
\left[{\mathbb{M}}_{\mathbb{A}_1}\right]_{11}
  \\
\end{array}
\right).
\label{eq:A1Dic4adj}
\end{eqnarray}
We obtain the overlap factor for the $\mathbb{A}_1$ irrep for the $\text{Dic}_4$ group, 
\begin{eqnarray}
|\langle 0|\mathcal{O}_{\mathbb{A}_1\mu,0,\textbf{P}}|E_{\mathbb{A}_1,n}\textbf{P};L\rangle|
&=&
{L^{3/2}}
~\frac{
\left(\mathbb{Y}_{\mathbb{A}_1\mu,n}~ \text{adj}[{\mathbb{M}}_{\mathbb{A}_1}]~\mathbb{Y}^\dag_{\mathbb{A}_1\mu,n}
\right)^{1/2}
}
{\text{tr}\left[ \text{adj}[{\mathbb{M}}_{\mathbb{A}_1}]\frac{\partial {\mathbb{M}}_{\mathbb{A}_1}}{\partial E_{\mathbb{A}_1,n}}\right]^{1/2}} .
\end{eqnarray}
Similar expressions can be found for the $\mathbb{A}_1$ irreps of the $\text{Dic}_2$ and  $\text{Dic}_3$ groups, the only differences would be the values of the finite volume functions and the K-matrix appearing in Eqs.~\ref{eq:A1Dic4} and \ref{eq:FVA1Dic4}. For example, the $\mathbb{A}_1$ irrep of the $\text{Dic}_2$ mixes the $(l,m)=\{(0,0),(1,-1),(1,1),\ldots\}$ partial waves,
 \begin{eqnarray}
 \label{eq:FA1Dic2}
 \text{Dic}_2 ~\mathbb{A}_1:~~~~~~
\mathbb{F}^{V}_{\mathbb{A}_1}&=&
\frac{q^*_{\mathbb{A}_1,on}}{8\pi E^*_{\mathbb{A}_1}}
\left(\begin{array}{ccc}\cot\phi^\textbf{d}_{00}  
& i^{3/2}~\text{Re}[\cot\phi^\textbf{d}_{11}] 
& i^{1/2}~\text{Re}[\cot\phi^\textbf{d}_{11}] 
\\
 -i^{1/2}~\text{Re}[\cot\phi^\textbf{d}_{11}]  
&\cot\phi^\textbf{d}_{00} - \cot\phi^\textbf{d}_{20} /\sqrt{{5}}  
& - \sqrt{6/5} \cot\phi^\textbf{d}_{22}  
\\
-i^{3/2}~\text{Re}[\cot\phi^\textbf{d}_{11}] 
&
 \sqrt{6/5} \cot\phi^\textbf{d}_{22}
 &\cot\phi^\textbf{d}_{00} - \cot\phi^\textbf{d}_{20} /\sqrt{{5}}  
  \\\end{array}\right)
,\\ 
 \mathbb{K}_{on,on;\mathbb{A}_1}&=&
\frac{8\pi E^*_{\mathbb{A}_1}}{q^*_{\mathbb{A}_1,on}}
\left(
\begin{array}{ccc}
[{\cot\delta_S}]^{-1}  & 
  0& 
  0 \\
0&
[{\cot\delta_P}]^{-1} & 
  0
  \\
0& 0&
[{\cot\delta_P}]^{-1}  \\
\end{array}
\right). 
\end{eqnarray}
The final piece needed is the evaluation of the adjugate of a three-dimensional matrix
\begin{eqnarray}
\rm{adj}[{\mathbb{M}}_{\mathbb{A}_1}] = \begin{pmatrix}
\left| \begin{matrix} \left[{\mathbb{M}}_{\mathbb{A}_1}\right]_{22} & \left[{\mathbb{M}}_{\mathbb{A}_1}\right]_{23} \\ \left[{\mathbb{M}}_{\mathbb{A}_1}\right]_{32} & \left[{\mathbb{M}}_{\mathbb{A}_1}\right]_{33} \end{matrix} \right| &
-\left| \begin{matrix} \left[{\mathbb{M}}_{\mathbb{A}_1}\right]_{12} & \left[{\mathbb{M}}_{\mathbb{A}_1}\right]_{13} \\ \left[{\mathbb{M}}_{\mathbb{A}_1}\right]_{32} & \left[{\mathbb{M}}_{\mathbb{A}_1}\right]_{33}  \end{matrix} \right| &
\left| \begin{matrix} \left[{\mathbb{M}}_{\mathbb{A}_1}\right]_{12} & \left[{\mathbb{M}}_{\mathbb{A}_1}\right]_{13} \\ \left[{\mathbb{M}}_{\mathbb{A}_1}\right]_{22} & \left[{\mathbb{M}}_{\mathbb{A}_1}\right]_{23} \end{matrix} \right| \\
 & & \\
-\left| \begin{matrix} \left[{\mathbb{M}}_{\mathbb{A}_1}\right]_{21} & \left[{\mathbb{M}}_{\mathbb{A}_1}\right]_{23} \\ \left[{\mathbb{M}}_{\mathbb{A}_1}\right]_{31} & \left[{\mathbb{M}}_{\mathbb{A}_1}\right]_{33} \end{matrix} \right| &
\left| \begin{matrix} \left[{\mathbb{M}}_{\mathbb{A}_1}\right]_{11} & \left[{\mathbb{M}}_{\mathbb{A}_1}\right]_{13} \\ \left[{\mathbb{M}}_{\mathbb{A}_1}\right]_{31} & \left[{\mathbb{M}}_{\mathbb{A}_1}\right]_{33} \end{matrix} \right| &
-\left| \begin{matrix} \left[{\mathbb{M}}_{\mathbb{A}_1}\right]_{11} & \left[{\mathbb{M}}_{\mathbb{A}_1}\right]_{13} \\ \left[{\mathbb{M}}_{\mathbb{A}_1}\right]_{21} & \left[{\mathbb{M}}_{\mathbb{A}_1}\right]_{23}  \end{matrix} \right| \\
 & & \\
\left| \begin{matrix} \left[{\mathbb{M}}_{\mathbb{A}_1}\right]_{21} & \left[{\mathbb{M}}_{\mathbb{A}_1}\right]_{22} \\ \left[{\mathbb{M}}_{\mathbb{A}_1}\right]_{31} & \left[{\mathbb{M}}_{\mathbb{A}_1}\right]_{32} \end{matrix} \right| &
-\left| \begin{matrix} \left[{\mathbb{M}}_{\mathbb{A}_1}\right]_{11} & \left[{\mathbb{M}}_{\mathbb{A}_1}\right]_{12} \\ \left[{\mathbb{M}}_{\mathbb{A}_1}\right]_{31} & \left[{\mathbb{M}}_{\mathbb{A}_1}\right]_{32} \end{matrix} \right| &
\left| \begin{matrix} \left[{\mathbb{M}}_{\mathbb{A}_1}\right]_{11} & \left[{\mathbb{M}}_{\mathbb{A}_1}\right]_{12} \\ \left[{\mathbb{M}}_{\mathbb{A}_1}\right]_{21} & \left[{\mathbb{M}}_{\mathbb{A}_1}\right]_{22} \end{matrix} \right|
\end{pmatrix} .
\end{eqnarray}
 These two examples explicitly illustrate {how to correctly handle partial wave mixing} in numerical studies of the two-point correlation function. Similarly, one can consider the scenario where the scattering amplitudes couple different on-shell channels, in Section~\ref{sec:2Dcase} we discuss how to determine the LL-factor for such systems. }

\section{Three-point correlation functions \\
and the generalized Lellouch-L\"uscher formula\label{sec:corr3}}
 Having discussed two-point correlation functions extensively in the previous section, we now proceed to the main focus of this work, three-point correlation functions. In particular, we are interested in processes where an external current annihilates a single-particle state and creates a two-particle state. Such a transition was first considered in this context by Lellouch and L\"uscher, who derived a relation between a finite-volume matrix element and the physical decay rate for $K\rightarrow \pi\pi$~\cite{Lellouch:2000pv}. The weak Hamiltonian is the external current in this process, and thus the analysis is restricted to scalar currents which insert zero momentum. Here we extend the result by allowing the external current to inject arbitrary four momentum and to be in any irrep of the finite-volume symmetry group. This is particularly relevant for meson photoproduction processes such as $\pi\gamma^*\rightarrow\pi\pi$ as well as meson decays of the form $\phi_1 \rightarrow \phi_2\phi_3 X$, where $X$ denotes an arbitrary leptonic current. Even the relatively simple example of $\pi\gamma^*\rightarrow\pi\pi$ illustrates that the finite-volume final state mixes different angular momenta, due to the reduction of rotational symmetry. In addition, the finite-volume matrix element is related to multiple infinite-volume matrix elements, defined via asymptotic states with different particle content. For example the $\pi\pi$ state mixes with $ K \bar{K}$.\footnote{This mixing is also present for the coupled-channel generalization of Lellouch-Lüscher presented in Ref.~\cite{Hansen:2012tf}.} Following the discussion of the previous section, we accommodate any number of strongly-coupled channels, but restrict attention to energies for which only two-particle states can go on-shell.

\subsection{Construction of currents in irreps of LG(\textbf{Q})\label{sec:current}}
In order to construct the three-point correlation function, we must first define currents in irreps of the finite-volume symmetry groups. We begin by defining a current which transforms as a representation of the infinite-volume symmetry group. As a specific example, consider a four-vector current which couples an incoming single-particle state, with momentum $P_i$, to an outgoing (asymptotic) two-particle state, where one particle has momentum $k$ and the other $P_f-k$. Defining $h_\nu(P_i,P_f-k,k)$ as the LO transition amplitude for this process, we introduce
\begin{eqnarray}
\label{eq:vec_curr}
\mathcal{J}_\nu(x)&=&\frac{\xi}{L^9}\sum_{\textbf{P}_f,\textbf{k},\textbf{P}_i}
\int\frac{dP_{f,0}}{2\pi}\frac{dP_{i,0}}{2\pi}\frac{dk_0}{2\pi} e^{i(P_i-P_f)\cdot x} 
~\bar\varphi^\dag (-P_f+k)
~\tilde\varphi^\dag(-{k})  ~\varphi({P_i})~
h_\nu(P_i,P_f-k,k).
\end{eqnarray}
Here \(\xi=1/2\) if \(\bar \varphi = \tilde \varphi\) and otherwise \(\xi=1\). The zero component of this four-vector current transforms trivially under rotations, also within the finite-volume subgroups. By contrast, the spatial vector (or pseudovector) is in the $J=1$ irrep of \(SO(3)\), and thus transforms under multiple irreps of the finite-volume groups. 

In order to discuss the subduction of the vector current onto irreps of the octahedral group and the little groups, it is convenient to first Fourier transform
\begin{eqnarray}
\tilde{\mathcal{J}}_j(x_0,\textbf{Q})&=&\int d \textbf x e^{-i\textbf{Q}\cdot \textbf{x}} \mathcal{J}_j(x)\nn\\
&=&\frac{\xi}{L^6}\sum_{\textbf{P}_f,\textbf{k},\textbf{P}_i}
\int\frac{dP_{f,0}}{2\pi}\frac{dP_{i,0}}{2\pi}\frac{dk_0}{2\pi} e^{i(P_{i,0}-P_{f,0}) x_0} 
~\bar\varphi^\dag (-P_f+k)
~\tilde\varphi^\dag(-{k})  ~\varphi({P_i})~
h_j(P_i,P_f-k,k)~\delta_{\textbf{P}_i,\textbf{Q}+\textbf{P}_f},~~~~~~
\label{eq:vec_currQ}
\end{eqnarray}
and also to switch from the Cartesian to the spherical-harmonic basis
\begin{eqnarray}
\tilde {\mathcal J}_{\pm 1}=\mp\frac{1}{\sqrt{2}}(\tilde {\mathcal J}_{x} \pm i \tilde {\mathcal J}_{y}),\hspace{1cm}
\tilde {\mathcal J}_{0} = \tilde {\mathcal J}_{z} \,.
\end{eqnarray}

For non-zero \textbf{Q}, the azimuthal component of the vector current is only a good quantum number if the $\hat{z}$ axis and the momentum axis coincide. It is thus convenient to instead use operators in the helicity basis. These are found by defining $R$ as an active rotation from $(0,0,|\textbf{Q}|)$ to $\textbf{Q}$ and $\mathcal{D}^{(J)}_{m_1m_2}(R)$ as the $m_1m_2$ component of the corresponding Wigner-$\mathcal{D}$ matrix in the $J$ representation. With this, one can rotate from the spherical-harmonic to the helicity basis%
\begin{eqnarray}
\tilde{\mathcal{J}}_\lambda(y_0,\textbf{Q})&=&\sum_m \mathcal{D}^{(1)*}_{m\lambda}(R)~\tilde{\mathcal{J}}_m(y_0,\textbf{Q}).
\end{eqnarray}

We are now in position to decompose the current into irreps of the finite-volume symmetry groups. First restricting attention to $\tilde{\mathcal{J}}_\lambda(y_0,\textbf{0})$, we comment that the current can be {subduced} onto the $\Lambda$ irrep of $O_h$ using the \emph{subduction} coefficients, $[{C}^{J}_{\Lambda}]_{\mu,\lambda}$~\cite{Thomas:2011rh}. As can be seen in Table~\ref{table:irreps}\ref{table:subduca}, for this case the subduction is trivial. The \(J=1\) irrep becomes the \(\mathbb T_1\) irrep of the octahedral group, with each element of the helicity basis equal to one of the three \(\mu\) values labeling the finite-volume counterpart. For systems in flight, one may define a similar subduction. In this case nontrivial linear combinations arise, given by    
\begin{eqnarray}
\tilde{\mathcal{J}}^{[J,P,|\lambda|]}_{\Lambda\mu}(y_0,\textbf{Q})&=&\sum_{\hat{\lambda}=\pm|\lambda|} \mathcal{S}_{\Lambda\mu}^{\tilde{\eta}\hat{\lambda}}~
\tilde{\mathcal{J}}_{\hat \lambda}^{[J,P]}(y_0,\textbf{Q}),
\label{eq:projboost}
\end{eqnarray}
where now $J$ and $P$ specify the angular momentum and parity of the system in the c.m.~frame. Table~\ref{table:irreps}\ref{table:subducb} shows the values of $\mathcal{S}_{\Lambda\mu}^{\tilde{\eta}{\lambda}}$ for systems with  integer $J\leq 2$ and $L\textbf{Q}/2\pi=\{(0,0,n),(n,n,0),(n,n,n)\}$ and other possible cubic rotations are determined in Ref.~\cite{Thomas:2011rh}. 

The subduction of the vector current onto a definite irrep of LG(\textbf{Q}) can be easily generalized to currents of any rank,
\begin{eqnarray}
\tilde{\mathcal{J}}_{\alpha,\beta,\ldots,\omega}(x_0,\textbf{Q})\longrightarrow\tilde{\mathcal{J}}^{[J,P,|\lambda|]}_{\Lambda\mu}(x_0,\textbf{Q}).
\end{eqnarray}
The discussion that follows is relevant for arbitrary rank currents with either positive or negative parity that have been properly subduced onto the irreps of LG(\textbf{Q}). The key point is that, by taking appropriate linear combinations, one can relate an operator in any basis to one that transforms as an irrep of the finite-volume group. The linear combinations of currents induce linear combinations of the transition amplitudes so that both \(\tilde {\mathcal J}\) and \(h\) may be reexpressed in terms of finite-volume irreps, and the form of Eq.~\ref{eq:vec_curr} is preserved in the new basis
\begin{eqnarray}
\tilde{\mathcal{J}}^{[J,P,|\lambda|]}_{\Lambda\mu}(x_0,\textbf{Q})&=&\frac{\xi}{L^6}\sum_{\textbf{P}_f,\textbf{k},\textbf{P}_i}
\int\frac{dP_{f,0}}{2\pi}\frac{dP_{i,0}}{2\pi}\frac{dk_0}{2\pi} e^{i(P_{i,0}-P_{f,0}) x_0}
~\bar\varphi^\dag (-P_f+k)
~\tilde\varphi^\dag(-{k})  ~\varphi(P_i)~\nn\\
&&\times
h_{\Lambda\mu}^{[J,P,|\lambda|]}(P_i,P_f-k,k)~\delta_{\textbf{P}_i,\textbf{Q}+\textbf{P}_f} \,.
\label{eq:vec_currLambda}
\end{eqnarray}
Finally, to consider scenarios with \(N>1\) open two-particle channels, one need only generalize this expression to 
\begin{eqnarray}
\tilde{\mathcal{J}}^{[J,P,|\lambda|]}_{\Lambda\mu}(x_0,\textbf{Q})&=&\sum_{a=1}^{N}~\frac{\xi_a}{L^6}\sum_{\textbf{P}_f,\textbf{k},\textbf{P}_i}
\int\frac{dP_{f,0}}{2\pi}\frac{dP_{i,0}}{2\pi}\frac{dk_0}{2\pi} e^{i(P_{i,0}-P_{f,0}) x_0}
~\bar\varphi_a^\dag (-P_f+k)
~\tilde\varphi_a^\dag(-{k})  ~\varphi(P_i)~\nn\\
&&\times
h_{\Lambda\mu}^{[J,P,|\lambda|]}(P_i,P_f-k,k,a)~\delta_{\textbf{P}_i,\textbf{Q}+\textbf{P}_f},
\label{eq:vec_currLambdaa}
\end{eqnarray}
where $\bar\varphi_a^\dag$ and $\tilde\varphi_a^\dag$ each create one of the two particles in the $ath$ channel and $h_{\Lambda\mu}^{[J,P,|\lambda|]}(P_i,P_f-k,k,a)$ is the LO transition amplitude for that channel.


\begin{figure*}[t]
\begin{center}
\subfigure[]{
\label{fig:FVcorr3}
\includegraphics[scale=0.45]{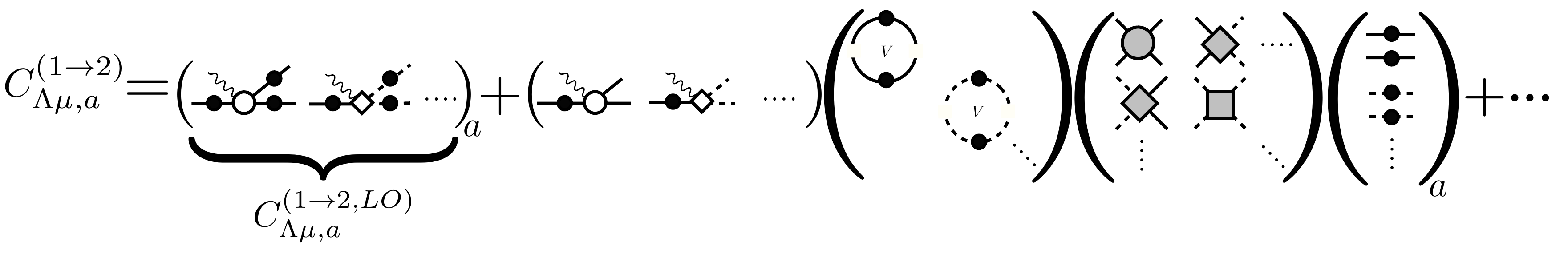}}\\
\subfigure[]{
\label{fig:FLambda3}
\includegraphics[scale=0.45]{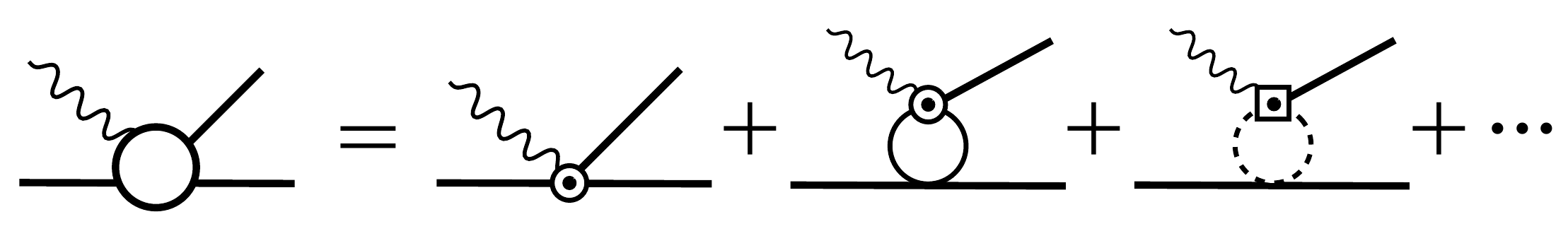}} 

\caption{a) Diagramatic representation for the three-point correlation function for processes involving a single incoming particle and outgoing two-particle state. This is written in terms of the LO transition amplitudes, one of which is explicit shown in (b), and the Bethe-Salpeter kernel, depicted in Fig.~\ref{fig:kernel}. The wiggly line is meant to depict an integer spin external current that can inject arbitrary four-momenta. Note that disconnected diagrams appearing in the LO transition amplitudes vanish except in the case where the current has the same quantum numbers as one of the outgoing external legs.}\label{fig:corr3}
\end{center}
\end{figure*}
 
\subsection{Three-point correlation function~\label{sec:corr3diagram}} 
Having properly defined the current of interest, we proceed to evaluate three-point correlation functions. Arriving at the result with an arbitrary number of open two-particle states is straightforward after one determines the single-channel result. We thus suppress the channel index for the time being and use Eq.~\ref{eq:vec_currLambda} for the current. We begin by giving the expression analogous to Eq.~\ref{eq:corr21}, with a current of arbitrary momentum  inserted at time $t=y_0$,
\begin{eqnarray} 
 C^{(1\rightarrow2)}_{\Lambda_f\mu_f;\Lambda\mu}(x_{f,0}-y_0;y_0-x_{i,0})&=&
\langle 0|\mathcal{O}_{\Lambda_f\mu_f}(x_{f,0},\textbf{P}_f)
~\tilde{\mathcal{J}}_{\Lambda\mu}^{[J,P,|\lambda|]}(y_0,\textbf{Q})
~\varphi^\dag (x_{i,0},-\textbf{P}_i)
|0\rangle\nn\\
&=&
 \sum_{n_f}
e^{-E_{\Lambda_f,n_f}(x_{f,0}-y_0)}
e^{-E_{\Lambda_i,0}(y_0-x_{i,0})}
\langle 0|\mathcal{O}_{\Lambda_f\mu_f}(0,\textbf{P}_f)|E_{\Lambda_f,n_f}\textbf{P}_f;L\rangle\nn\\
&&\times
\langle E_{\Lambda_f,n_f}\textbf{P}_f;L|\tilde{\mathcal{J}}_{\Lambda\mu}^{[J,P,|\lambda|]}(0,\textbf{Q})| E_{\Lambda_i,0}\textbf{P}_i;L\rangle
\langle E_{\Lambda_i,0}\textbf{P}_i;L|\varphi^\dag({0,-\textbf{P}_i})|0\rangle.~~~~~~~
\label{eq:corr31}
\end{eqnarray}
In the second line we have assumed \(x_{i,0} < y_0 < x_{f,0}\).

In order to get insight as to how one can interpret $\langle E_{\Lambda_f,n_f} \textbf P_f; L|\tilde{\mathcal{J}}_{\Lambda\mu}^{[J,P,|\lambda|]}(0,\textbf{Q})| E_{\Lambda_i,0}\textbf{P}_i;L\rangle$, we also evaluate the correlation function diagrammatically, as depicted in Fig.~\ref{fig:corr3}.\footnote{Note that Fig.~\ref{fig:corr3} shows the expression for the correlation function when an arbitrary number of final two-particle states are present. The single channel scenario is recovered by suppressing the dependence on the $a$ index and reducing all matrices in the channel space to scalars.} First observe that the transition amplitude, shown in Fig.~\ref{fig:FLambda3}, is defined in analogy to the Bethe-Salpeter kernel as the sum of all amputated diagrams that are two particle irreducible in the s-channel. The object differs from the Bethe-Salpeter kernel only in the form of external legs and in the insertion of a new contact interaction associated with the electroweak process of interest. To evaluate the three-point correlator we must sum all diagrams that appear when the external legs of the transition amplitude are contracted with the single incoming particle and outgoing two-particle state. We perform the calculation of the three-point correlator in two steps. First we consider the contraction of the incoming state with the current%
\begin{eqnarray}
\mathcal{D}^{(1)}(y_0-x_{i,0})
&=&\frac{1}{L^3}\sum_{\textbf{P}_{i'}}\int\frac{dP_{i',0}}{2\pi}e^{iP_{i',0} y_0}
~\langle \varphi({{P}_{i'}})
\varphi^\dag(x_{i,0},-\textbf{P}_i)
\rangle~
h_{\Lambda\mu}^{[J,P,|\lambda|]}(P_{i'},P_{f}-k_{f'},k_{f'})~\delta_{\textbf{P}_{i'},\textbf{Q}+\textbf{P}_{f}} \nn\\ 
&=&
\left(\frac{e^{-(y_0-x_{i,0})E_{\Lambda_i,0}}}{2E_{\Lambda_i,0}}\right)
h_{\Lambda\mu}^{[J,P,|\lambda|]}(P_{i},P_{f}-k_{f'},k_{f'})~\delta_{\textbf{P}_{i},\textbf{Q}+\textbf{P}_{f}}
+\mathcal{O}\left(\frac{e^{-E_{3,th}(y_0-x_{i,0})}}{E_{3,th}}\right),
\end{eqnarray}
where ${P_{i,0}=iE_{\Lambda_i,0}}$. The remaining contractions, between the current and the final two-particle operator, give 
\begin{eqnarray}
\mathcal{D}^{(2)}(x_{f,0}-y_0)&=&
\frac{\xi}{L^3}\sum_{\textbf{P}_f,\textbf{k}_{f}} \int\frac{dP_{f,0}}{2\pi}\frac{dk_{f,0}}{2\pi} e^{-i P_{f} y_0}
\langle 
\mathcal{O}_{\Lambda_f\mu_f}({x_f,\textbf{P}_f})
~\bar\varphi^\dag (-{{P}_{f}+k_{f}})
~\tilde\varphi^\dag (-{k_{f}})\rangle\nn\\
&&\times~h_{\Lambda\mu}^{[J,P,|\lambda|]}(P_{i},P_{f}-k_{f},k_{f})~\delta_{\textbf{P}_{i},\textbf{Q}+\textbf{P}_f}.
\end{eqnarray}
The LO contribution of this term is found to be
\begin{eqnarray}
\mathcal{D}^{(2,LO)}(x_{f,0}-y_0) 
&=&
\frac{L^3}{\eta}~ \int\frac{dP_{f,0}}{2\pi} e^{i P_{f,0}(x_{f,0}-y_0)}
{\sum_{R \in {\rm LG(\textbf{P}_f)}}}~\mathcal{C}(\textbf{P}_f\Lambda_f\mu_f; R (\textbf{P}_f-\textbf{k}_f); R \textbf{k}_f) \nn\\
&&~~~~\times~\frac{h_{\Lambda\mu}^{[J,P,|\lambda|]}(P_{i},P_{f}-k_{f},k_{f})}{{4~\omega_{1}~\omega_{2}(iP_{f,0}+(\omega_{1}+\omega_{2}))}}~\delta_{\textbf{P}_i,\textbf{Q}+\textbf{P}_f}+\cdots,
\label{eq:DLO}
\end{eqnarray}
where the ellipses denote contributions associated with higher energy poles of the two-particle propagator. Note that the symmetry factor cancels. 

 To complete our calculation of \(C^{(1\rightarrow2)}\), it remains only to include all higher order corrections to \(\mathcal D^{(2)}\). These arrise from insertions of the Bethe-Salpeter kernel between the current and the two-body operator. All contributions are included by making the substitution
 \begin{multline}
h_{\Lambda\mu}^{[J,P,|\lambda|]}(P_{i},P_{f}-k_{f},k_{f}) \rightarrow 
h_{\Lambda\mu}^{[J,P,|\lambda|]}(P_{i},P_{f}-k_{f},k_{f})\\
-\xi \, \frac{1}{L^3}\sum_{\textbf k_{f'}} \int \frac{d k_{f',0}}{2 \pi} 
  ~{\mathbb T}_{L}(P_{f},k_{f'},k_{f}) G(k_{f'}) G(P_f - k_{f'}) h_{\Lambda\mu}^{[J,P,|\lambda|]}(P_{i},P_{f}-k_{f'},k_{f'})   + \cdots \,,
\end{multline}
where the ellipses again denote higher energy poles.

\begin{figure*}[t]
\begin{center}
\subfigure[]{ 
\includegraphics[scale=0.45]{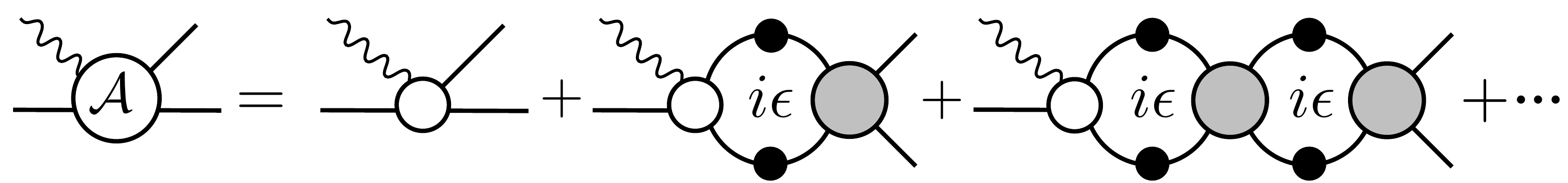}
\label{fig:A_mat}}
\subfigure[]{ 
\includegraphics[scale=0.45]{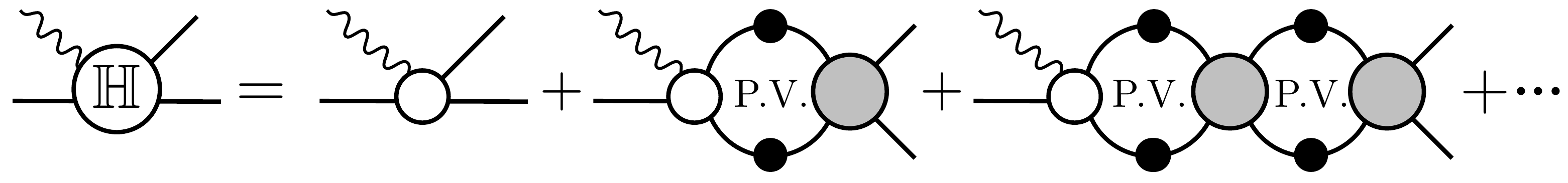}
\label{fig:H_mat}}\\
\caption{ 
In order to illustrate the differences and similarities between the transition amplitudes a) $\mathcal{A}$ and b) $\mathbb{H}$, we show their diagrammatic representation for the single-channel case in terms of the LO transition amplitude (defined in Fig.~\ref{fig:FLambda3}), Bethe-Salpeter kernel (defined in Fig.~\ref{fig:kernel}) and in terms of infinite volume loops. The infinite volume loops of $\mathcal{A}$ are evaluated using the $i\epsilon$ prescription, while those of $\mathbb{H}$ are evaluated using the principal value. For multichannel scenarios, the kernels and two-particle loops become matrices in the space of open channels and the LO transition amplitude becomes a vector in the space, as depicted in Fig.~\ref{fig:corr3}. Single particle propagators are fully dressed as defined in Fig.~\ref{fig:1bodyprop}.}\label{fig:AH_mat}
\end{center}
\end{figure*}

To give the final result we must first define ${\mathbb H}_{\Lambda\mu}^{[J,P,|\lambda|]}(P_{i},P_{f}-k_{f},k_{f})$ as the sum over all infinite-volume diagrams contributing to the transition amplitude, evaluated using the principal-value prescription (depicted in Fig.~\ref{fig:H_mat}). This is also given by
\begin{multline}
{\mathbb H}_{\Lambda\mu}^{[J,P,|\lambda|]}(P_{i},P_{f}-k_{f},k_{f}) \equiv 
h_{\Lambda\mu}^{[J,P,|\lambda|]}(P_{i},P_{f}-k_{f},k_{f})\\
+\xi \, \text{P.V.}   \int \frac{d \textbf k_{f'}}{(2 \pi)^3}   \int \frac{d k_{f',0}}{2 \pi}
  ~{\mathbb K}_{off,off}(P_{f},k_{f'},k_{f}) G(k_{f'}) G(P_f - k_{f'}) h_{\Lambda\mu}^{[J,P,|\lambda|]}(P_{i},P_{f}-k_{f'},k_{f'})   \,.
  \label{eq:Hlambda}
\end{multline}
In addition, we define {$\mathbb{H}^{[J,P,|\lambda|]}_{lm_l;\Lambda\mu}=\int d\Omega ~ Y^{*}_{lm_{l}}(\hat{\textbf{k}}^*_f)~{\mathbb H}_{\Lambda\mu}^{[J,P,|\lambda|]}(P_{i},P_{f'}-k_{f},k_{f})$, which is the projection of this amplitude onto the spherical harmonic basis of the outgoing state. Note that this requires evaluating the transition amplitude in the frame where the final two-particle state is at rest. In order to minimize excess in labels, from now on we suppress the superscripts on $\mathbb{H}^{[J,P,|\lambda|]}_{lm_l;\Lambda\mu}$ and simply denote it as $\mathbb{H}_{lm_l;\Lambda\mu}$}. %

Putting all the pieces together and performing the integral over $P_{f,0}$, one finds the following expressions for the three-point correlation function
\begin{eqnarray}
 C^{(1\rightarrow2)}_{\Lambda_f\mu_f;\Lambda\mu}(x_{f,0}-y_0;y_0-x_{i,0})&=&  
  \left(\frac{e^{-(y_0-x_{i,0})E_{\Lambda_i,0}}}{2E_{\Lambda_i,0}}\right)
 \int\frac{dP_{f,0}}{2\pi} e^{i P_{f,0}(x_{f,0}-y_0)}~\delta_{\textbf{P}_i,\textbf{Q}+\textbf{P}_f}
 \nn\\
  &&\hspace{-3.5cm}\times
 \left\{ \frac{1}{\eta} \frac{ \mathcal{C}(\textbf{P}\Lambda_{f}\mu_{f}; (\textbf{P}_{f}-\textbf{k}_f); \textbf{k}_f)~{\mathbb H}_{\Lambda\mu}^{[J,P,|\lambda|]}(P_{i},P_{f}-k_{f},k_{f})}{4~\omega_{1,\textbf{P}_{f}-\textbf{k}_{f'}}~\omega_{2,\textbf{k}_{f'}}(\omega_{1,\textbf{P}_{f}-\textbf{k}_{f'}}+\omega_{2,\textbf{k}_{f'}}+iP_{f,0})}
-
\mathbb{Y}_{\Lambda_f\mu_f}
~~\frac{1}{\mathbb{K}+\left(\mathbb{F}^{V}\right)^{-1}}~\mathbb{H}_{\Lambda\mu,on}~
~L^3+\cdots\right\}\nn \\
 &&\hspace{-3.5cm}=
 \left(\frac{e^{-(y_0-x_{i,0})E_{\Lambda_i,0}}}{2E_{\Lambda_i,0}}\right)~L^3~
\sum_{n_f}
e^{-E_{\Lambda_f,n_f}(x_{f,0}-y_0)}~
~\mathbb{Y}_{\Lambda_f\mu_f,n_f}
~R_{\Lambda_f,n_f}~~\mathbb{H}_{\Lambda_f,n_f;\Lambda\mu}~\delta_{\textbf{P}_i,\textbf{Q}+\textbf{P}_f}
+\cdots,~~~~~
\label{eq:corr32}
\end{eqnarray}
where the ellipses denote contribution from higher energy poles. Note that, just like in the two-point correlation function, the free-particle poles do not contribute due to the careful cancelation of the two objects inside the braces. 

By comparing Eqs.~\ref{eq:corr31} and \ref{eq:corr32} and multiplying with the complex conjugate expression, one finds an identity for the finite-volume matrix element 
\begin{align}
\begin{split}
\left|\langle E_{\Lambda_f,n_f} \textbf P_f; L|\tilde{\mathcal{J}}_{\Lambda\mu}^{[J,P,|\lambda|]}(0,\textbf{P}_i-\textbf{P}_f)| E_{\Lambda_i,0}\textbf{P}_i;L\rangle\right |
&  \\
& \hspace{-90pt} = \left(\frac{L^3}{2E_{\Lambda_i,0}}\right)
\frac{\sqrt{
~(\mathbb{Y}_{\Lambda_f\mu_f,n_f}
~R_{\Lambda_f,n_f}~\mathbb{H}_{\Lambda_f,n_f;\Lambda\mu})
(\mathbb{H}^\dag_{\Lambda_f,n_f;\Lambda\mu}
~R_{\Lambda_f,n_f}
~\mathbb{Y}^\dag_{\Lambda_f\mu_f,n_f} )
}
}
{
\large|
\langle 0|\mathcal{O}_{\Lambda\mu}({0,\textbf{P}_f})|E_{\Lambda_f,n_f}\textbf{P}_f;L\rangle
\large|~
\large|
\langle E_{\Lambda_i,0}\textbf{P}_i;L|\varphi^\dag({0,-\textbf{P}_i})|0\rangle
\large|
}
\end{split}\\[10pt]
& \hspace{-90pt}=
 \frac{1}{\sqrt{2E_{\Lambda_i,0}}}
\sqrt{\frac{
(\mathbb{Y}_{\Lambda_f\mu_f,n_f}
~R_{\Lambda_f,n_f}~\mathbb{H}_{\Lambda_f,n_f;\Lambda\mu})
(\mathbb{H}^\dag_{\Lambda_f,n_f;\Lambda\mu}
~R_{\Lambda_f,n_f}
~\mathbb{Y}^\dag_{\Lambda_f\mu_f,n_f}~)
}
{~{\mathbb{Y}}_{\Lambda_f\mu_f,n_f}~R_{\Lambda_f,n_f}~\mathbb{Y}^\dag_{\Lambda_f\mu_f,n_f}}},
\label{eq:matJ}
\end{align}
where we have used Eqs.~\ref{eq:phiLambda} and \ref{eq:On} to write the second equality. It is important to emphasize that the value of $\mathbb Y$ depends on the two-body interpolators used, and it is essential to use the same interpolators in the two-point and three-point functions for the second equality to follow. Indeed, although we constructed our two-body interpolators from scalar fields (with residue one at the single-particle pole), this result holds for any interpolating field with the desired quantum numbers. Any nontrivial overlap factors cancel between numerator and denominator.

For multichannel systems, one needs to evaluate the three-point correlation function using a current that couples to all open channels, as defined in Eq.~\ref{eq:vec_currLambdaa}. In this case one has the freedom to choose which flavor of two-particle operator is used in evaluating the correlation function. We define  
\begin{eqnarray} 
 C^{(1\rightarrow2)}_{\Lambda_f\mu_f,a;\Lambda\mu}(x_{f,0}-y_0;y_0-x_{i,0})&=&
\langle 0|\mathcal{O}_{\Lambda_f\mu_f,a}(x_{f,0},\textbf{P}_f)
~\tilde{\mathcal{J}}_{\Lambda\mu}^{[J,P,|\lambda|]}(y_0,\textbf{Q})
~\varphi^\dag (x_{i,0},-\textbf{P}_i)
|0\rangle\nn\\
&=&
 \sum_{n_f}
e^{-E_{\Lambda_f,n_f}(x_{f,0}-y_0)}
e^{-E_{\Lambda_i,0}(y_0-x_{i,0})}
\langle 0|\mathcal{O}_{\Lambda\mu,a}(0,\textbf{P}_f)|E_{\Lambda_f,n_f}\textbf{P}_f;L\rangle\nn\\
&&\times
\langle E_{\Lambda_f,n_f}\textbf{P}_f;L|\tilde{\mathcal{J}}_{\Lambda\mu}^{[J,P,|\lambda|]}(0,\textbf{Q})| E_{\Lambda_i,0}\textbf{P}_i;L\rangle
\langle E_{\Lambda_i,0}\textbf{P}_i;L|\varphi^\dag({0,-\textbf{P}_i})|0\rangle.~~~~~~~
\label{eq:corr31a}
\end{eqnarray}
This generic representation of the three-point function is diagrammatically depicted in Fig.~\ref{fig:FVcorr3}. Following the steps above, it is straightforward to see that Eq.~\ref{eq:matJ} generalizes to
\begin{multline}
\left|\langle E_{\Lambda_f,n_f}\textbf{P}_f;L|\tilde{\mathcal{J}}_{\Lambda\mu}^{[J,P,|\lambda|]}(0,\textbf{P}_i-\textbf{P}_f)| E_{\Lambda_i,0}\textbf{P}_i;L\rangle\right| = \\
 \frac{1}{\sqrt{2E_{\Lambda_i,0}}}
\sqrt{\frac{
[\mathbb{Y}_{\Lambda_f\mu_f,n_f}
~R_{\Lambda_f,n_f}~\mathbb{H}_{\Lambda_f,n_f;\Lambda\mu}]_a
[\mathbb{H}^\dag_{\Lambda_f,n_f;\Lambda\mu}
~R_{\Lambda_f,n_f}
~\mathbb{Y}^\dag_{\Lambda_f\mu_f,n_f}]_a
}
{[{\mathbb{Y}}_{\Lambda_f\mu_f,n_f}~R_{\Lambda_f,n_f}~\mathbb{Y}^\dag_{\Lambda_f\mu_f,n_f}]_{aa}}} \,,
\label{eq:matJa0}
\end{multline}
where the repeated channel indices on the right-hand side are \emph{not} summed. 

We now show that this result is equivalent to the main result of this work, Eq.~\ref{eq:matJaieps} above. To do so we define
\begin{equation}
\mathbb V^{(a)}_b \equiv \mathbb{Y}^\dag_{\Lambda_f\mu_f,n_f; a, b}\,,
\label{eq:Vab}
\end{equation}
where \(a\) and \(b\) are channel indices. We stress that, for each fixed value of \(a\), \(\mathbb V^{(a)}_b\) is a column in angular-momentum/channel space. Suppressing the channel index, \(b\), this notation allows us to rewrite Eq.~\ref{eq:matJa0} as
\begin{equation}
\left|\langle E_{\Lambda_f,n_f} ;L|\tilde{\mathcal{J}}_{\Lambda\mu}^{[J,P,|\lambda|]}| E_{\Lambda_i,0};L \rangle \right| = \\
 \frac{1}{\sqrt{2E_{\Lambda_i,0}}}
\sqrt{\frac{
[\mathbb V^{(a)\dagger}
~R_{\Lambda_f,n_f}~\mathbb{H}_{\Lambda_f,n_f;\Lambda\mu}]
[\mathbb{H}^\dag_{\Lambda_f,n_f;\Lambda\mu}
~R_{\Lambda_f,n_f}
~\mathbb V^{(a)}] 
}
{[\mathbb V^{(a)\dagger}
~R_{\Lambda_f,n_f}~\mathbb V^{(a)}] 
}} \,.
\end{equation}
Here we have dropped all momentum and time labels for compactness of notation.

We next observe that \(R_{\Lambda_f,n_f}\), which is Hermitian and therefore diagonalizable, has only one non-zero eigenvalue. To see this, recall that \(R_{\Lambda_f,n_f}\) is equal to a scalar prefactor times \(\text{adj}[\mathbb{M}(P_{0,M}=E_{\Lambda,n})]_{\Lambda_f}\). The adjugate here is understood as a matrix in angular-momentum/channel space, that has been projected onto the $\Lambda_f$ subspace. We now consider the adjugate as a function of \(\epsilon_n \equiv P_{0,M} - E_{\Lambda,n}\), and show that all but one of its eigenvalues vanishes as \(\epsilon_n \rightarrow 0\). 

Recall the defining relation
\begin{equation}
\text{adj}[\mathbb{M}(\epsilon_n)] = \det[\mathbb{M}(\epsilon_n)]~[\mathbb{M}(\epsilon_n)]^{-1}\,.
\end{equation} 
Formally diagonalizing both sides, we argue that exactly one of the eigenvalues of \([\mathbb{M}(\epsilon_n)]^{-1}\) scales as \(1/\epsilon_n\) and the rest are finite. Note that the divergence of two eigenvalues would imply the existence two orthogonal states that are exactly degenerate in finite volume. This represents two possibilities. The first is that distinct energies coincide only at certain values of \(L\). This would imply a level crossing, which does not occur unless the Hilbert space divides into distinct, non-interacting subspaces. The second possibility is that the finite volume spectrum includes states that are degenerate for all values of \(L\). This occurs whenever there is a symmetry relating the finite-volume states. However, in the present context the matrix has been projected to a particular irrep and row. It follows that, within the subspace that we consider, exactly one of the eigenvalues of \([\mathbb{M}(\epsilon_n)]^{-1}\) scales as \(1/\epsilon_n\). This in turn implies that the determinant of \(\mathbb M(\epsilon_n)\) vanishes as \(\epsilon_n\) or faster, and thus all but one of the adjugate's eigenvalues vanishes.

We denote the nonzero eigenvalue of \(R_{\Lambda_f, \mu_f}\) by \(\lambda\) and the corresponding eigenvector, \(\mathbb E\). We also introduce \(\mathbb E_1, \mathbb E_2, \cdots\) as the remaining orthonormal set that is annihilated by \(R_{\Lambda_f,n_f}\). These eigenvectors span the space, so we may substitute
 \(\mathbb V^{(a)} = c \mathbb E + \sum c_i \mathbb E_i\) and deduce
\begin{align}
\left|\langle E_{\Lambda_f,n_f} ;L|\tilde{\mathcal{J}}_{\Lambda\mu}^{[J,P,|\lambda|]}| E_{\Lambda_i,0};L \rangle \right| &= 
 \frac{1}{\sqrt{2E_{\Lambda_i,0}}}
\sqrt{\frac{
[c^* \lambda \mathbb E^\dagger
~\mathbb{H}_{\Lambda_f,n_f;\Lambda\mu}]
[\mathbb{H}^\dag_{\Lambda_f,n_f;\Lambda\mu}
~c \lambda \mathbb E] 
}
{\lambda c^* \mathbb E^\dagger
~\mathbb E c 
}} \,,\\
&= \frac{1}{\sqrt{2E_{\Lambda_i,0}}} \sqrt{\mathrm{Tr} \left[   \lambda \mathbb E^\dagger
~\mathbb{H}_{\Lambda_f,n_f;\Lambda\mu} \mathbb{H}^\dag_{\Lambda_f,n_f;\Lambda\mu}
~  \mathbb E\right ] }
 \,, \\
 &= \frac{1}{\sqrt{2E_{\Lambda_i,0}}} \sqrt{\mathrm{Tr} \left[ \mathbb{H}^\dag_{\Lambda_f,n_f;\Lambda\mu}~\lambda \mathbb E \mathbb E^\dagger
~\mathbb{H}_{\Lambda_f,n_f;\Lambda\mu} 
   \right ] }
 \,,
\end{align}
where in the first line we acted \(R_{\Lambda_f,n_f}\) on each eigenvector, in the second line we canceled common factors and inserted a redundant trace, and in the third we used the cyclic property of the trace. Observing finally that
\begin{equation}
R_{\Lambda_f,n_f} = \lambda \mathbb E \mathbb E^\dagger \,,
\label{eq:REE}
\end{equation}
we conclude
\begin{equation}
\left|\langle E_{\Lambda_f,n_f}\textbf{P}_f;L|\tilde{\mathcal{J}}_{\Lambda\mu}^{[J,P,|\lambda|]}(0,\textbf{P}_i-\textbf{P}_f)| E_{\Lambda_i,0}\textbf{P}_i;L\rangle\right| =
 \frac{1}{\sqrt{2E_{\Lambda_i,0}}}
\sqrt{
\left[\mathbb{H}^\dag_{\Lambda_f,n_f;\Lambda\mu}
~R_{\Lambda_f,n_f}~
\mathbb{H}_{\Lambda_f,n_f;\Lambda\mu}\right]
}.
\label{eq:matJa}
\end{equation}


 \subsection{Relation of $\mathbb H$ to infinite-volume matrix elements~\label{sec:relations}}


In this section we relate $\mathbb H_{\Lambda_f, n_f; \Lambda \mu; J_fm_{J_f}}=\mathbb H_{\Lambda_f; \Lambda \mu; J_fm_{J_f}}(E^*_{\Lambda_f,n_f})$ to infinite-volume matrix elements. Here we have given the full set of indices including $J_fm_{J_f}=l_fm_{f}$, which are the angular momentum indices of the final two-particle state and were suppressed in the derivation above. We have also emphasized that the label $n_f$ only refers to the particular two-particle pole at which the transition amplitude is evaluated. Finally, we stress that the subscript $\Lambda_f$ on $\mathbb H$ indicates that the angular momentum space has been projected onto a finite-volume irrep. For example in the case of $\Lambda_f=\mathbb A_1^+$ the transition amplitude will include $J_f=0$, $J_f=4$ and certain higher waves, but not $J_f=2, J_f=3$. However, by considering different irreps one can in principal sample all partial waves, and so construct an unprojected vector $\mathbb H_{ \Lambda \mu; J_fm_{J_f}}$.

 To give the relation to physical matrix elements, we first connect this transition amplitude, defined using principal-value prescription, to the amplitude defined via $i\epsilon$ prescription. We label the latter $\mathcal{A}_{ \Lambda \mu; J_fm_{J_f}}$. Both amplitudes are explicitly shown in Fig.~\ref{fig:AH_mat} and the relationship between the two is found by noting that the difference in each two-particle loop is a simple kinematic factor, determined by the residue of each loop at the poles. This is very similar to the relation between $\mathbb K$ and $\mathcal M$ discussed above. We find 
\label{eq:MJ}
\begin{eqnarray}
\mathcal{A} &=&\mathbb{H}+\mathbb{K}~\left(i\mathbb{P}^{2}/2\right)~\mathbb{H}
+\mathbb{K}~\left(i\mathbb{P}^{2}/2\right)~\mathbb{K}~\left(i\mathbb{P}^{2}/2\right)~\mathbb{H} +\cdots
= \left[\frac{1}{1-\mathbb{K}~\left(i\mathbb{P}^{2}/2\right)}\right] \mathbb{H} \nn\\
&=& \left[\frac{1}{\mathbb{K}^{-1}-\left(i\mathbb{P}^{2}/2\right)}\right]~\mathbb{K}^{-1}~\mathbb{H}
=\mathcal M~\mathbb{K}^{-1}~\mathbb{H}.
\label{AjHjtildematrix}
\end{eqnarray}
For systems with only a single channel present $\mathbb H$ and $\mathcal A$ are columns and $\mathbb P$, $\mathbb K$ and $\mathcal M$ are diagonal matrices in angular momentum space, otherwise these objects are defined on the direct product space of angular momenta and open channels. Note that $\mathbb H$ only has complex values from the spherical harmonics, the function $\mathbb H$ before decomposition is pure real. Thus the non-trivial complex phases of \(\mathcal A\) are determined entirely by the strong interaction, as encoded in $\mathbb{K}^{-1} \mathcal M$. In the single channel case we see that the phase of \(\mathcal A\) is equal to the elastic scattering phase of the two-particle channel considered. Thus Eq.~\ref{AjHjtildematrix} is simply the generalization of Watson's theorem for multichannel systems. 

This relation motivates the definition 
\begin{equation}
\mathcal R_{\Lambda_f,n_f} =  [{\mathcal M}^{-1\dagger}~\mathbb K~R~\mathbb K~{\mathcal M}^{-1}]_{\Lambda_f,n_f}\,,
\label{eq:calRdef}
\end{equation}
which allows us to compactly display our main result in terms of $\mathcal A$, as in Eq.~\ref{eq:matJaieps} above.

Finally we comment that $\mathcal{A}_{ \Lambda \mu; J_fm_{J_f}}$ is trivially related to the infinite-volume matrix element of the current. To see this, we first rewrite the current $\tilde{\mathcal{J}}^{[J,P,|\lambda|]}_{\Lambda\mu}(x_0,\textbf{Q})$, Eq.~\ref{eq:vec_currLambdaa}, in infinite volume and set $x_0=0$,
\begin{eqnarray}
\tilde{\mathcal{J}}^{[J,P,|\lambda|]}_{\Lambda\mu}(0,\textbf{Q};\infty)&=&\sum_a^{N}~{\xi_a} 
\int\frac{d^4P_{f}}{(2\pi)^4}
\frac{d^4P_{i}}{(2\pi)^4}
\frac{d^4k}{(2\pi)^4}  ~\bar\varphi_a^\dag (-P_f+k)
~\tilde\varphi_a^\dag(-{k}) ~\varphi(P_i)~h_{\Lambda\mu}^{[J,P,|\lambda|]}(P_i,P_f-k,k,a)~\nn\\
&&\hspace{9cm}\times
(2\pi)^3\delta^3(\textbf{P}_i-\textbf{P}_f-\textbf{Q}).
\label{eq:vec_currLambdinfty}
\end{eqnarray}
Note that we still label the current by $\Lambda \mu$. The linear combinations that relate this basis to more standard infinite-volume bases are discussed above. Requiring only that states are normalized according to the standard infinite-volume relativistic convention (Eq.~\ref{eq:infiniteVnorm}) and also that the single-particle operators have propagators with unit residue (Eq.~\ref{eq:infiniteVopnorm}) one arrives at Eq.~\ref{eq:JaIV}.

 


 \subsection{Examples of applications of Eq.~\ref{eq:matJaieps} \label{sec:apps}}
 \subsubsection{ $K\rightarrow \pi\pi$ decay amplitude \label{sec:Kpipi}}
First, we demonstrate that this formalism properly recovers the well known result for $K\rightarrow \pi\pi$ weak decay. In this case, the initial state is a single kaon and the external current is a pseudoscalar. The current cannot inject any momentum, so we set $\textbf{P}_f=\textbf{P}_i$. By conservation of angular momentum, the infinite-volume current can only create a two-pion state in an S-wave. For this scenario our master equation gives the following relationship between the infinite-volume transition amplitude and the finite-volume matrix element
\begin{eqnarray}
\frac{|\mathcal{A}_{S,n_f}|^2}
{|\langle \pi\pi,E_{n_f}\textbf{P},\Lambda_f\mu_f;L|\tilde{\mathcal{J}}_{\Lambda\mu}^{[0,- 1,|0|]}(0,\textbf{0})| K,E_{K}\textbf{P};L\rangle|^2}=
\frac{16\pi E_{K}~E^{*}_{n_f}}{q^{*}_{n_f}\xi}
\left.~\frac{\partial (\delta_S+\phi^\textbf{d}_{00})}
{\partial P_{0,M}}\right|
_{ P_{0,M}=E_{n_f}}.
\label{eq:LLfactor}
\end{eqnarray}
For the problem at hand $E_{K}$ is equal to the energy of the incoming kaon and the symmetry factor $\xi$ is equal to $1/2$. If one wishes, it is straight forward to replace the derivative with respect to total energy with a derivative with respect to relative momentum. Doing so, one finds agreement with Refs.~\cite{Lellouch:2000pv, Kim:2005gf, Christ:2005gi, Lin:2001ek} in the limit that the initial and final state are exactly degenerate. Note that, since the current is evaluated at a specific time slice, the current need not conserve energy and Eq.~\ref{eq:LLfactor} reflects this fact. For a process such as $K\rightarrow \pi\pi$ this is an artifact, and one must set the $\pi\pi$ energy to be degenerate with the kaon in order to extract the physical decay amplitude.

 \subsubsection{$\pi\gamma^*\rightarrow \pi\pi$ transition amplitude \label{sec:pigammapipi}}
 
Unlike the previous example, for a process such as $\pi\gamma^*\rightarrow \pi\pi$ the external current can inject arbitrary momentum. For such a process, the lowest energy configuration of the final state is a P-wave. Therefore, it is expected that the Lellouch-L\"uscher factor gets modified. Since the two particles in the final state are degenerate, odd and even partial waves cannot mix. By ignoring contamination from the F-wave and using the results of Section~\ref{sec:pipiPwave} one finds the generalization of the previous result for two particles in a P-wave,
\begin{eqnarray}
\frac{|\mathcal{A}_{\Lambda_f\mu_f,n_f;\Lambda\mu; J_f=1}|^2}
{|\langle \pi\pi,E_{n_f}\textbf{P}_f,\Lambda_f\mu_f;L|\tilde{\mathcal{J}}_{\Lambda\mu}^{[1,-1,|\lambda|]}(0,\textbf{P}_i-\textbf{P}_f)| \pi,E_{i}\textbf{P}_i;L\rangle|^2}
&=&
\frac{16\pi E_{i}~E^{*}_{n_f}}{ q^{*}_{n_f} \xi}
\left.~\frac{\partial (\delta_P+\phi^\textbf{d}_{P})}
{\partial P_{0,M}}\right|
_{ P_{0,M}=E_{n_f}},
\label{eq:LLfactorP}
\end{eqnarray}
where $\phi^\textbf{d}_{P}$ has been defined in Eq.~\ref{eq:Ppseudophase}. Also as already discussed after that equation, $\xi=1/2$ must be used if $\mathcal A$ is defined in the isospin basis and $\xi=1$ must be used if $\mathcal A$ is defined with unsymmetrized two pion states. The two choices are consistent since the two definitions of the transition amplitude differ by a factor of $\sqrt{2}$. The $J_f=1$ superscript on the transition amplitude means that we have integrated it against one of the $l_f=1$ spherical harmonics. As discussed above, this projection is performed in the two-particle center of mass frame. Observe that the right-hand side does not depend on the representation of the current or the single-particle state.

The right-hand side effectively corrects for the large finite-volume artifacts associated with the two-particle state. This gives a one-to-one mapping between the finite-volume matrix element and infinite-volume transition amplitude for this process. The result thus allows one to determine, using LQCD, the same quantity that is extracted from experiments. If one wants to evaluate this transition amplitude at the $\rho$ pole, in order to study processes such as $\pi\gamma^*\rightarrow\rho$, then it is necessary to analytically continue into the complex plane~\cite{Bernard:2012bi}. This requires parameterizing the transition amplitude as a function of the exchange momentum as well as the relative momentum between the two pions in the P-wave. By fitting this function to the LQCD results, one can study the behavior of the transition amplitude as a function of the exchange momentum at the resonance pole. 

 \subsubsection{ Two-dimensional case \label{sec:2Dcase}}

As we have already stressed, partial wave mixing is inevitable when performing calculations in a finite volume and this mixing is quantified by our main result, Eq.~\ref{eq:matJaieps}. In addition, the final two-particle state may in general have overlap with more than one infinite-volume state. This leads us to consider a generic scenario where the matrix $\mathcal R$ in Eq.~\ref{eq:matJaieps} is two dimensional. In order to avoid introducing additional notation we consider the form of the main result using infinite-volume quantities that are defined via principal-value prescription, namely Eq.~\ref{eq:matJa}. 

In Section~\ref{sec:piKSPwave} we discussed one explicit example for a $\pi K$ boosted state, where we neglected contributions from $J_f\geq2$ partial waves. We could also consider a system with two open channels where we ignore partial wave mixing, e.g., $\pi \pi-K\bar{K}$. In the first case, the finite volume matrix $\mathbb{F}^V_{\Lambda_f}$ will have off-diagonal terms but the K-matrix will be diagonal. In the second case this is reversed; the K-matrix has non-zero off diagonal terms while $\mathbb{F}^V_{\Lambda_f}$ is diagonal. In order to accommodate these two scenarios simultaneously, we allow the K-matrix and the $\mathbb{F}^V_{\Lambda_f}$ matrix to have off diagonal terms. The spectrum of this system must satisfy 
\begin{eqnarray}
{\det[\mathbb{M}_{\Lambda_f}]}={\det\left[\mathbb{K}_{\Lambda_f}+ \left(\mathbb{F}^V_{\Lambda_f}\right)^{-1}\right]}=0.
\label{eq:2DimQC}
\end{eqnarray}
By restricting $\mathbb{M}_{\Lambda_f}$  to be a two-dimensional matrix, its adjugate can be written as
\begin{eqnarray}
\rm{adj}[{\mathbb{M}}_{\Lambda_f}]\large|_{ P_{0,M}=E_{\Lambda_f,n_f}}=
\left(
\begin{array}{cc}
~~\left[{\mathbb{M}}_{\Lambda_f}\right]_{22}
 & 
-\left[{\mathbb{M}}_{\Lambda_f}\right]_{12} \\
- \left[{\mathbb{M}}_{\Lambda_f}\right]_{21}&
~~\left[{\mathbb{M}}_{\Lambda_f}\right]_{11}
  \\
\end{array}
\right).
\label{eq:2Dimadj}
\end{eqnarray}
By requiring $\mathbb{M}$ to satisfy Eq.~\ref{eq:2DimQC}, we note that not all the elements of its adjugate are independent.

Inserting the above expression into Eq.~\ref{eq:matJa}, one finds
\begin{multline}
\left|\langle E_{\Lambda_f,n_f} \textbf P_f; L|\tilde{\mathcal{J}}_{\Lambda\mu}^{[J,P,|\lambda|]}(0,\textbf{P}_i-\textbf{P}_f)| E_{\Lambda_i,0} \textbf P_i; L\rangle\right|^2
=\\
 \frac{1}
 {{2E_{\Lambda_i,0}}}\left.\left(\frac{
 |[\mathbb{H}]_1|^2~{\left[{\mathbb{M}}_{\Lambda_f}\right]_{22}}
 +|[\mathbb{H}]_2|^2~{\left[{\mathbb{M}}_{\Lambda_f}\right]_{11}} 
 - ~[\mathbb{H}]_1^*~[\mathbb{H}]_2~\left[{\mathbb{M}}_{\Lambda_f}\right]_{12} 
 - ~[\mathbb{H}]_2^*~[\mathbb{H}]_1~\left[{\mathbb{M}}_{\Lambda_f}\right]_{21} 
 }
 {\text{tr}\left[ \text{adj}[{\mathbb{M}}_{\Lambda_f}]\frac{\partial {\mathbb{M}}_{\Lambda_f}}{\partial P_{0,M}}\right]}\right)\right|
_{ P_{0,M}=E_{\Lambda_f,n_f}}
  \label{eq:LLfactor2D}
\end{multline}
where the subscripts of $\mathbb{H}_{\Lambda_f\mu_f,n_f;\Lambda\mu}$ have been suppressed in the last line for compactness. Here we emphasize, that although the full transition amplitude is real, the spherical harmonic decomposition may in general be complex. This is due to the fact that the spherical harmonics are themselves complex. This result illustrates the power of Eq.~\ref{eq:matJaieps}.  


 \subsubsection{$D \rightarrow \left\{\pi\pi,K\bar{K}\right\}$ decays \label{sec:DpipiKK}}
Assuming sufficiently heavy pion masses, such that the multi-particle threshold lies above the energy of the $D$ meson, Eq.~\ref{eq:LLfactor2D} allows one to study $D \rightarrow \left\{\pi\pi,K\bar{K}\right\}$ decays. To find the equivalence between the result presented in the previous section and the result presented in Ref.~\cite{Hansen:2012tf}, we rederive the result of Ref.~\cite{Hansen:2012tf} using notation presented here. This allows for a more compact representation of the result. In Ref.~\cite{Hansen:2012tf}, the authors follow the trick first utilized by Lellouch and L\"uscher to describe $K\rightarrow\pi\pi$ decays. We present this method in the context of the two-channel system. 

The argument proceeds by modifying the $\pi\pi-K\bar{K}$ correlation function, by including a contribution to the Hamiltonian density due to the weak interaction. We denote this perturbative shift by $\lambda \mathcal{H}_W(x)$, where $\lambda$ is a free parameter that allows us to organize an expansion. The modified Hamiltonian density couples $\pi\pi-K\bar{K}$ with the $D$ state, both in a finite and an infinite volume. Considering first the finite-volume theory, we tune the box size $L$ such that the $D$ state and some $\pi\pi-K\bar{K}$ finite-volume state are exactly degenerate (for a given total momentum). The presence of the weak interaction will break the degeneracy and result in two nearly degenerate states with energies
\begin{eqnarray}
\label{eq:dEw}
E^{(1)}= E_{D} \pm\lambda L^3~
|\langle E_{D}\textbf{P};L|\mathcal{H}_W(0)| D,E_{D}\textbf{P};L\rangle|,
\end{eqnarray}
where we have only kept the leading order contribution in $\lambda$ and where $E_{D} = \sqrt{M_D^2 + \textbf P^2}$ with $M_D$ the $D$ meson mass. 

Turning to the infinite-volume theory, the weak perturbation has the effect of modifying the scattering amplitude. This modification is due to the additional interaction that couples the $D$ to the two-particle states. The shift in the scattering amplitude contains two insertions of the weak Hamiltonian, one for transitioning from two particles to the $D$ and one for transitioning back to two particles. Thus the shift is generically $\mathcal O(\lambda^2)$, but in the present case we evaluate the amplitude at an energy which is shifted by $\mathcal O(\lambda)$ from $E_{D}$. This enhances the shift in the scattering amplitude to be $\mathcal O(\lambda)$. One finds~\cite{Hansen:2012tf}
\begin{eqnarray}
\mathcal{M}^{(1)}=\mathcal{M}^{(0)}\mp\lambda \Delta~\mathcal{M}
\end{eqnarray}
where
\begin{eqnarray}
\label{eq:dmathcalM}
\Delta\mathcal{M}
=\frac{1}{2E_{D} \Delta E}
\left(
\begin{array}{cc}
|\mathcal{A}_{D\rightarrow\pi\pi}|^2& \mathcal{A}_{D\rightarrow\pi\pi} \mathcal{A}_{D\rightarrow K\bar{K}}^\dag\\
\mathcal{A}_{D\rightarrow\pi\pi}^\dag \mathcal{A}_{D\rightarrow K\bar{K}}&|\mathcal{A}_{D\rightarrow K\bar{K}}|^2
  \\
\end{array}
\right)\,,
\end{eqnarray}
and where we have defined $\Delta E \equiv L^3 |\langle E_{D}\textbf{P};L|\mathcal{H}_W(0)| D,E_{D}\textbf{P};L\rangle|$.

We next find it convenient to rewrite this perturbation to the scattering amplitude as a perturbation to the K-matrix. To do this, we follow the reasoning of Eq.~\ref{AjHjtildematrix} and observe that the only difference between the transition amplitude and the scattering amplitude is that for the latter we need to include the imaginary part of the diagrams associated with both incoming as well as outgoing two particle states. This leads to the following relation between $\Delta\mathcal{M}$ and $\Delta\mathbb{K}$, 
\begin{eqnarray}
\label{eq:dmathbbK}
\Delta\mathbb{K}=\mathbb{K}~\mathcal{M}^{-1}~\Delta\mathcal{M}~\mathcal{M}^{-1}~\mathbb{K}.
\end{eqnarray}
At this point we can combine the shift in the finite-volume spectrum with the shift in the infinite-volume K-matrix to determine the leading order modification to $\mathbb{M}$, defined in Eq.~\ref{eq:mathbbM}. We find that the matrix is shifted by an amount
\begin{eqnarray}
\label{eq:dmathbbM}
\lambda\Delta\mathbb{M}=\lambda\Delta E\left.\frac{\partial{\mathbb{M}}}{\partial P_{0,M}}\right|
_{ P_{0,M}=E_D}\mp\lambda\Delta \mathbb{K} \,.
\end{eqnarray}
Of course, the quantization condition must also be valid for the perturbed theory. We thus deduce that the linear shift to the determinant of $\mathbb M$ should vanish
\begin{eqnarray}
{\det[ \mathbb{M}(\lambda)]}|&=&
{\det[ \mathbb{M}(0)]}
+\lambda\text{tr}\left[ \text{adj}[\mathbb{M}(0)]~\Delta\mathbb{M}\right] =\lambda\left.\text{tr}\left[ \text{adj}[\mathbb{M}(0)]~\Delta\mathbb{M}\right]\right|
_{ P_{0,M}=E_D} =0,
\label{eq:altLLtwoch}
\end{eqnarray}
where we have used the fact that $\mathbb M(0)$ also has vanishing determinant, since this defines the quantization condition of the unperturbed theory. 

Showing that this result is equivalent to Eq.~\ref{eq:LLfactor2D} requires some algebra. First we substitute Eq.~\ref{eq:dmathbbM} into Eq.~\ref{eq:altLLtwoch} and solve for $\Delta E$
\begin{equation}
 \Delta E =\left.\frac{\text{tr}\left[ \text{adj}[\mathbb{M}] \Delta\mathbb{K}\right]}{\text{tr}\left[ \text{adj}[\mathbb{M}]  \frac{\partial{\mathbb{M}}}{\partial P_{0,M}}\right]}
\right|
_{ P_{0,M}=E_D} \,.
\end{equation}
Next we substitute the specific two-channel form, Eq.~\ref{eq:2Dimadj}, and also use Eqs.~\ref{AjHjtildematrix}, \ref{eq:dmathcalM} and \ref{eq:dmathbbK} to simplify the result. We conclude
\begin{equation}
\Delta E^2 = L^6
|\langle E_D\textbf{P};L|\mathcal{H}_W(0)| D,E_{D}\textbf{P};L\rangle|^2
=
 \frac{1}
 {{2E_D}}\left.\left(\frac{
 [\mathbb{H}]_1^2~{\left[{\mathbb{M}}_{\Lambda_f}\right]_{22}}
 +[\mathbb{H}]_2^2~{\left[{\mathbb{M}}_{\Lambda_f}\right]_{11}} 
 - 2~[\mathbb{H}]_1~[\mathbb{H}]_2~\left[{\mathbb{M}}_{\Lambda_f}\right]_{12} 
 }
 {\text{tr}\left[ \text{adj}[{\mathbb{M}}]\frac{\partial {\mathbb{M}}}{\partial P_{0,M}}\right]}\right)\right|
_{ P_{0,M}=E_D} \,, 
\end{equation}
which is equivalent to Eq.~\ref{eq:LLfactor2D} for the special case where the initial and final states are exactly degenerate and have the same total momentum. Furthermore, since the outgoing two-particle state is in an S-wave, all of the elements in the right hands side of the equation above are real. Note that the left hand side of the above equation contains an extra factor of $L^6$. This is because the current in Eq.~\ref{eq:LLfactor2D} is in momentum-space. 

We note that although it might seem that this result is sensitive to the relative sign between $[\mathbb{H}]_1$ and $[\mathbb{H}]_2$, our result only allows one to determine the sign of $[\mathbb{H}]_1~[\mathbb{H}]_2~\left[{\mathbb{M}}_{\Lambda_f}\right]_{12}$. The determinant condition describing the spectrum is only sensitive to the magnitude of $\left[{\mathbb{M}}_{\Lambda_f}\right]_{12}$. Therefore, we find no method here for determining the relative sign of $[\mathbb{H}]_1$ and $[\mathbb{H}]_2$.

 \subsubsection{$B \rightarrow \pi K$ transition amplitudes \label{sec:BKellell}}

One example where partial wave mixing may in general not be small is in studies of $B\rightarrow \pi K$ transition amplitudes. This is due to the fact that for boosted systems the final state will be an admixture of even and odd partial waves. In particular, if one is interested in the case where the infinite volume final state has overlap with the $K^*(892)$ resonance, then one must consider irreps that have strong overlap with the $\pi K$ P-wave. If the final state is at rest or if it is in the $\mathbb{E}$ irrep for $\textbf{d}=(00n)$, $\mathbb{B}_1$ and $\mathbb{B}_2$ for $\textbf{d}=(nn0)$, or $\mathbb{E}$ for $\textbf{d}=(nnn)$, and if one neglects the contribution from the D and higher partial waves by following the discussion of Section~\ref{sec:piKSPwave}, then one finds that the ratio of the infinite transition amplitudes and finite volume matrix elements of vector or pseudo vector currents satisfies Eq.~\ref{eq:LLfactorP}.

For the $\mathbb{A}_1$ irrep of the Dic$_4$ group, one simply needs to insert the expressions for the on-shell K-matrix in Eq.~\ref{eq:KmatpiK} along with $\mathbb{F}_{\mathbb{A}_1}$ in Eq.~\ref{eq:FVA1Dic4} onto Eq.~\ref{eq:LLfactor2D} to find the relation between the finite volume matrix elements of currents and infinite volume transition amplitudes. Because of the symmetries of the infinite volume, only one of the transition amplitudes is non vanishing. For example, if we consider the case where the current is subduced from $J_f=1$ with odd parity, then $\mathbb{H}_{S,n_f;\Lambda\mu}$ must exactly vanish. Therefore, for vector currents Eq.~\ref{eq:LLfactor2D} simplifies to
\begin{eqnarray}
\frac{|\mathcal{A}_{P0,n_f;\Lambda\mu}|^2}
{|\langle \pi K,E_{n_f}\textbf{P}_f,\Lambda_f\mu_f;L|\tilde{\mathcal{J}}_{\Lambda\mu}^{[1,-1,|\lambda|]}(0,\textbf{Q})| B^0,E_{B^0}\textbf{P}_i;L\rangle|^2}
&=&
2E_{\Lambda_i,0}~\cos^2\delta_P~
 \left|
 \frac{\text{tr}\left[ \text{adj}[{\mathbb{M}}_{\Lambda_f}]\frac{\partial {\mathbb{M}}_{\Lambda_f}}{\partial P_{0,M}}\right]}
 {\left[{\mathbb{M}}_{\Lambda_f}\right]_{11}}\right|
_{ P_{0,M}=E_{\Lambda_f,n_f}}.
 \label{eq:LLfactorvecA1D}
\end{eqnarray}
where $\mathcal{A}_{P0,n_f;\Lambda\mu}$ denotes the P-wave transition amplitude with zero helicity. This follows from the helicity decomposition of the $\mathbb{A}_1$ irrep of the Dic$_4$ group as shown in Table~\ref{table:irreps}\ref{table:subducb}. For a pseudovector current or for rank two tensor currents neither $\mathbb{H}_{S,n_f;\Lambda\mu}$ nor $\mathbb{H}_{Pm,n_f;\Lambda\mu}$ need vanish. Therefore one necessarily must use Eq.~\ref{eq:LLfactor2D}. For the $\mathbb{A}_1$ irreps of the Dic$_2$ group one must input the finite volume function and scattering matrices defined in Section~\ref{sec:piKSPwave} into the general result for the matrix element of currents, Eq.~\ref{eq:matJa}.

As discussed in the previous section, this result does not require that the initial and final state are exactly degenerate. For studies of B meson decays on the lattice is a necessity since the formalism does not currently support multi-particle states. Therefore this result is of most significance for studies of B meson decays with large energy exchange, while the momentum exchange could be arbitrarily small.

Finally, it is important to remember that if one is interested in studying transition amplitudes involving the isospin-1/2 $K\pi$ final state, one necessarily must consider the admixture of this with $K\eta$. Although the inelasticity is seen to be small at physical values, this will depend on the quark masses used to perform the calculation.\footnote{{It is important to remember that at the physical point, the $\eta$ is a resonance that decays approximately one third of time to $3\pi^0$~\cite{Beringer:1900zz}. However the width of $1.31\pm 0.05$~keV~\cite{Beringer:1900zz} is sufficiently narrow that treating the resonance as stable would likely be a good approximation. In this approach, resolving the finite volume spectrum where the $\eta$ resides will most likely require having three-body operators. Additionally, at the physical point one can no longer neglect mixing between $K\pi$ and $K\pi\pi$, which our formalism does not accommodate.}} Furthermore, for unphysically large quark masses, such as those in used in Refs.~\cite{Horgan:2013hoa, Horgan:2013pva, Dudek:2014qha}, the $K\eta$ threshold is significantly closer to the $K\pi$ threshold than it is in nature. In order to include this mixing between the channels one will have to use Eq.~\ref{eq:LLfactor2D} when there are two open channels with negligible partial wave mixing or in general Eq.~\ref{eq:matJaieps}. 


\section{Conclusion}

In this work we present a non-perturbative derivation of two and three-point functions in the mesonic sector.  In Section~\ref{sec:corr2} we explicitly demonstrate how to construct operators with the appropriate symmetries of a finite volume system. This allowed us to write down the correlator as a function of time and energy, Eq.~\ref{eq:corrP1}. We find that although the spectrum solely depends on the on-shell scattering amplitudes, the correlation function also depends on off-shell scattering amplitudes. Furthermore, the result presented explains why if one constructs an operator with a particular set of discrete momenta, then the resulting correlation function will be dominated by the nearest eigenstate. This is because the overlap of an operator with a state, Eq.~\ref{eq:On}, scales as $\sim {|E_{\Lambda,n}-E_{free}|}^{-1}$, where $E_{free}$ stands for the free energy of the two-particle system and $E_{\Lambda,n}$ is the $nth$ eigenstate of the $\Lambda$ irrep of the corresponding symmetry group. 
 
In Section~\ref{sec:corr3} we discuss the construction and interpretation of three-point correlation functions in the mesonic sector. Section~\ref{sec:current} reviews the work of  Ref.~\cite{Thomas:2011rh} in the construction of currents that have been properly subduced onto an irrep of the symmetry group of the system. Having defined the subduced currents, in Section~\ref{sec:corr3diagram} we evaluate the three-point correlation function diagrammatically to all orders in perturbation theory, Eq.~\ref{eq:corr32}. By comparing the expression of the three-point function with Eq.~\ref{eq:On}, we find a master equation for the matrix element of currents between a one and two-particle finite-volume state, Eq.~\ref{eq:matJaieps}. This result is the generalization of the Lellouch-L\"uscher formula, relating matrix elements of currents in finite and infinite volume, to processes where the external current can inject arbitrary total momentum into the system and the final state can be in an arbitrary partial wave. The generalization also includes an arbitrary number of strongly coupled two-particle channels. The result is exact up to exponentially suppressed volume corrections that are governed by $L m_\pi$. In Section~\ref{sec:Kpipi} we demonstrate that this result recovers the well known $K\rightarrow\pi\pi$ result. Section~\ref{sec:pigammapipi} demonstrates how one determines the $\pi\gamma^*\rightarrow \pi\pi$ transition amplitude. Section~\ref{sec:2Dcase} gives a generic expression for the determination of finite-volume matrix elements where there are two coupled channels open, Eq.~\ref{eq:LLfactor2D}. Equation~\ref{eq:LLfactor2D} is relevant for two channel systems, regardless of whether the mixing is physical or an artifact of the reduction of rotational symmetry in a finite volume. Section~\ref{sec:BKellell} demonstrates how to implement this formalism for future studies of $B\rightarrow \pi K$ transition amplitudes, where the final state is properly treated as a scattering state. Finally, we remark that although we have chosen to perform the derivation using a current that has been subduced onto an irrep of the symmetry group of the system, one can implement the formalism derived here to currents that do not satisfy this criteria. This is due to the fact that one can also write any current as a linear combination of subduced currents.

\noindent
\subsection*{Acknowledgments}
\noindent

R.B. and A.W.L. acknowledge support from the U.S. Department of Energy contract DE-AC05-06OR23177, under which Jefferson Science Associates, LLC, manages and operates the Jefferson Lab.  M.T.H. was supported in part by DOE grant No. DE-FG02-96ER40956. R.B. would like to thank Robert Edwards, Kostas Orginos, Christian Shultz, Jozef Dudek, David Wilson, Christopher Thomas, David Richards, Michael Peardon, Zohreh Davoudi, Igor Danilkin, Roman Zwicky, Akaki Rusetsky, and Stefan Meinel for many useful discussions. MH would like Stephen Sharpe for useful discussions. The authors would like to thank Robert Edwards and Jozef Dudek for motivating the work by posing the question as to whether it is possible to rigorously study transition amplitudes via lattice QCD.

\appendix


\section{Cancelation of free poles\label{sec:freepoles}}
In arriving at the final expression for the two-point correlation, Eq.~\ref{eq:corrP1}, we argued that the free particle poles of the integrand of Eq.~\ref{eq:corrP0} do not contribute. Here we give a proof of this statement. In Sections~\ref{sec:corr2} and \ref{sec:corr3}, we constructed operators that are in the irrep of the symmetry group of the system, but the cancelation of free poles cannot depend on this fact. It must only depend on the fact that the particle interactions are not exactly zero. If one would choose to not define an operator with good quantum numbers, then Eq.~\ref{eq:corr21} would acquire an additional sum over all possible irreps that have overlap with the operator of interest. This in turn would lead to a far less reliable extraction of the spectrum since multiple irreps could in principle have nearly degenerate eigenstates. With this caveat in mind, we decide to illustrate the cancelation of free particle poles using a set of generic operators with no restrictions on quantum numbers
\begin{equation}
\mathcal A(x_0, \textbf{P}) = \sum_{\textbf{k}} a(\textbf{k}) \varphi({x_0,\textbf{P}-\textbf{k}})~\tilde{\varphi}({x_0,\textbf{k}}) \,,
\hspace{2cm}
 \mathcal B({x_0, \textbf{P}}) = \sum_{\textbf{k}}  b(\textbf{k})\varphi^\dag({x_0,-\textbf{P}+\textbf{k}})~\tilde{\varphi}^\dag ({x_0,-\textbf{k}})\,,
\end{equation}
where $a(\textbf{k})$ and $b(\textbf{k})$ are generic functions of $\textbf{k}$. Note that we have not specified wether the sum is over all possible values of $\textbf{k}$ or one specific shell; this distinction does not matter. 

\begin{figure*}[t]
\begin{center} 
\includegraphics[scale=0.45]{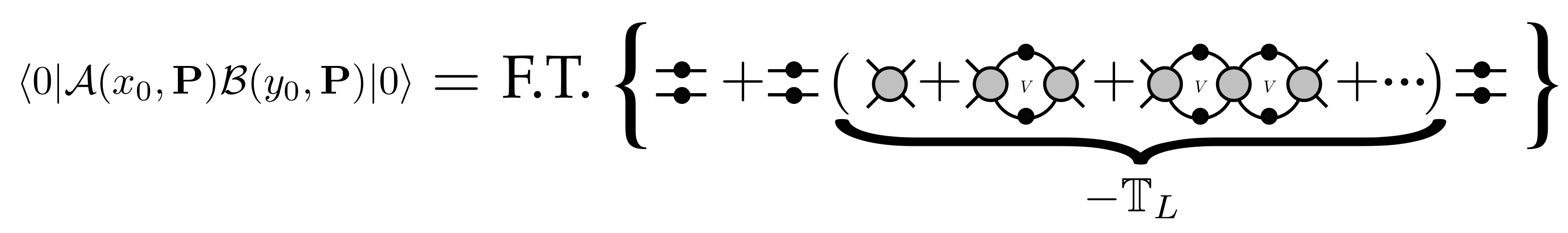} 
\caption{ 
Shown is the diagrammatic representation of the correlation function defined in Eq.~\ref{eq:ABcorr} in terms of the kernels (defined in Fig.~\ref{fig:kernel}), the fully dressed single particle propagators (defined in Fig.~\ref{fig:1bodyprop}) and the finite volume loops. The ``F. T." label around the braces reminds the reader that one must Fourier transform the energy-momentum correlation function to obtain the correct exponential dependence in time. The $T$ function, which is explicitly labeled, is defined in Eq.~\ref{eq:TLdef}.}\label{fig:TL}
\end{center}
\end{figure*}

Since the cancelation of free poles is not affected by the number of open channels, we restrict the discussion here to the case of only one open channel. It is straightforward to write down the two-point correlation function (depicted in Fig.~\ref{fig:TL}) in the vicinity of the free poles 
\begin{multline}
\langle 0 \vert \mathcal A({x_0, \textbf{P}})  \mathcal B({y_0, \textbf{P}}) \vert 0 \rangle =L^3  \int \frac{dP_0}{2 \pi}   e^{i P_0 (x_0-y_0)}   \Bigg[ \sum_{\textbf{k}}  \frac{- i L^3}{4 \omega_{1,\textbf{P}-\textbf{k}} \omega_{2,\textbf{k}}} \frac{a(\textbf{k}) b(\textbf{k})}{P_0 - i (\omega_{1,\textbf{P}-\textbf{k}}  + \omega_{2,\textbf{k}} )}\\
 - \sum_{\textbf{k}, \textbf{k}'} \frac{ a(\textbf{k}) b(\textbf{k}')~\mathbb T_L(P,k,k')}{4 \omega_{1,\textbf{P}-\textbf{k}} \omega_{2,\textbf{k}}  [  P_0 - i(\omega_{1,\textbf{P}-\textbf{k}}  + \omega_{2,\textbf{k}})]4 \omega_{1,\textbf{P}-\textbf{k}'} \omega_{2,\textbf{k}'}[  P_0 - i(\omega_{1,\textbf{P}-\textbf{k}'}  + \omega_{2,\textbf{k}'})]}      +\cdots \Bigg ] \,,
\label{eq:ABcorr}
\end{multline}
where the ellipses denote finite contributions to the correlation function near the free poles. In writing the correlation function we have introduce a function $\mathbb T_L(P,k,k')$, which is related to the K-matrix via Eq.~\ref{eq:TLdef}. Near the free particle poles this can be written as 
\begin{equation}
\label{eq:Tmatrix}
 \mathbb T_L(P,k,k') =  -\mathbb K(P,k,k') + i \bigg[ \frac{1}{L^3} \sum_{\textbf{l}} - \int_{\textbf{l}} \bigg ] \frac{ \mathbb K(P,k,l)  \mathbb T_L(P,l,k') }{4 \omega_{1,\textbf{P}-\textbf{l}} \omega_{2,\textbf{l}}[  P_0 - i(\omega_{1,\textbf{P}-\textbf{l}}  + \omega_{2,\textbf{l}})]} ,
\end{equation} 
where we have neglected contributions which are exponentially suppressed in $m_\pi L$. The free particle poles satisfy $P_0 = i(\omega_{1,\textbf{P}-\textbf{k}}  + \omega_{2,\textbf{k}})$ and in order to obtain the contribution of these, we investigate the leading \(\epsilon\) behavior, where \(\epsilon\) is defined via
\begin{equation}
P_0 = i(\omega_{1,\textbf{P}-\textbf{k}}  + \omega_{2,\textbf{k}}) + \epsilon \,.
\end{equation}
To do so, we upgrade these functions to matrices in momentum space. It is important to observe that, in general, there will be multiple values of $\textbf{k}$ and $\textbf{P}-\textbf{k}$ that  satisfy the free energy condition, specifically all elements of are $\{\textbf{k}\}_\Pstar$ and $\{\textbf{P}-\textbf{k}\}_\Pstar$. Defining $\omega_{1}$ and $\omega_{2}$ as the free energies that satisfy $P_0 = i(\omega_{1}  + \omega_{2})$, at leading order in $\epsilon$, Eq.~\ref{eq:ABcorr} simplifies to 
\begin{equation}
\label{eq:ABmatform}
\langle 0 \vert \mathcal A({x_0, \textbf{P}})  \mathcal B({y_0, \textbf{P}}) \vert 0 \rangle = - i L^3 a  \left[ \frac{1}{4 \omega_1 \omega_2 \epsilon}\right]^\Nstar b  -  a \left[ \frac{1}{4 \omega_1 \omega_2 \epsilon}\right]^\Nstar \mathbb T_L \left[ \frac{1}{4 \omega_1 \omega_2 \epsilon}\right]^\Nstar b \,.
\end{equation}
Here \(a\) is understood as a row and \(b\) as a column vector, \([1/(4 \omega_1 \omega_2 \epsilon)]^\Nstar\) is a matrix that acts in the restricted space of $\{\textbf{k}\}_\Pstar$ and $\{\textbf{P}-\textbf{k}\}_\Pstar$ with value equal to $1/(4 \omega_1 \omega_2 \epsilon)$, while \(\mathbb T_L\) is a matrix with off-diagonal entries. If we next restrict attention to the set of momenta that satisfy the free energy conditions, then the $T$-matrix, $Eq.~\ref{eq:Tmatrix}$, satisfies
\begin{equation}
 \mathbb T_L =  - \mathbb K + i \frac{1}{L^3} \mathbb K \left[ \frac{1}{4 \omega_1 \omega_2 \epsilon}\right]^\Nstar \mathbb T_L \,.
\end{equation}
At this stage we observe that, since \(\mathbb K = \mathcal O(1)\), one can shown that 
\begin{equation}
\mathbb T_L = -i L^3 \left[ \frac{1}{4 \omega_1 \omega_2 \epsilon}\right]^{-1} + \mathcal O(\epsilon^2) \,. 
\end{equation}
Substituting this into Eq.~\ref{eq:ABmatform} gives perfect cancellation of the \(\mathcal O(1/\epsilon)\) terms independent of the values of \(a\) and \(b\). This justifies the cancellation of free particle poles in Eq.~\ref{eq:corrP0}, which is recovered by setting \(a\) and \(b\) equal to the Clebsch-Gordan coefficients.

However, it is common practice to restrict the scattering amplitude to a particular partial wave when obtaining the finite volume spectrum. Here we demonstrate how this approximation can lead to spurious free poles in the correlation function. Let $\mathbb K_S(n,P_0)$ and $ \mathbb T_S(n,P_0)$ be the S-wave K-matrix and $T$ functions at the $nth$ free particle pole which has a degeneracy of  $N$. From Eq.~\ref{eq:Tmatrix}, we see that these satisfy 
\begin{eqnarray}
 \mathbb T_S(n,P_0) &=&  -\mathbb K_S(n,P_0) + i \frac{N}{L^3}    \frac{\mathbb K_S(n,P_0)  \mathbb T_S(n,P_0) }{4 \omega_1 \omega_2  \epsilon}  +\mathcal{O}(\epsilon^2) \,,\\
\Rightarrow 
\label{eq:Tswavedone}
 \mathbb T_S(n,P_0) &=& - i \frac{4 \omega_1  \omega_2  L^3 }{N} \epsilon +\mathcal{O}(\epsilon^2)\,.
\label{eq:Tswavetwo}
\end{eqnarray} 
Substituting Eq.~\ref{eq:Tswavedone} into the S-wave reduction of Eq.~\ref{eq:ABcorr}, we deduce that free particle poles only cancel when
\begin{equation}
(1/N) 
\sum_{R,R' \in {\rm LG(\textbf{P})}}
 a(R\textbf{k}) b(R\textbf{k}') = 
 \sum_{R \in {\rm LG(\textbf{P})}} a(R\textbf{k}) b(R\textbf{k}) \,.
\end{equation}
If one chooses \(a\) and \(b\) to be Kronecker deltas in momentum, as has been done in previous work, the cancellation in Eq.~\ref{eq:ABcorr} is lost, unless $N=1$. But this is a contradiction to the statement above, that free particle poles should not appear regardless of the values of \(a\) and \(b\) for any momentum. The apparent contradiction here is resolved by noting that the matrix \(\mathbb K\) is only invertible if each row is linearly independent. However, in the case of S-wave amplitude the matrix is proportional to a matrix which has \(1\) in every single entry. Thus the matrix argument fails and the alternative argument shows that cancellation does not occur for all \(a\) and \(b\). Furthermore, we argue that imposing a scattering amplitude to exactly vanish for all but one partial waves at all values of momentum is unnatural. The only way to achieve this is to require all shape parameters of the partial waves not included to be equal to zero. Restricting the final results of quantization condition, the matrix elements of the two-particle interpolating operator and the matrix elements of the currents, Eqs.~\ref{eq:QC}, \ref{eq:Onab}, \ref{eq:On}, \ref{eq:matJ} \&  \ref{eq:matJa}, to a single partial wave can be done if the contribution from higher partial waves is seen to be significantly suppressed at low energies. This is to say that the order of operations in studying finite volume physics is relevant and can lead to significantly different results. 

From this discussion it is clear that if one is solely interested in obtaining the spectrum and is not in arriving at a nonperturbative expression for the correlation functions, it suffices to look at the poles of \(  \mathbb T_L\). As is evident from Fig.~\ref{fig:TL}, the free particle poles correspond to zeros of \(  \mathbb T_L\), and consequently one does not need to worry about any spurious poles. Furthermore, the subtlety regarding the order of operations does not play a role when studying the pole structure of \(  \mathbb T_L\). Therefore, as was done in Ref.~\cite{Beane:2003da}, one may first proceed to set the angular momentum to any partial wave desired and then obtain the quantization condition from the pole structure of  \(  \mathbb T_L\).

 \section{Generalization for twisted boundary conditions in asymmetric volumes\label{sec:SpinFVTBCs}}

In the derivation of the master equations of this work, namely Eqs.~\ref{eq:QC}, \ref{eq:Onab}, \ref{eq:On}, \ref{eq:matJ} \&  \ref{eq:matJa}, periodic boundary condition on the spatial extent of the cubic volume have been assumed. The periodicity constraint is encoded in the expression for the $\mathcal{Z}$ functions shown in Eq.~\ref{eq:clm}, and this is generally true for arbitrary boundary conditions, and Ref.~\cite{Briceno:2014oea} demonstrated how to compactly write the $\mathcal{Z}$ functions in such a way that they accommodate the different geometries and boundary conditions. For relevant work that lead to this result, see Refs.~\cite{Li:2003jn, Detmold:2004qn, Feng:2004ua, Bedaque:2004kc, Bedaque:2004ax, Bernard:2010fp, Doring:2011vk, Ozaki:2012ce, Briceno:2013hya, Agadjanov:2013wqa}. TBCs require that fields in general satisfy 
\begin{eqnarray}
{\psi}(\mathbf{x}+\mathbf{n}{{L}})=e^{i{\bm{\theta}} \cdot \mathbf{n}}{\psi}(\mathbf{x}),
\end{eqnarray}
where $\theta$ is a three-dimensional real angle. Therefore, the free momentum of the $ith$ particle in the $jth$ channel will be equal to $\textbf{p}_{j,i}=\frac{2\pi\textbf{n}_i}{L}+\frac{\bm{\phi}_{j,i}}{L}$.

 For asymmetric volumes, let $L$ to be the spatial extent of the z-axis and $\eta_i$ be defined such that $L_x=\eta_xL$ and  $L_y=\eta_yL$. Using the notation $\tilde{\bm{\chi}}=(\chi_x/\eta_x,\chi_y/\eta_y,\chi_z)$, one can readily find the most general form of the $c_{lm}$ and $\mathcal{Z}$ functions with arbitrary twist and asymmetric volumes
\begin{eqnarray}
c^{\textbf{d},\bm{\phi}_{j,1},\bm{\phi}_{j,2}}_{lm}(k^{*2};{L};\eta_x,\eta_y)
\ &=&\ \frac{\sqrt{4\pi}}{\eta_x\eta_y\gamma {L}^3}\left(\frac{2\pi}{{L}}\right)^{l-2}\times\mathcal{Z}^{\mathbf{d},\bm{\phi}_{j,1},\bm{\phi}_{j,2}}_{lm}[1;(k^*{L}/2\pi)^2;\eta_x,\eta_y],~~
\label{clmasymTBCs}\\
\mathcal{Z}^{\mathbf{d},\bm{\phi}_{j,1},\bm{\phi}_{j,2}}_{lm}[s;x^2;\eta_x,\eta_y]
\ &=&\ \sum_{\mathbf r \in \mathcal{P}_{\mathbf{d};\eta_x,\eta_y}^{\bm{\phi}_1,\bm{\phi}_2;}}
\frac{ |{\bf r}|^l \ Y_{lm}(\mathbf{r})}{(\mathbf{r}^2-x^2)^s},~~~
\label{ZlmasymTBCs}
\end{eqnarray}
where $ \mathcal{P}_{\mathbf{d};\eta_x,\eta_y}^{\bm{\phi}_1,\bm{\phi}_2}=\left\{\mathbf{r}\in \textbf{R}^3\hspace{.1cm} | \hspace{.1cm}\mathbf{r}={\hat{\gamma}}^{-1}(\tilde{\mathbf m}-\alpha_j \tilde{\mathbf d} +\frac{\tilde{\bm{\Delta}}^{(j)}}{2\pi})\right\}$, where $\textbf{m}$ is a triplet integer, $\tilde{\bm{\Delta}}^{(j)}=-({\alpha}_{j}-\frac{1}{2})(\tilde{\bm{\phi}}_{j,1}+\tilde{\bm{\phi}}_{j,2})+\frac{1}{2}(\tilde{\bm{\phi}}_{j,1}-\tilde{\bm{\phi}}_{j,2})$ and $\tilde{\mathbf d}={\mathbf P}L/2\pi$.  Additionally, one obtained an overall factor of $\sqrt{\eta_x\eta_y}$ in Eqs.~\ref{eq:phiLambda}, \ref{eq:Onab} \& \ref{eq:On}, i.e. one must make the following replacements
\begin{eqnarray}
|\langle 0|\varphi(0,\textbf{k})|E^{(1)}_{k}\rangle|
&\longrightarrow&\sqrt{\frac{\eta_x\eta_y L^3}{2\omega_k}}.
\label{eq:phiLambdaetaL}
\\
\label{eq:OnetaL}
|\langle 0|\mathcal{O}_{\Lambda\mu,a}(0,\textbf{P})|E_{\Lambda,n}\rangle|
&\longrightarrow&
\sqrt{\eta_x\eta_y L^3}~\sqrt{\left[
~{\mathbb{Y}}_{\Lambda\mu,n}~R_{\Lambda,n}~\mathbb{Y}^\dag_{\Lambda\mu,n}~\right]_{aa}}.
\end{eqnarray}

 
 
\bibliography{bibi}

\begin{thebibliography}{132}%
\makeatletter
\providecommand \@ifxundefined [1]{%
 \@ifx{#1\undefined}
}%
\providecommand \@ifnum [1]{%
 \ifnum #1\expandafter \@firstoftwo
 \else \expandafter \@secondoftwo
 \fi
}%
\providecommand \@ifx [1]{%
 \ifx #1\expandafter \@firstoftwo
 \else \expandafter \@secondoftwo
 \fi
}%
\providecommand \natexlab [1]{#1}%
\providecommand \enquote  [1]{``#1''}%
\providecommand \bibnamefont  [1]{#1}%
\providecommand \bibfnamefont [1]{#1}%
\providecommand \citenamefont [1]{#1}%
\providecommand \href@noop [0]{\@secondoftwo}%
\providecommand \href [0]{\begingroup \@sanitize@url \@href}%
\providecommand \@href[1]{\@@startlink{#1}\@@href}%
\providecommand \@@href[1]{\endgroup#1\@@endlink}%
\providecommand \@sanitize@url [0]{\catcode `\\12\catcode `\$12\catcode
  `\&12\catcode `\#12\catcode `\^12\catcode `\_12\catcode `\%12\relax}%
\providecommand \@@startlink[1]{}%
\providecommand \@@endlink[0]{}%
\providecommand \url  [0]{\begingroup\@sanitize@url \@url }%
\providecommand \@url [1]{\endgroup\@href {#1}{\urlprefix }}%
\providecommand \urlprefix  [0]{URL }%
\providecommand \Eprint [0]{\href }%
\providecommand \doibase [0]{http://dx.doi.org/}%
\providecommand \selectlanguage [0]{\@gobble}%
\providecommand \bibinfo  [0]{\@secondoftwo}%
\providecommand \bibfield  [0]{\@secondoftwo}%
\providecommand \translation [1]{[#1]}%
\providecommand \BibitemOpen [0]{}%
\providecommand \bibitemStop [0]{}%
\providecommand \bibitemNoStop [0]{.\EOS\space}%
\providecommand \EOS [0]{\spacefactor3000\relax}%
\providecommand \BibitemShut  [1]{\csname bibitem#1\endcsname}%
\let\auto@bib@innerbib\@empty
\bibitem [{\citenamefont {Andreev}\ \emph {et~al.}(2010)\citenamefont {Andreev}
  \emph {et~al.}}]{Andreev:2010wd}%
  \BibitemOpen
  \bibfield  {author} {\bibinfo {author} {\bibfnamefont {V.}~\bibnamefont
  {Andreev}} \emph {et~al.} (\bibinfo {collaboration} {MuSun Collaboration}),\
  }\href@noop {} {\  (\bibinfo {year} {2010})},\ \Eprint
  {http://arxiv.org/abs/1004.1754} {arXiv:1004.1754 [nucl-ex]} \BibitemShut
  {NoStop}%
\bibitem [{\citenamefont {Ando}\ \emph {et~al.}(2002)\citenamefont {Ando},
  \citenamefont {Park}, \citenamefont {Kubodera},\ and\ \citenamefont
  {Myhrer}}]{Ando:2001es}%
  \BibitemOpen
  \bibfield  {author} {\bibinfo {author} {\bibfnamefont {S.}~\bibnamefont
  {Ando}}, \bibinfo {author} {\bibfnamefont {T.}~\bibnamefont {Park}}, \bibinfo
  {author} {\bibfnamefont {K.}~\bibnamefont {Kubodera}}, \ and\ \bibinfo
  {author} {\bibfnamefont {F.}~\bibnamefont {Myhrer}},\ }\href {\doibase
  10.1016/S0370-2693(02)01619-2} {\bibfield  {journal} {\bibinfo  {journal}
  {Phys.Lett.}\ }\textbf {\bibinfo {volume} {B533}},\ \bibinfo {pages} {25}
  (\bibinfo {year} {2002})},\ \Eprint {http://arxiv.org/abs/nucl-th/0109053}
  {arXiv:nucl-th/0109053 [nucl-th]} \BibitemShut {NoStop}%
\bibitem [{\citenamefont {Bardin}\ \emph {et~al.}(1986)\citenamefont {Bardin},
  \citenamefont {Duclos}, \citenamefont {Martino}, \citenamefont {Bertin},
  \citenamefont {Capponi}, \citenamefont {Piccinini},\ and\ \citenamefont
  {Vitale}}]{Bardin}%
  \BibitemOpen
  \bibfield  {author} {\bibinfo {author} {\bibfnamefont {G.}~\bibnamefont
  {Bardin}}, \bibinfo {author} {\bibfnamefont {J.}~\bibnamefont {Duclos}},
  \bibinfo {author} {\bibfnamefont {J.}~\bibnamefont {Martino}}, \bibinfo
  {author} {\bibfnamefont {A.}~\bibnamefont {Bertin}}, \bibinfo {author}
  {\bibfnamefont {M.}~\bibnamefont {Capponi}}, \bibinfo {author} {\bibfnamefont
  {M.}~\bibnamefont {Piccinini}}, \ and\ \bibinfo {author} {\bibfnamefont
  {A.}~\bibnamefont {Vitale}},\ }\href@noop {} {\bibfield  {journal} {\bibinfo
  {journal} {Nucl. Phys.}\ }\textbf {\bibinfo {volume} {A453}},\ \bibinfo
  {pages} {591} (\bibinfo {year} {1986})}\BibitemShut {NoStop}%
\bibitem [{\citenamefont {Cargnelli}()}]{Cargnelli}%
  \BibitemOpen
  \bibfield  {author} {\bibinfo {author} {\bibfnamefont {M.~t.}\ \bibnamefont
  {Cargnelli}},\ }\href@noop {} {}\bibinfo {note} {In \textit{Proceedings of
  the XXIII Yamada Conf. on Nuclear Weak Processes and Nuclear Structure,
  Osaka, Japan}, 1989}\BibitemShut {NoStop}%
\bibitem [{\citenamefont {Chen}\ \emph {et~al.}(2005)\citenamefont {Chen},
  \citenamefont {Inoue}, \citenamefont {Ji},\ and\ \citenamefont
  {Li}}]{Chen:2005ak}%
  \BibitemOpen
  \bibfield  {author} {\bibinfo {author} {\bibfnamefont {J.-W.}\ \bibnamefont
  {Chen}}, \bibinfo {author} {\bibfnamefont {T.}~\bibnamefont {Inoue}},
  \bibinfo {author} {\bibfnamefont {X.-d.}\ \bibnamefont {Ji}}, \ and\ \bibinfo
  {author} {\bibfnamefont {Y.-c.}\ \bibnamefont {Li}},\ }\href {\doibase
  10.1103/PhysRevC.72.061001} {\bibfield  {journal} {\bibinfo  {journal}
  {Phys.Rev.}\ }\textbf {\bibinfo {volume} {C72}},\ \bibinfo {pages} {061001}
  (\bibinfo {year} {2005})},\ \Eprint {http://arxiv.org/abs/nucl-th/0506001}
  {arXiv:nucl-th/0506001 [nucl-th]} \BibitemShut {NoStop}%
\bibitem [{\citenamefont {Wei}\ \emph {et~al.}(2009)\citenamefont {Wei} \emph
  {et~al.}}]{Wei:2009zv}%
  \BibitemOpen
  \bibfield  {author} {\bibinfo {author} {\bibfnamefont {J.-T.}\ \bibnamefont
  {Wei}} \emph {et~al.} (\bibinfo {collaboration} {BELLE Collaboration}),\
  }\href {\doibase 10.1103/PhysRevLett.103.171801} {\bibfield  {journal}
  {\bibinfo  {journal} {Phys.Rev.Lett.}\ }\textbf {\bibinfo {volume} {103}},\
  \bibinfo {pages} {171801} (\bibinfo {year} {2009})},\ \Eprint
  {http://arxiv.org/abs/0904.0770} {arXiv:0904.0770 [hep-ex]} \BibitemShut
  {NoStop}%
\bibitem [{\citenamefont {Aaltonen}\ \emph {et~al.}(2012)\citenamefont
  {Aaltonen} \emph {et~al.}}]{Aaltonen:2011ja}%
  \BibitemOpen
  \bibfield  {author} {\bibinfo {author} {\bibfnamefont {T.}~\bibnamefont
  {Aaltonen}} \emph {et~al.} (\bibinfo {collaboration} {CDF Collaboration}),\
  }\href {\doibase 10.1103/PhysRevLett.108.081807} {\bibfield  {journal}
  {\bibinfo  {journal} {Phys.Rev.Lett.}\ }\textbf {\bibinfo {volume} {108}},\
  \bibinfo {pages} {081807} (\bibinfo {year} {2012})},\ \Eprint
  {http://arxiv.org/abs/1108.0695} {arXiv:1108.0695 [hep-ex]} \BibitemShut
  {NoStop}%
\bibitem [{\citenamefont {Lees}\ \emph {et~al.}(2012)\citenamefont {Lees} \emph
  {et~al.}}]{Lees:2012tva}%
  \BibitemOpen
  \bibfield  {author} {\bibinfo {author} {\bibfnamefont {J.}~\bibnamefont
  {Lees}} \emph {et~al.} (\bibinfo {collaboration} {BaBar Collaboration}),\
  }\href {\doibase 10.1103/PhysRevD.86.032012} {\bibfield  {journal} {\bibinfo
  {journal} {Phys.Rev.}\ }\textbf {\bibinfo {volume} {D86}},\ \bibinfo {pages}
  {032012} (\bibinfo {year} {2012})},\ \Eprint {http://arxiv.org/abs/1204.3933}
  {arXiv:1204.3933 [hep-ex]} \BibitemShut {NoStop}%
\bibitem [{\citenamefont {Aaij}\ \emph
  {et~al.}(2013{\natexlab{a}})\citenamefont {Aaij} \emph
  {et~al.}}]{Aaij:2013iag}%
  \BibitemOpen
  \bibfield  {author} {\bibinfo {author} {\bibfnamefont {R.}~\bibnamefont
  {Aaij}} \emph {et~al.} (\bibinfo {collaboration} {LHCb Collaboration}),\
  }\href {\doibase 10.1007/JHEP08(2013)131} {\bibfield  {journal} {\bibinfo
  {journal} {JHEP}\ }\textbf {\bibinfo {volume} {1308}},\ \bibinfo {pages}
  {131} (\bibinfo {year} {2013}{\natexlab{a}})},\ \Eprint
  {http://arxiv.org/abs/1304.6325} {arXiv:1304.6325} \BibitemShut {NoStop}%
\bibitem [{\citenamefont {Aaij}\ \emph
  {et~al.}(2013{\natexlab{b}})\citenamefont {Aaij} \emph
  {et~al.}}]{Aaij:2013qta}%
  \BibitemOpen
  \bibfield  {author} {\bibinfo {author} {\bibfnamefont {R.}~\bibnamefont
  {Aaij}} \emph {et~al.} (\bibinfo {collaboration} {LHCb collaboration}),\
  }\href {\doibase 10.1103/PhysRevLett.111.191801} {\bibfield  {journal}
  {\bibinfo  {journal} {Phys.Rev.Lett.}\ }\textbf {\bibinfo {volume} {111}},\
  \bibinfo {pages} {191801} (\bibinfo {year} {2013}{\natexlab{b}})},\ \Eprint
  {http://arxiv.org/abs/1308.1707} {arXiv:1308.1707 [hep-ex]} \BibitemShut
  {NoStop}%
\bibitem [{\citenamefont {Bobeth}\ \emph {et~al.}(2013)\citenamefont {Bobeth},
  \citenamefont {Hiller},\ and\ \citenamefont {van Dyk}}]{Bobeth:2012vn}%
  \BibitemOpen
  \bibfield  {author} {\bibinfo {author} {\bibfnamefont {C.}~\bibnamefont
  {Bobeth}}, \bibinfo {author} {\bibfnamefont {G.}~\bibnamefont {Hiller}}, \
  and\ \bibinfo {author} {\bibfnamefont {D.}~\bibnamefont {van Dyk}},\ }\href
  {\doibase 10.1103/PhysRevD.87.034016} {\bibfield  {journal} {\bibinfo
  {journal} {Phys.Rev.}\ }\textbf {\bibinfo {volume} {D87}},\ \bibinfo {pages}
  {034016} (\bibinfo {year} {2013})},\ \Eprint {http://arxiv.org/abs/1212.2321}
  {arXiv:1212.2321 [hep-ph]} \BibitemShut {NoStop}%
\bibitem [{\citenamefont {Descotes-Genon}\ \emph {et~al.}(2013)\citenamefont
  {Descotes-Genon}, \citenamefont {Matias},\ and\ \citenamefont
  {Virto}}]{Descotes-Genon:2013wba}%
  \BibitemOpen
  \bibfield  {author} {\bibinfo {author} {\bibfnamefont {S.}~\bibnamefont
  {Descotes-Genon}}, \bibinfo {author} {\bibfnamefont {J.}~\bibnamefont
  {Matias}}, \ and\ \bibinfo {author} {\bibfnamefont {J.}~\bibnamefont
  {Virto}},\ }\href {\doibase 10.1103/PhysRevD.88.074002} {\bibfield  {journal}
  {\bibinfo  {journal} {Phys.Rev.}\ }\textbf {\bibinfo {volume} {D88}},\
  \bibinfo {pages} {074002} (\bibinfo {year} {2013})},\ \Eprint
  {http://arxiv.org/abs/1307.5683} {arXiv:1307.5683 [hep-ph]} \BibitemShut
  {NoStop}%
\bibitem [{\citenamefont {Hambrock}\ \emph {et~al.}(2013)\citenamefont
  {Hambrock}, \citenamefont {Hiller}, \citenamefont {Schacht},\ and\
  \citenamefont {Zwicky}}]{Hambrock:2013zya}%
  \BibitemOpen
  \bibfield  {author} {\bibinfo {author} {\bibfnamefont {C.}~\bibnamefont
  {Hambrock}}, \bibinfo {author} {\bibfnamefont {G.}~\bibnamefont {Hiller}},
  \bibinfo {author} {\bibfnamefont {S.}~\bibnamefont {Schacht}}, \ and\
  \bibinfo {author} {\bibfnamefont {R.}~\bibnamefont {Zwicky}},\ }\href@noop {}
  {\  (\bibinfo {year} {2013})},\ \Eprint {http://arxiv.org/abs/1308.4379}
  {arXiv:1308.4379 [hep-ph]} \BibitemShut {NoStop}%
\bibitem [{\citenamefont {Beaujean}\ \emph {et~al.}(2013)\citenamefont
  {Beaujean}, \citenamefont {Bobeth},\ and\ \citenamefont {van
  Dyk}}]{Beaujean:2013soa}%
  \BibitemOpen
  \bibfield  {author} {\bibinfo {author} {\bibfnamefont {F.}~\bibnamefont
  {Beaujean}}, \bibinfo {author} {\bibfnamefont {C.}~\bibnamefont {Bobeth}}, \
  and\ \bibinfo {author} {\bibfnamefont {D.}~\bibnamefont {van Dyk}},\
  }\href@noop {} {\  (\bibinfo {year} {2013})},\ \Eprint
  {http://arxiv.org/abs/1310.2478} {arXiv:1310.2478 [hep-ph]} \BibitemShut
  {NoStop}%
\bibitem [{\citenamefont {Bowler}\ \emph {et~al.}(1994)\citenamefont {Bowler}
  \emph {et~al.}}]{Bowler:1993rz}%
  \BibitemOpen
  \bibfield  {author} {\bibinfo {author} {\bibfnamefont {K.}~\bibnamefont
  {Bowler}} \emph {et~al.} (\bibinfo {collaboration} {UKQCD Collaboration}),\
  }\href {\doibase 10.1103/PhysRevLett.72.1398} {\bibfield  {journal} {\bibinfo
   {journal} {Phys.Rev.Lett.}\ }\textbf {\bibinfo {volume} {72}},\ \bibinfo
  {pages} {1398} (\bibinfo {year} {1994})},\ \Eprint
  {http://arxiv.org/abs/hep-lat/9311004} {arXiv:hep-lat/9311004 [hep-lat]}
  \BibitemShut {NoStop}%
\bibitem [{\citenamefont {Bernard}\ \emph {et~al.}(1994)\citenamefont
  {Bernard}, \citenamefont {Hsieh},\ and\ \citenamefont
  {Soni}}]{Bernard:1993yt}%
  \BibitemOpen
  \bibfield  {author} {\bibinfo {author} {\bibfnamefont {C.~W.}\ \bibnamefont
  {Bernard}}, \bibinfo {author} {\bibfnamefont {P.}~\bibnamefont {Hsieh}}, \
  and\ \bibinfo {author} {\bibfnamefont {A.}~\bibnamefont {Soni}},\ }\href
  {\doibase 10.1103/PhysRevLett.72.1402} {\bibfield  {journal} {\bibinfo
  {journal} {Phys.Rev.Lett.}\ }\textbf {\bibinfo {volume} {72}},\ \bibinfo
  {pages} {1402} (\bibinfo {year} {1994})},\ \Eprint
  {http://arxiv.org/abs/hep-lat/9311010} {arXiv:hep-lat/9311010 [hep-lat]}
  \BibitemShut {NoStop}%
\bibitem [{\citenamefont {Burford}\ \emph {et~al.}(1995)\citenamefont {Burford}
  \emph {et~al.}}]{Burford:1995fc}%
  \BibitemOpen
  \bibfield  {author} {\bibinfo {author} {\bibfnamefont {D.}~\bibnamefont
  {Burford}} \emph {et~al.} (\bibinfo {collaboration} {UKQCD Collaboration}),\
  }\href {\doibase 10.1016/0550-3213(95)00223-F} {\bibfield  {journal}
  {\bibinfo  {journal} {Nucl.Phys.}\ }\textbf {\bibinfo {volume} {B447}},\
  \bibinfo {pages} {425} (\bibinfo {year} {1995})},\ \Eprint
  {http://arxiv.org/abs/hep-lat/9503002} {arXiv:hep-lat/9503002 [hep-lat]}
  \BibitemShut {NoStop}%
\bibitem [{\citenamefont {Abada}\ \emph {et~al.}(1996)\citenamefont {Abada}
  \emph {et~al.}}]{Abada:1995fa}%
  \BibitemOpen
  \bibfield  {author} {\bibinfo {author} {\bibfnamefont {A.}~\bibnamefont
  {Abada}} \emph {et~al.} (\bibinfo {collaboration} {APE Collaboration}),\
  }\href {\doibase 10.1016/0370-2693(95)01236-2} {\bibfield  {journal}
  {\bibinfo  {journal} {Phys.Lett.}\ }\textbf {\bibinfo {volume} {B365}},\
  \bibinfo {pages} {275} (\bibinfo {year} {1996})},\ \Eprint
  {http://arxiv.org/abs/hep-lat/9503020} {arXiv:hep-lat/9503020 [hep-lat]}
  \BibitemShut {NoStop}%
\bibitem [{\citenamefont {Abada}\ \emph {et~al.}(2003)\citenamefont {Abada}
  \emph {et~al.}}]{Abada:2002ie}%
  \BibitemOpen
  \bibfield  {author} {\bibinfo {author} {\bibfnamefont {A.}~\bibnamefont
  {Abada}} \emph {et~al.} (\bibinfo {collaboration} {SPQcdR collaboration}),\
  }\href {\doibase 10.1016/S0920-5632(03)01643-8} {\bibfield  {journal}
  {\bibinfo  {journal} {Nucl.Phys.Proc.Suppl.}\ }\textbf {\bibinfo {volume}
  {119}},\ \bibinfo {pages} {625} (\bibinfo {year} {2003})},\ \Eprint
  {http://arxiv.org/abs/hep-lat/0209116} {arXiv:hep-lat/0209116 [hep-lat]}
  \BibitemShut {NoStop}%
\bibitem [{\citenamefont {Bowler}\ \emph {et~al.}(2004)\citenamefont {Bowler},
  \citenamefont {Gill}, \citenamefont {Maynard},\ and\ \citenamefont
  {Flynn}}]{Bowler:2004zb}%
  \BibitemOpen
  \bibfield  {author} {\bibinfo {author} {\bibfnamefont {K.}~\bibnamefont
  {Bowler}}, \bibinfo {author} {\bibfnamefont {J.}~\bibnamefont {Gill}},
  \bibinfo {author} {\bibfnamefont {C.}~\bibnamefont {Maynard}}, \ and\
  \bibinfo {author} {\bibfnamefont {J.}~\bibnamefont {Flynn}} (\bibinfo
  {collaboration} {UKQCD Collaboration}),\ }\href {\doibase
  10.1088/1126-6708/2004/05/035} {\bibfield  {journal} {\bibinfo  {journal}
  {JHEP}\ }\textbf {\bibinfo {volume} {0405}},\ \bibinfo {pages} {035}
  (\bibinfo {year} {2004})},\ \Eprint {http://arxiv.org/abs/hep-lat/0402023}
  {arXiv:hep-lat/0402023 [hep-lat]} \BibitemShut {NoStop}%
\bibitem [{\citenamefont {Becirevic}\ \emph {et~al.}(2007)\citenamefont
  {Becirevic}, \citenamefont {Lubicz},\ and\ \citenamefont
  {Mescia}}]{Becirevic:2006nm}%
  \BibitemOpen
  \bibfield  {author} {\bibinfo {author} {\bibfnamefont {D.}~\bibnamefont
  {Becirevic}}, \bibinfo {author} {\bibfnamefont {V.}~\bibnamefont {Lubicz}}, \
  and\ \bibinfo {author} {\bibfnamefont {F.}~\bibnamefont {Mescia}},\ }\href
  {\doibase 10.1016/j.nuclphysb.2007.01.032} {\bibfield  {journal} {\bibinfo
  {journal} {Nucl.Phys.}\ }\textbf {\bibinfo {volume} {B769}},\ \bibinfo
  {pages} {31} (\bibinfo {year} {2007})},\ \Eprint
  {http://arxiv.org/abs/hep-ph/0611295} {arXiv:hep-ph/0611295 [hep-ph]}
  \BibitemShut {NoStop}%
\bibitem [{\citenamefont {Horgan}\ \emph
  {et~al.}(2014{\natexlab{a}})\citenamefont {Horgan}, \citenamefont {Liu},
  \citenamefont {Meinel},\ and\ \citenamefont {Wingate}}]{Horgan:2013hoa}%
  \BibitemOpen
  \bibfield  {author} {\bibinfo {author} {\bibfnamefont {R.~R.}\ \bibnamefont
  {Horgan}}, \bibinfo {author} {\bibfnamefont {Z.}~\bibnamefont {Liu}},
  \bibinfo {author} {\bibfnamefont {S.}~\bibnamefont {Meinel}}, \ and\ \bibinfo
  {author} {\bibfnamefont {M.}~\bibnamefont {Wingate}},\ }\href {\doibase
  10.1103/PhysRevD.89.094501} {\bibfield  {journal} {\bibinfo  {journal}
  {Phys.Rev.}\ }\textbf {\bibinfo {volume} {D89}},\ \bibinfo {pages} {094501}
  (\bibinfo {year} {2014}{\natexlab{a}})},\ \Eprint
  {http://arxiv.org/abs/1310.3722} {arXiv:1310.3722 [hep-lat]} \BibitemShut
  {NoStop}%
\bibitem [{\citenamefont {Horgan}\ \emph
  {et~al.}(2014{\natexlab{b}})\citenamefont {Horgan}, \citenamefont {Liu},
  \citenamefont {Meinel},\ and\ \citenamefont {Wingate}}]{Horgan:2013pva}%
  \BibitemOpen
  \bibfield  {author} {\bibinfo {author} {\bibfnamefont {R.~R.}\ \bibnamefont
  {Horgan}}, \bibinfo {author} {\bibfnamefont {Z.}~\bibnamefont {Liu}},
  \bibinfo {author} {\bibfnamefont {S.}~\bibnamefont {Meinel}}, \ and\ \bibinfo
  {author} {\bibfnamefont {M.}~\bibnamefont {Wingate}},\ }\href {\doibase
  10.1103/PhysRevLett.112.212003} {\bibfield  {journal} {\bibinfo  {journal}
  {Phys.Rev.Lett.}\ }\textbf {\bibinfo {volume} {112}},\ \bibinfo {pages}
  {212003} (\bibinfo {year} {2014}{\natexlab{b}})},\ \Eprint
  {http://arxiv.org/abs/1310.3887} {arXiv:1310.3887 [hep-ph]} \BibitemShut
  {NoStop}%
\bibitem [{\citenamefont {Maiani}\ and\ \citenamefont
  {Testa}(1990)}]{Maiani:1990ca}%
  \BibitemOpen
  \bibfield  {author} {\bibinfo {author} {\bibfnamefont {L.}~\bibnamefont
  {Maiani}}\ and\ \bibinfo {author} {\bibfnamefont {M.}~\bibnamefont {Testa}},\
  }\href {\doibase 10.1016/0370-2693(90)90695-3} {\bibfield  {journal}
  {\bibinfo  {journal} {Phys.Lett.}\ }\textbf {\bibinfo {volume} {B245}},\
  \bibinfo {pages} {585} (\bibinfo {year} {1990})}\BibitemShut {NoStop}%
\bibitem [{\citenamefont {Lellouch}\ and\ \citenamefont
  {Luscher}(2001)}]{Lellouch:2000pv}%
  \BibitemOpen
  \bibfield  {author} {\bibinfo {author} {\bibfnamefont {L.}~\bibnamefont
  {Lellouch}}\ and\ \bibinfo {author} {\bibfnamefont {M.}~\bibnamefont
  {Luscher}},\ }\href@noop {} {\bibfield  {journal} {\bibinfo  {journal}
  {Commun.Math.Phys.}\ }\textbf {\bibinfo {volume} {219}},\ \bibinfo {pages}
  {31} (\bibinfo {year} {2001})},\ \Eprint
  {http://arxiv.org/abs/hep-lat/0003023} {arXiv:hep-lat/0003023 [hep-lat]}
  \BibitemShut {NoStop}%
\bibitem [{\citenamefont {Lin}\ \emph {et~al.}(2001)\citenamefont {Lin},
  \citenamefont {Martinelli}, \citenamefont {Sachrajda},\ and\ \citenamefont
  {Testa}}]{Lin:2001ek}%
  \BibitemOpen
  \bibfield  {author} {\bibinfo {author} {\bibfnamefont {C.~D.}\ \bibnamefont
  {Lin}}, \bibinfo {author} {\bibfnamefont {G.}~\bibnamefont {Martinelli}},
  \bibinfo {author} {\bibfnamefont {C.~T.}\ \bibnamefont {Sachrajda}}, \ and\
  \bibinfo {author} {\bibfnamefont {M.}~\bibnamefont {Testa}},\ }\href
  {\doibase 10.1016/S0550-3213(01)00495-3} {\bibfield  {journal} {\bibinfo
  {journal} {Nucl.Phys.}\ }\textbf {\bibinfo {volume} {B619}},\ \bibinfo
  {pages} {467} (\bibinfo {year} {2001})},\ \Eprint
  {http://arxiv.org/abs/hep-lat/0104006} {arXiv:hep-lat/0104006 [hep-lat]}
  \BibitemShut {NoStop}%
\bibitem [{\citenamefont {Kim}\ \emph {et~al.}(2005)\citenamefont {Kim},
  \citenamefont {Sachrajda},\ and\ \citenamefont {Sharpe}}]{Kim:2005gf}%
  \BibitemOpen
  \bibfield  {author} {\bibinfo {author} {\bibfnamefont {C.}~\bibnamefont
  {Kim}}, \bibinfo {author} {\bibfnamefont {C.}~\bibnamefont {Sachrajda}}, \
  and\ \bibinfo {author} {\bibfnamefont {S.~R.}\ \bibnamefont {Sharpe}},\
  }\href {\doibase 10.1016/j.nuclphysb.2005.08.029} {\bibfield  {journal}
  {\bibinfo  {journal} {Nucl.Phys.}\ }\textbf {\bibinfo {volume} {B727}},\
  \bibinfo {pages} {218} (\bibinfo {year} {2005})},\ \Eprint
  {http://arxiv.org/abs/hep-lat/0507006} {arXiv:hep-lat/0507006 [hep-lat]}
  \BibitemShut {NoStop}%
\bibitem [{\citenamefont {Christ}\ \emph {et~al.}(2005)\citenamefont {Christ},
  \citenamefont {Kim},\ and\ \citenamefont {Yamazaki}}]{Christ:2005gi}%
  \BibitemOpen
  \bibfield  {author} {\bibinfo {author} {\bibfnamefont {N.~H.}\ \bibnamefont
  {Christ}}, \bibinfo {author} {\bibfnamefont {C.}~\bibnamefont {Kim}}, \ and\
  \bibinfo {author} {\bibfnamefont {T.}~\bibnamefont {Yamazaki}},\ }\href
  {\doibase 10.1103/PhysRevD.72.114506} {\bibfield  {journal} {\bibinfo
  {journal} {Phys.Rev.}\ }\textbf {\bibinfo {volume} {D72}},\ \bibinfo {pages}
  {114506} (\bibinfo {year} {2005})},\ \Eprint
  {http://arxiv.org/abs/hep-lat/0507009} {arXiv:hep-lat/0507009 [hep-lat]}
  \BibitemShut {NoStop}%
\bibitem [{\citenamefont {Blum}\ \emph {et~al.}(2011)\citenamefont {Blum},
  \citenamefont {Boyle}, \citenamefont {Christ}, \citenamefont {Garron},
  \citenamefont {Goode} \emph {et~al.}}]{Blum:2011pu}%
  \BibitemOpen
  \bibfield  {author} {\bibinfo {author} {\bibfnamefont {T.}~\bibnamefont
  {Blum}}, \bibinfo {author} {\bibfnamefont {P.}~\bibnamefont {Boyle}},
  \bibinfo {author} {\bibfnamefont {N.}~\bibnamefont {Christ}}, \bibinfo
  {author} {\bibfnamefont {N.}~\bibnamefont {Garron}}, \bibinfo {author}
  {\bibfnamefont {E.}~\bibnamefont {Goode}},  \emph {et~al.},\ }\href {\doibase
  10.1103/PhysRevD.84.114503} {\bibfield  {journal} {\bibinfo  {journal}
  {Phys.Rev.}\ }\textbf {\bibinfo {volume} {D84}},\ \bibinfo {pages} {114503}
  (\bibinfo {year} {2011})},\ \Eprint {http://arxiv.org/abs/1106.2714}
  {arXiv:1106.2714 [hep-lat]} \BibitemShut {NoStop}%
\bibitem [{\citenamefont {Blum}\ \emph
  {et~al.}(2012{\natexlab{a}})\citenamefont {Blum}, \citenamefont {Boyle},
  \citenamefont {Christ}, \citenamefont {Garron}, \citenamefont {Goode} \emph
  {et~al.}}]{Blum:2011ng}%
  \BibitemOpen
  \bibfield  {author} {\bibinfo {author} {\bibfnamefont {T.}~\bibnamefont
  {Blum}}, \bibinfo {author} {\bibfnamefont {P.}~\bibnamefont {Boyle}},
  \bibinfo {author} {\bibfnamefont {N.}~\bibnamefont {Christ}}, \bibinfo
  {author} {\bibfnamefont {N.}~\bibnamefont {Garron}}, \bibinfo {author}
  {\bibfnamefont {E.}~\bibnamefont {Goode}},  \emph {et~al.},\ }\href {\doibase
  10.1103/PhysRevLett.108.141601} {\bibfield  {journal} {\bibinfo  {journal}
  {Phys.Rev.Lett.}\ }\textbf {\bibinfo {volume} {108}},\ \bibinfo {pages}
  {141601} (\bibinfo {year} {2012}{\natexlab{a}})},\ \Eprint
  {http://arxiv.org/abs/1111.1699} {arXiv:1111.1699 [hep-lat]} \BibitemShut
  {NoStop}%
\bibitem [{\citenamefont {Blum}\ \emph
  {et~al.}(2012{\natexlab{b}})\citenamefont {Blum}, \citenamefont {Boyle},
  \citenamefont {Christ}, \citenamefont {Garron}, \citenamefont {Goode} \emph
  {et~al.}}]{Blum:2012uk}%
  \BibitemOpen
  \bibfield  {author} {\bibinfo {author} {\bibfnamefont {T.}~\bibnamefont
  {Blum}}, \bibinfo {author} {\bibfnamefont {P.}~\bibnamefont {Boyle}},
  \bibinfo {author} {\bibfnamefont {N.}~\bibnamefont {Christ}}, \bibinfo
  {author} {\bibfnamefont {N.}~\bibnamefont {Garron}}, \bibinfo {author}
  {\bibfnamefont {E.}~\bibnamefont {Goode}},  \emph {et~al.},\ }\href {\doibase
  10.1103/PhysRevD.86.074513} {\bibfield  {journal} {\bibinfo  {journal}
  {Phys.Rev.}\ }\textbf {\bibinfo {volume} {D86}},\ \bibinfo {pages} {074513}
  (\bibinfo {year} {2012}{\natexlab{b}})},\ \Eprint
  {http://arxiv.org/abs/1206.5142} {arXiv:1206.5142 [hep-lat]} \BibitemShut
  {NoStop}%
\bibitem [{\citenamefont {Boyle}\ \emph {et~al.}(2013)\citenamefont {Boyle}
  \emph {et~al.}}]{Boyle:2012ys}%
  \BibitemOpen
  \bibfield  {author} {\bibinfo {author} {\bibfnamefont {P.}~\bibnamefont
  {Boyle}} \emph {et~al.} (\bibinfo {collaboration} {RBC, UKQCD}),\ }\href
  {\doibase 10.1103/PhysRevLett.110.152001} {\bibfield  {journal} {\bibinfo
  {journal} {Phys.Rev.Lett.}\ }\textbf {\bibinfo {volume} {110}},\ \bibinfo
  {pages} {152001} (\bibinfo {year} {2013})},\ \Eprint
  {http://arxiv.org/abs/1212.1474} {arXiv:1212.1474 [hep-lat]} \BibitemShut
  {NoStop}%
\bibitem [{\citenamefont {Ciuchini}\ \emph {et~al.}(1996)\citenamefont
  {Ciuchini}, \citenamefont {Franco}, \citenamefont {Martinelli},\ and\
  \citenamefont {Silvestrini}}]{Ciuchini:1996mq}%
  \BibitemOpen
  \bibfield  {author} {\bibinfo {author} {\bibfnamefont {M.}~\bibnamefont
  {Ciuchini}}, \bibinfo {author} {\bibfnamefont {E.}~\bibnamefont {Franco}},
  \bibinfo {author} {\bibfnamefont {G.}~\bibnamefont {Martinelli}}, \ and\
  \bibinfo {author} {\bibfnamefont {L.}~\bibnamefont {Silvestrini}},\ }\href
  {\doibase 10.1016/0370-2693(96)00529-1} {\bibfield  {journal} {\bibinfo
  {journal} {Phys.Lett.}\ }\textbf {\bibinfo {volume} {B380}},\ \bibinfo
  {pages} {353} (\bibinfo {year} {1996})},\ \Eprint
  {http://arxiv.org/abs/hep-ph/9604240} {arXiv:hep-ph/9604240 [hep-ph]}
  \BibitemShut {NoStop}%
\bibitem [{\citenamefont {Testa}(1998)}]{Testa:1997ne}%
  \BibitemOpen
  \bibfield  {author} {\bibinfo {author} {\bibfnamefont {M.}~\bibnamefont
  {Testa}},\ }\href {\doibase 10.1016/S0920-5632(97)00694-4} {\bibfield
  {journal} {\bibinfo  {journal} {Nucl.Phys.Proc.Suppl.}\ }\textbf {\bibinfo
  {volume} {63}},\ \bibinfo {pages} {38} (\bibinfo {year} {1998})},\ \Eprint
  {http://arxiv.org/abs/hep-lat/9709044} {arXiv:hep-lat/9709044 [hep-lat]}
  \BibitemShut {NoStop}%
\bibitem [{\citenamefont {Dawson}\ \emph {et~al.}(1998)\citenamefont {Dawson},
  \citenamefont {Martinelli}, \citenamefont {Rossi}, \citenamefont {Sachrajda},
  \citenamefont {Sharpe} \emph {et~al.}}]{Dawson:1997ic}%
  \BibitemOpen
  \bibfield  {author} {\bibinfo {author} {\bibfnamefont {C.}~\bibnamefont
  {Dawson}}, \bibinfo {author} {\bibfnamefont {G.}~\bibnamefont {Martinelli}},
  \bibinfo {author} {\bibfnamefont {G.}~\bibnamefont {Rossi}}, \bibinfo
  {author} {\bibfnamefont {C.~T.}\ \bibnamefont {Sachrajda}}, \bibinfo {author}
  {\bibfnamefont {S.~R.}\ \bibnamefont {Sharpe}},  \emph {et~al.},\ }\href
  {\doibase 10.1016/S0550-3213(97)00756-6} {\bibfield  {journal} {\bibinfo
  {journal} {Nucl.Phys.}\ }\textbf {\bibinfo {volume} {B514}},\ \bibinfo
  {pages} {313} (\bibinfo {year} {1998})},\ \Eprint
  {http://arxiv.org/abs/hep-lat/9707009} {arXiv:hep-lat/9707009 [hep-lat]}
  \BibitemShut {NoStop}%
\bibitem [{\citenamefont {Golterman}\ and\ \citenamefont
  {Leung}(1997)}]{Golterman:1997wb}%
  \BibitemOpen
  \bibfield  {author} {\bibinfo {author} {\bibfnamefont {M.~F.}\ \bibnamefont
  {Golterman}}\ and\ \bibinfo {author} {\bibfnamefont {K.~C.}\ \bibnamefont
  {Leung}},\ }\href {\doibase 10.1103/PhysRevD.56.2950} {\bibfield  {journal}
  {\bibinfo  {journal} {Phys.Rev.}\ }\textbf {\bibinfo {volume} {D56}},\
  \bibinfo {pages} {2950} (\bibinfo {year} {1997})},\ \Eprint
  {http://arxiv.org/abs/hep-lat/9702015} {arXiv:hep-lat/9702015 [hep-lat]}
  \BibitemShut {NoStop}%
\bibitem [{\citenamefont {Rossi}(1998)}]{Rossi:1998kc}%
  \BibitemOpen
  \bibfield  {author} {\bibinfo {author} {\bibfnamefont {G.}~\bibnamefont
  {Rossi}},\ }\href@noop {} {\  (\bibinfo {year} {1998})},\ \Eprint
  {http://arxiv.org/abs/hep-lat/9811009} {arXiv:hep-lat/9811009 [hep-lat]}
  \BibitemShut {NoStop}%
\bibitem [{\citenamefont {Buras}\ \emph {et~al.}(2000)\citenamefont {Buras},
  \citenamefont {Ciuchini}, \citenamefont {Franco}, \citenamefont {Isidori},
  \citenamefont {Martinelli} \emph {et~al.}}]{Buras:2000kx}%
  \BibitemOpen
  \bibfield  {author} {\bibinfo {author} {\bibfnamefont {A.}~\bibnamefont
  {Buras}}, \bibinfo {author} {\bibfnamefont {M.}~\bibnamefont {Ciuchini}},
  \bibinfo {author} {\bibfnamefont {E.}~\bibnamefont {Franco}}, \bibinfo
  {author} {\bibfnamefont {G.}~\bibnamefont {Isidori}}, \bibinfo {author}
  {\bibfnamefont {G.}~\bibnamefont {Martinelli}},  \emph {et~al.},\ }\href
  {\doibase 10.1016/S0370-2693(00)00362-2} {\bibfield  {journal} {\bibinfo
  {journal} {Phys.Lett.}\ }\textbf {\bibinfo {volume} {B480}},\ \bibinfo
  {pages} {80} (\bibinfo {year} {2000})},\ \Eprint
  {http://arxiv.org/abs/hep-ph/0002116} {arXiv:hep-ph/0002116 [hep-ph]}
  \BibitemShut {NoStop}%
\bibitem [{\citenamefont {Pallante}(2001)}]{Pallante:2000ut}%
  \BibitemOpen
  \bibfield  {author} {\bibinfo {author} {\bibfnamefont {E.}~\bibnamefont
  {Pallante}},\ }\href {\doibase 10.1016/S0920-5632(01)01150-1} {\bibfield
  {journal} {\bibinfo  {journal} {Nucl.Phys.Proc.Suppl.}\ }\textbf {\bibinfo
  {volume} {96}},\ \bibinfo {pages} {336} (\bibinfo {year} {2001})},\ \Eprint
  {http://arxiv.org/abs/hep-ph/0010011} {arXiv:hep-ph/0010011 [hep-ph]}
  \BibitemShut {NoStop}%
\bibitem [{\citenamefont {Colangelo}(2002)}]{Colangelo:2001uv}%
  \BibitemOpen
  \bibfield  {author} {\bibinfo {author} {\bibfnamefont {G.}~\bibnamefont
  {Colangelo}},\ }\href {\doibase 10.1016/S0920-5632(01)01643-7} {\bibfield
  {journal} {\bibinfo  {journal} {Nucl.Phys.Proc.Suppl.}\ }\textbf {\bibinfo
  {volume} {106}},\ \bibinfo {pages} {53} (\bibinfo {year} {2002})},\ \Eprint
  {http://arxiv.org/abs/hep-lat/0111003} {arXiv:hep-lat/0111003 [hep-lat]}
  \BibitemShut {NoStop}%
\bibitem [{\citenamefont {Lin}\ \emph {et~al.}(2002)\citenamefont {Lin},
  \citenamefont {Martinelli}, \citenamefont {Sachrajda},\ and\ \citenamefont
  {Testa}}]{Lin:2001fi}%
  \BibitemOpen
  \bibfield  {author} {\bibinfo {author} {\bibfnamefont {C.~D.}\ \bibnamefont
  {Lin}}, \bibinfo {author} {\bibfnamefont {G.}~\bibnamefont {Martinelli}},
  \bibinfo {author} {\bibfnamefont {C.}~\bibnamefont {Sachrajda}}, \ and\
  \bibinfo {author} {\bibfnamefont {M.}~\bibnamefont {Testa}},\ }\href
  {\doibase 10.1016/S0920-5632(02)01419-6} {\bibfield  {journal} {\bibinfo
  {journal} {Nucl.Phys.Proc.Suppl.}\ }\textbf {\bibinfo {volume} {109A}},\
  \bibinfo {pages} {218} (\bibinfo {year} {2002})},\ \Eprint
  {http://arxiv.org/abs/hep-lat/0111033} {arXiv:hep-lat/0111033 [hep-lat]}
  \BibitemShut {NoStop}%
\bibitem [{\citenamefont {Laiho}\ and\ \citenamefont
  {Soni}(2002)}]{Laiho:2002jq}%
  \BibitemOpen
  \bibfield  {author} {\bibinfo {author} {\bibfnamefont {J.}~\bibnamefont
  {Laiho}}\ and\ \bibinfo {author} {\bibfnamefont {A.}~\bibnamefont {Soni}},\
  }\href {\doibase 10.1103/PhysRevD.65.114020} {\bibfield  {journal} {\bibinfo
  {journal} {Phys.Rev.}\ }\textbf {\bibinfo {volume} {D65}},\ \bibinfo {pages}
  {114020} (\bibinfo {year} {2002})},\ \Eprint
  {http://arxiv.org/abs/hep-ph/0203106} {arXiv:hep-ph/0203106 [hep-ph]}
  \BibitemShut {NoStop}%
\bibitem [{\citenamefont {Aoki}\ \emph {et~al.}(1998)\citenamefont {Aoki} \emph
  {et~al.}}]{Aoki:1997ev}%
  \BibitemOpen
  \bibfield  {author} {\bibinfo {author} {\bibfnamefont {S.}~\bibnamefont
  {Aoki}} \emph {et~al.} (\bibinfo {collaboration} {JLQCD Collaboration}),\
  }\href {\doibase 10.1103/PhysRevD.58.054503} {\bibfield  {journal} {\bibinfo
  {journal} {Phys.Rev.}\ }\textbf {\bibinfo {volume} {D58}},\ \bibinfo {pages}
  {054503} (\bibinfo {year} {1998})},\ \Eprint
  {http://arxiv.org/abs/hep-lat/9711046} {arXiv:hep-lat/9711046 [hep-lat]}
  \BibitemShut {NoStop}%
\bibitem [{\citenamefont {Bijnens}\ \emph {et~al.}(1998)\citenamefont
  {Bijnens}, \citenamefont {Pallante},\ and\ \citenamefont
  {Prades}}]{Bijnens:1998mb}%
  \BibitemOpen
  \bibfield  {author} {\bibinfo {author} {\bibfnamefont {J.}~\bibnamefont
  {Bijnens}}, \bibinfo {author} {\bibfnamefont {E.}~\bibnamefont {Pallante}}, \
  and\ \bibinfo {author} {\bibfnamefont {J.}~\bibnamefont {Prades}},\ }\href
  {\doibase 10.1016/S0550-3213(98)00202-8} {\bibfield  {journal} {\bibinfo
  {journal} {Nucl.Phys.}\ }\textbf {\bibinfo {volume} {B521}},\ \bibinfo
  {pages} {305} (\bibinfo {year} {1998})},\ \Eprint
  {http://arxiv.org/abs/hep-ph/9801326} {arXiv:hep-ph/9801326 [hep-ph]}
  \BibitemShut {NoStop}%
\bibitem [{\citenamefont {Pekurovsky}\ and\ \citenamefont
  {Kilcup}(2001)}]{Pekurovsky:1998jd}%
  \BibitemOpen
  \bibfield  {author} {\bibinfo {author} {\bibfnamefont {D.}~\bibnamefont
  {Pekurovsky}}\ and\ \bibinfo {author} {\bibfnamefont {G.}~\bibnamefont
  {Kilcup}},\ }\href {\doibase 10.1103/PhysRevD.64.074502} {\bibfield
  {journal} {\bibinfo  {journal} {Phys.Rev.}\ }\textbf {\bibinfo {volume}
  {D64}},\ \bibinfo {pages} {074502} (\bibinfo {year} {2001})},\ \Eprint
  {http://arxiv.org/abs/hep-lat/9812019} {arXiv:hep-lat/9812019 [hep-lat]}
  \BibitemShut {NoStop}%
\bibitem [{\citenamefont {Pekurovsky}\ and\ \citenamefont
  {Kilcup}(1999)}]{Pekurovsky:1999pn}%
  \BibitemOpen
  \bibfield  {author} {\bibinfo {author} {\bibfnamefont {D.}~\bibnamefont
  {Pekurovsky}}\ and\ \bibinfo {author} {\bibfnamefont {G.}~\bibnamefont
  {Kilcup}},\ }\href@noop {} {\  (\bibinfo {year} {1999})},\ \Eprint
  {http://arxiv.org/abs/hep-lat/9903025} {arXiv:hep-lat/9903025 [hep-lat]}
  \BibitemShut {NoStop}%
\bibitem [{\citenamefont {Pekurovsky}(1999)}]{Pekurovsky:1999er}%
  \BibitemOpen
  \bibfield  {author} {\bibinfo {author} {\bibfnamefont {D.}~\bibnamefont
  {Pekurovsky}},\ }\href@noop {} {\  (\bibinfo {year} {1999})},\ \Eprint
  {http://arxiv.org/abs/hep-lat/9909141} {arXiv:hep-lat/9909141 [hep-lat]}
  \BibitemShut {NoStop}%
\bibitem [{\citenamefont {Mawhinney}(2001)}]{Mawhinney:2000hk}%
  \BibitemOpen
  \bibfield  {author} {\bibinfo {author} {\bibfnamefont {R.~D.}\ \bibnamefont
  {Mawhinney}} (\bibinfo {collaboration} {RBC Collaboration}),\ }\href
  {\doibase 10.1016/S0920-5632(01)00949-5} {\bibfield  {journal} {\bibinfo
  {journal} {Nucl.Phys.Proc.Suppl.}\ }\textbf {\bibinfo {volume} {94}},\
  \bibinfo {pages} {315} (\bibinfo {year} {2001})},\ \Eprint
  {http://arxiv.org/abs/hep-lat/0010030} {arXiv:hep-lat/0010030 [hep-lat]}
  \BibitemShut {NoStop}%
\bibitem [{\citenamefont {Capitani}\ and\ \citenamefont
  {Giusti}(2001)}]{Capitani:2000bm}%
  \BibitemOpen
  \bibfield  {author} {\bibinfo {author} {\bibfnamefont {S.}~\bibnamefont
  {Capitani}}\ and\ \bibinfo {author} {\bibfnamefont {L.}~\bibnamefont
  {Giusti}},\ }\href {\doibase 10.1103/PhysRevD.64.014506} {\bibfield
  {journal} {\bibinfo  {journal} {Phys.Rev.}\ }\textbf {\bibinfo {volume}
  {D64}},\ \bibinfo {pages} {014506} (\bibinfo {year} {2001})},\ \Eprint
  {http://arxiv.org/abs/hep-lat/0011070} {arXiv:hep-lat/0011070 [hep-lat]}
  \BibitemShut {NoStop}%
\bibitem [{\citenamefont {Kim}\ and\ \citenamefont
  {Christ}(2003)}]{Kim:2002np}%
  \BibitemOpen
  \bibfield  {author} {\bibinfo {author} {\bibfnamefont {C.-h.}\ \bibnamefont
  {Kim}}\ and\ \bibinfo {author} {\bibfnamefont {N.~H.}\ \bibnamefont
  {Christ}},\ }\href {\doibase 10.1016/S0920-5632(03)01553-6} {\bibfield
  {journal} {\bibinfo  {journal} {Nucl.Phys.Proc.Suppl.}\ }\textbf {\bibinfo
  {volume} {119}},\ \bibinfo {pages} {365} (\bibinfo {year} {2003})},\ \Eprint
  {http://arxiv.org/abs/hep-lat/0210003} {arXiv:hep-lat/0210003 [hep-lat]}
  \BibitemShut {NoStop}%
\bibitem [{\citenamefont {Laiho}\ and\ \citenamefont
  {Soni}(2005)}]{Laiho:2003uy}%
  \BibitemOpen
  \bibfield  {author} {\bibinfo {author} {\bibfnamefont {J.}~\bibnamefont
  {Laiho}}\ and\ \bibinfo {author} {\bibfnamefont {A.}~\bibnamefont {Soni}},\
  }\href {\doibase 10.1103/PhysRevD.71.014021} {\bibfield  {journal} {\bibinfo
  {journal} {Phys.Rev.}\ }\textbf {\bibinfo {volume} {D71}},\ \bibinfo {pages}
  {014021} (\bibinfo {year} {2005})},\ \Eprint
  {http://arxiv.org/abs/hep-lat/0306035} {arXiv:hep-lat/0306035 [hep-lat]}
  \BibitemShut {NoStop}%
\bibitem [{\citenamefont {Yamazaki}(2006)}]{Yamazaki:2006ce}%
  \BibitemOpen
  \bibfield  {author} {\bibinfo {author} {\bibfnamefont {T.}~\bibnamefont
  {Yamazaki}} (\bibinfo {collaboration} {RIKEN-BNL-Columbia Collaboration}),\
  }\href@noop {} {\bibfield  {journal} {\bibinfo  {journal} {PoS}\ }\textbf
  {\bibinfo {volume} {LAT2006}},\ \bibinfo {pages} {100} (\bibinfo {year}
  {2006})},\ \Eprint {http://arxiv.org/abs/hep-lat/0610051}
  {arXiv:hep-lat/0610051 [hep-lat]} \BibitemShut {NoStop}%
\bibitem [{\citenamefont {Yamazaki}(2009)}]{Yamazaki:2008hg}%
  \BibitemOpen
  \bibfield  {author} {\bibinfo {author} {\bibfnamefont {T.}~\bibnamefont
  {Yamazaki}} (\bibinfo {collaboration} {RBC Collaboration, UKQCD
  Collaboration}),\ }\href {\doibase 10.1103/PhysRevD.79.094506} {\bibfield
  {journal} {\bibinfo  {journal} {Phys.Rev.}\ }\textbf {\bibinfo {volume}
  {D79}},\ \bibinfo {pages} {094506} (\bibinfo {year} {2009})},\ \Eprint
  {http://arxiv.org/abs/0807.3130} {arXiv:0807.3130 [hep-lat]} \BibitemShut
  {NoStop}%
\bibitem [{\citenamefont {Hansen}\ and\ \citenamefont
  {Sharpe}(2012)}]{Hansen:2012tf}%
  \BibitemOpen
  \bibfield  {author} {\bibinfo {author} {\bibfnamefont {M.~T.}\ \bibnamefont
  {Hansen}}\ and\ \bibinfo {author} {\bibfnamefont {S.~R.}\ \bibnamefont
  {Sharpe}},\ }\href {\doibase 10.1103/PhysRevD.86.016007} {\bibfield
  {journal} {\bibinfo  {journal} {Phys.Rev.}\ }\textbf {\bibinfo {volume}
  {D86}},\ \bibinfo {pages} {016007} (\bibinfo {year} {2012})},\ \Eprint
  {http://arxiv.org/abs/1204.0826} {arXiv:1204.0826 [hep-lat]} \BibitemShut
  {NoStop}%
\bibitem [{\citenamefont {Meyer}(2013)}]{Meyer:2013dxa}%
  \BibitemOpen
  \bibfield  {author} {\bibinfo {author} {\bibfnamefont {H.~B.}\ \bibnamefont
  {Meyer}},\ }\href {\doibase 10.1140/epja/i2013-13084-9} {\bibfield  {journal}
  {\bibinfo  {journal} {Eur.Phys.J.}\ }\textbf {\bibinfo {volume} {A49}},\
  \bibinfo {pages} {84} (\bibinfo {year} {2013})},\ \Eprint
  {http://arxiv.org/abs/1303.0138} {arXiv:1303.0138 [hep-lat]} \BibitemShut
  {NoStop}%
\bibitem [{\citenamefont {Agadjanov}\ \emph {et~al.}(2014)\citenamefont
  {Agadjanov}, \citenamefont {Bernard}, \citenamefont {Mei§ner},\ and\
  \citenamefont {Rusetsky}}]{Agadjanov:2014kha}%
  \BibitemOpen
  \bibfield  {author} {\bibinfo {author} {\bibfnamefont {A.}~\bibnamefont
  {Agadjanov}}, \bibinfo {author} {\bibfnamefont {V.}~\bibnamefont {Bernard}},
  \bibinfo {author} {\bibfnamefont {U.-G.}\ \bibnamefont {Mei§ner}}, \ and\
  \bibinfo {author} {\bibfnamefont {A.}~\bibnamefont {Rusetsky}},\ }\href@noop
  {} {\  (\bibinfo {year} {2014})},\ \Eprint {http://arxiv.org/abs/1405.3476}
  {arXiv:1405.3476 [hep-lat]} \BibitemShut {NoStop}%
\bibitem [{\citenamefont {Detmold}\ and\ \citenamefont
  {Savage}(2004)}]{Detmold:2004qn}%
  \BibitemOpen
  \bibfield  {author} {\bibinfo {author} {\bibfnamefont {W.}~\bibnamefont
  {Detmold}}\ and\ \bibinfo {author} {\bibfnamefont {M.~J.}\ \bibnamefont
  {Savage}},\ }\href {\doibase 10.1016/j.nuclphysa.2004.07.007} {\bibfield
  {journal} {\bibinfo  {journal} {Nucl.Phys.}\ }\textbf {\bibinfo {volume}
  {A743}},\ \bibinfo {pages} {170} (\bibinfo {year} {2004})},\ \Eprint
  {http://arxiv.org/abs/hep-lat/0403005} {arXiv:hep-lat/0403005 [hep-lat]}
  \BibitemShut {NoStop}%
\bibitem [{\citenamefont {Meyer}(2012)}]{Meyer:2012wk}%
  \BibitemOpen
  \bibfield  {author} {\bibinfo {author} {\bibfnamefont {H.~B.}\ \bibnamefont
  {Meyer}},\ }\href@noop {} {\  (\bibinfo {year} {2012})},\ \Eprint
  {http://arxiv.org/abs/1202.6675} {arXiv:1202.6675 [hep-lat]} \BibitemShut
  {NoStop}%
\bibitem [{\citenamefont {Bernard}\ \emph {et~al.}(2012)\citenamefont
  {Bernard}, \citenamefont {Hoja}, \citenamefont {Meissner},\ and\
  \citenamefont {Rusetsky}}]{Bernard:2012bi}%
  \BibitemOpen
  \bibfield  {author} {\bibinfo {author} {\bibfnamefont {V.}~\bibnamefont
  {Bernard}}, \bibinfo {author} {\bibfnamefont {D.}~\bibnamefont {Hoja}},
  \bibinfo {author} {\bibfnamefont {U.-G.}\ \bibnamefont {Meissner}}, \ and\
  \bibinfo {author} {\bibfnamefont {A.}~\bibnamefont {Rusetsky}},\ }\href
  {\doibase 10.1007/JHEP09(2012)023} {\bibfield  {journal} {\bibinfo  {journal}
  {JHEP}\ }\textbf {\bibinfo {volume} {1209}},\ \bibinfo {pages} {023}
  (\bibinfo {year} {2012})},\ \Eprint {http://arxiv.org/abs/1205.4642}
  {arXiv:1205.4642 [hep-lat]} \BibitemShut {NoStop}%
\bibitem [{\citenamefont {Briceno}\ and\ \citenamefont
  {Davoudi}(2013)}]{Briceno:2012yi}%
  \BibitemOpen
  \bibfield  {author} {\bibinfo {author} {\bibfnamefont {R.~A.}\ \bibnamefont
  {Briceno}}\ and\ \bibinfo {author} {\bibfnamefont {Z.}~\bibnamefont
  {Davoudi}},\ }\href {\doibase 10.1103/PhysRevD.88.094507} {\bibfield
  {journal} {\bibinfo  {journal} {Phys. Rev. D. 88,}\ }\textbf {\bibinfo
  {volume} {094507}},\ \bibinfo {pages} {094507} (\bibinfo {year} {2013})},\
  \Eprint {http://arxiv.org/abs/1204.1110} {arXiv:1204.1110 [hep-lat]}
  \BibitemShut {NoStop}%
\bibitem [{\citenamefont {Luscher}(1986{\natexlab{a}})}]{Luscher:1985dn}%
  \BibitemOpen
  \bibfield  {author} {\bibinfo {author} {\bibfnamefont {M.}~\bibnamefont
  {Luscher}},\ }\href {\doibase 10.1007/BF01211589} {\bibfield  {journal}
  {\bibinfo  {journal} {Commun.Math.Phys.}\ }\textbf {\bibinfo {volume}
  {104}},\ \bibinfo {pages} {177} (\bibinfo {year}
  {1986}{\natexlab{a}})}\BibitemShut {NoStop}%
\bibitem [{\citenamefont {Luscher}(1986{\natexlab{b}})}]{Luscher:1986pf}%
  \BibitemOpen
  \bibfield  {author} {\bibinfo {author} {\bibfnamefont {M.}~\bibnamefont
  {Luscher}},\ }\href {\doibase 10.1007/BF01211097} {\bibfield  {journal}
  {\bibinfo  {journal} {Commun.Math.Phys.}\ }\textbf {\bibinfo {volume}
  {105}},\ \bibinfo {pages} {153} (\bibinfo {year}
  {1986}{\natexlab{b}})}\BibitemShut {NoStop}%
\bibitem [{\citenamefont {Luscher}(1991)}]{Luscher:1990ux}%
  \BibitemOpen
  \bibfield  {author} {\bibinfo {author} {\bibfnamefont {M.}~\bibnamefont
  {Luscher}},\ }\href {\doibase 10.1016/0550-3213(91)90366-6} {\bibfield
  {journal} {\bibinfo  {journal} {Nucl.Phys.}\ }\textbf {\bibinfo {volume}
  {B354}},\ \bibinfo {pages} {531} (\bibinfo {year} {1991})}\BibitemShut
  {NoStop}%
\bibitem [{\citenamefont {Rummukainen}\ and\ \citenamefont
  {Gottlieb}(1995)}]{Rummukainen:1995vs}%
  \BibitemOpen
  \bibfield  {author} {\bibinfo {author} {\bibfnamefont {K.}~\bibnamefont
  {Rummukainen}}\ and\ \bibinfo {author} {\bibfnamefont {S.~A.}\ \bibnamefont
  {Gottlieb}},\ }\href {\doibase 10.1016/0550-3213(95)00313-H} {\bibfield
  {journal} {\bibinfo  {journal} {Nucl. Phys.}\ }\textbf {\bibinfo {volume}
  {B450}},\ \bibinfo {pages} {397} (\bibinfo {year} {1995})},\ \Eprint
  {http://arxiv.org/abs/hep-lat/9503028} {arXiv:hep-lat/9503028} \BibitemShut
  {NoStop}%
\bibitem [{\citenamefont {Beane}\ \emph {et~al.}(2005)\citenamefont {Beane},
  \citenamefont {Bedaque}, \citenamefont {Parreno},\ and\ \citenamefont
  {Savage}}]{Beane:2003yx}%
  \BibitemOpen
  \bibfield  {author} {\bibinfo {author} {\bibfnamefont {S.}~\bibnamefont
  {Beane}}, \bibinfo {author} {\bibfnamefont {P.}~\bibnamefont {Bedaque}},
  \bibinfo {author} {\bibfnamefont {A.}~\bibnamefont {Parreno}}, \ and\
  \bibinfo {author} {\bibfnamefont {M.}~\bibnamefont {Savage}},\ }\href
  {\doibase 10.1016/j.nuclphysa.2004.09.081} {\bibfield  {journal} {\bibinfo
  {journal} {Nucl.Phys.}\ }\textbf {\bibinfo {volume} {A747}},\ \bibinfo
  {pages} {55} (\bibinfo {year} {2005})},\ \Eprint
  {http://arxiv.org/abs/nucl-th/0311027} {arXiv:nucl-th/0311027 [nucl-th]}
  \BibitemShut {NoStop}%
\bibitem [{\citenamefont {Beane}\ \emph {et~al.}(2004)\citenamefont {Beane},
  \citenamefont {Bedaque}, \citenamefont {Parreno},\ and\ \citenamefont
  {Savage}}]{Beane:2003da}%
  \BibitemOpen
  \bibfield  {author} {\bibinfo {author} {\bibfnamefont {S.~R.}\ \bibnamefont
  {Beane}}, \bibinfo {author} {\bibfnamefont {P.}~\bibnamefont {Bedaque}},
  \bibinfo {author} {\bibfnamefont {A.}~\bibnamefont {Parreno}}, \ and\
  \bibinfo {author} {\bibfnamefont {M.}~\bibnamefont {Savage}},\ }\href
  {\doibase 10.1016/j.physletb.2004.02.007} {\bibfield  {journal} {\bibinfo
  {journal} {Phys.Lett.}\ }\textbf {\bibinfo {volume} {B585}},\ \bibinfo
  {pages} {106} (\bibinfo {year} {2004})},\ \Eprint
  {http://arxiv.org/abs/hep-lat/0312004} {arXiv:hep-lat/0312004 [hep-lat]}
  \BibitemShut {NoStop}%
\bibitem [{\citenamefont {Li}\ and\ \citenamefont {Liu}(2004)}]{Li:2003jn}%
  \BibitemOpen
  \bibfield  {author} {\bibinfo {author} {\bibfnamefont {X.}~\bibnamefont
  {Li}}\ and\ \bibinfo {author} {\bibfnamefont {C.}~\bibnamefont {Liu}},\
  }\href {\doibase 10.1016/j.physletb.2004.02.068} {\bibfield  {journal}
  {\bibinfo  {journal} {Phys.Lett.}\ }\textbf {\bibinfo {volume} {B587}},\
  \bibinfo {pages} {100} (\bibinfo {year} {2004})},\ \Eprint
  {http://arxiv.org/abs/hep-lat/0311035} {arXiv:hep-lat/0311035 [hep-lat]}
  \BibitemShut {NoStop}%
\bibitem [{\citenamefont {Bedaque}(2004)}]{Bedaque:2004kc}%
  \BibitemOpen
  \bibfield  {author} {\bibinfo {author} {\bibfnamefont {P.~F.}\ \bibnamefont
  {Bedaque}},\ }\href {\doibase 10.1016/j.physletb.2004.04.045} {\bibfield
  {journal} {\bibinfo  {journal} {Phys.Lett.}\ }\textbf {\bibinfo {volume}
  {B593}},\ \bibinfo {pages} {82} (\bibinfo {year} {2004})},\ \Eprint
  {http://arxiv.org/abs/nucl-th/0402051} {arXiv:nucl-th/0402051 [nucl-th]}
  \BibitemShut {NoStop}%
\bibitem [{\citenamefont {Feng}\ \emph {et~al.}(2004)\citenamefont {Feng},
  \citenamefont {Li},\ and\ \citenamefont {Liu}}]{Feng:2004ua}%
  \BibitemOpen
  \bibfield  {author} {\bibinfo {author} {\bibfnamefont {X.}~\bibnamefont
  {Feng}}, \bibinfo {author} {\bibfnamefont {X.}~\bibnamefont {Li}}, \ and\
  \bibinfo {author} {\bibfnamefont {C.}~\bibnamefont {Liu}},\ }\href {\doibase
  10.1103/PhysRevD.70.014505} {\bibfield  {journal} {\bibinfo  {journal}
  {Phys.Rev.}\ }\textbf {\bibinfo {volume} {D70}},\ \bibinfo {pages} {014505}
  (\bibinfo {year} {2004})},\ \Eprint {http://arxiv.org/abs/hep-lat/0404001}
  {arXiv:hep-lat/0404001 [hep-lat]} \BibitemShut {NoStop}%
\bibitem [{\citenamefont {Bernard}\ \emph {et~al.}(2008)\citenamefont
  {Bernard}, \citenamefont {Lage}, \citenamefont {Meissner},\ and\
  \citenamefont {Rusetsky}}]{Bernard:2008ax}%
  \BibitemOpen
  \bibfield  {author} {\bibinfo {author} {\bibfnamefont {V.}~\bibnamefont
  {Bernard}}, \bibinfo {author} {\bibfnamefont {M.}~\bibnamefont {Lage}},
  \bibinfo {author} {\bibfnamefont {U.-G.}\ \bibnamefont {Meissner}}, \ and\
  \bibinfo {author} {\bibfnamefont {A.}~\bibnamefont {Rusetsky}},\ }\href
  {\doibase 10.1088/1126-6708/2008/08/024} {\bibfield  {journal} {\bibinfo
  {journal} {JHEP}\ }\textbf {\bibinfo {volume} {0808}},\ \bibinfo {pages}
  {024} (\bibinfo {year} {2008})},\ \Eprint {http://arxiv.org/abs/0806.4495}
  {arXiv:0806.4495 [hep-lat]} \BibitemShut {NoStop}%
\bibitem [{\citenamefont {Bour}\ \emph {et~al.}(2011)\citenamefont {Bour},
  \citenamefont {Koenig}, \citenamefont {Lee}, \citenamefont {Hammer},\ and\
  \citenamefont {Meissner}}]{Bour:2011ef}%
  \BibitemOpen
  \bibfield  {author} {\bibinfo {author} {\bibfnamefont {S.}~\bibnamefont
  {Bour}}, \bibinfo {author} {\bibfnamefont {S.}~\bibnamefont {Koenig}},
  \bibinfo {author} {\bibfnamefont {D.}~\bibnamefont {Lee}}, \bibinfo {author}
  {\bibfnamefont {H.-W.}\ \bibnamefont {Hammer}}, \ and\ \bibinfo {author}
  {\bibfnamefont {U.-G.}\ \bibnamefont {Meissner}},\ }\href {\doibase
  10.1103/PhysRevD.84.091503} {\bibfield  {journal} {\bibinfo  {journal}
  {Phys.Rev.}\ }\textbf {\bibinfo {volume} {D84}},\ \bibinfo {pages} {091503}
  (\bibinfo {year} {2011})},\ \Eprint {http://arxiv.org/abs/1107.1272}
  {arXiv:1107.1272 [nucl-th]} \BibitemShut {NoStop}%
\bibitem [{\citenamefont {Davoudi}\ and\ \citenamefont
  {Savage}(2011)}]{Davoudi:2011md}%
  \BibitemOpen
  \bibfield  {author} {\bibinfo {author} {\bibfnamefont {Z.}~\bibnamefont
  {Davoudi}}\ and\ \bibinfo {author} {\bibfnamefont {M.~J.}\ \bibnamefont
  {Savage}},\ }\href {\doibase 10.1103/PhysRevD.84.114502} {\bibfield
  {journal} {\bibinfo  {journal} {Phys.Rev.}\ }\textbf {\bibinfo {volume}
  {D84}},\ \bibinfo {pages} {114502} (\bibinfo {year} {2011})},\ \Eprint
  {http://arxiv.org/abs/1108.5371} {arXiv:1108.5371 [hep-lat]} \BibitemShut
  {NoStop}%
\bibitem [{\citenamefont {Leskovec}\ and\ \citenamefont
  {Prelovsek}(2012)}]{Leskovec:2012gb}%
  \BibitemOpen
  \bibfield  {author} {\bibinfo {author} {\bibfnamefont {L.}~\bibnamefont
  {Leskovec}}\ and\ \bibinfo {author} {\bibfnamefont {S.}~\bibnamefont
  {Prelovsek}},\ }\href {\doibase 10.1103/PhysRevD.85.114507} {\bibfield
  {journal} {\bibinfo  {journal} {Phys.Rev.}\ }\textbf {\bibinfo {volume}
  {D85}},\ \bibinfo {pages} {114507} (\bibinfo {year} {2012})},\ \Eprint
  {http://arxiv.org/abs/1202.2145} {arXiv:1202.2145 [hep-lat]} \BibitemShut
  {NoStop}%
\bibitem [{\citenamefont {Gockeler}\ \emph {et~al.}(2012)\citenamefont
  {Gockeler}, \citenamefont {Horsley}, \citenamefont {Lage}, \citenamefont
  {Meissner}, \citenamefont {Rakow} \emph {et~al.}}]{Gockeler:2012yj}%
  \BibitemOpen
  \bibfield  {author} {\bibinfo {author} {\bibfnamefont {M.}~\bibnamefont
  {Gockeler}}, \bibinfo {author} {\bibfnamefont {R.}~\bibnamefont {Horsley}},
  \bibinfo {author} {\bibfnamefont {M.}~\bibnamefont {Lage}}, \bibinfo {author}
  {\bibfnamefont {U.-G.}\ \bibnamefont {Meissner}}, \bibinfo {author}
  {\bibfnamefont {P.}~\bibnamefont {Rakow}},  \emph {et~al.},\ }\href {\doibase
  10.1103/PhysRevD.86.094513} {\bibfield  {journal} {\bibinfo  {journal}
  {Phys.Rev.}\ }\textbf {\bibinfo {volume} {D86}},\ \bibinfo {pages} {094513}
  (\bibinfo {year} {2012})},\ \Eprint {http://arxiv.org/abs/1206.4141}
  {arXiv:1206.4141 [hep-lat]} \BibitemShut {NoStop}%
\bibitem [{\citenamefont {Ishizuka}(2009)}]{Ishizuka:2009bx}%
  \BibitemOpen
  \bibfield  {author} {\bibinfo {author} {\bibfnamefont {N.}~\bibnamefont
  {Ishizuka}},\ }\href@noop {} {\bibfield  {journal} {\bibinfo  {journal}
  {PoS}\ }\textbf {\bibinfo {volume} {LAT2009}},\ \bibinfo {pages} {119}
  (\bibinfo {year} {2009})},\ \Eprint {http://arxiv.org/abs/0910.2772}
  {arXiv:0910.2772 [hep-lat]} \BibitemShut {NoStop}%
\bibitem [{\citenamefont {Briceno}(2013)}]{Briceno:2013rwa}%
  \BibitemOpen
  \bibfield  {author} {\bibinfo {author} {\bibfnamefont {R.~A.}\ \bibnamefont
  {Briceno}},\ }\href@noop {} {\  (\bibinfo {year} {2013})},\ \Eprint
  {http://arxiv.org/abs/1311.6032} {arXiv:1311.6032 [hep-lat]} \BibitemShut
  {NoStop}%
\bibitem [{\citenamefont {Briceno}\ \emph
  {et~al.}(2013{\natexlab{a}})\citenamefont {Briceno}, \citenamefont
  {Davoudi},\ and\ \citenamefont {Luu}}]{Briceno:2013lba}%
  \BibitemOpen
  \bibfield  {author} {\bibinfo {author} {\bibfnamefont {R.~A.}\ \bibnamefont
  {Briceno}}, \bibinfo {author} {\bibfnamefont {Z.}~\bibnamefont {Davoudi}}, \
  and\ \bibinfo {author} {\bibfnamefont {T.~C.}\ \bibnamefont {Luu}},\ }\href
  {\doibase 10.1103/PhysRevD.88.034502} {\bibfield  {journal} {\bibinfo
  {journal} {Phys.Rev.}\ }\textbf {\bibinfo {volume} {D88}},\ \bibinfo {pages}
  {034502} (\bibinfo {year} {2013}{\natexlab{a}})},\ \Eprint
  {http://arxiv.org/abs/1305.4903} {arXiv:1305.4903 [hep-lat]} \BibitemShut
  {NoStop}%
\bibitem [{\citenamefont {Briceno}\ \emph
  {et~al.}(2013{\natexlab{b}})\citenamefont {Briceno}, \citenamefont {Davoudi},
  \citenamefont {Luu},\ and\ \citenamefont {Savage}}]{Briceno:2013bda}%
  \BibitemOpen
  \bibfield  {author} {\bibinfo {author} {\bibfnamefont {R.~A.}\ \bibnamefont
  {Briceno}}, \bibinfo {author} {\bibfnamefont {Z.}~\bibnamefont {Davoudi}},
  \bibinfo {author} {\bibfnamefont {T.}~\bibnamefont {Luu}}, \ and\ \bibinfo
  {author} {\bibfnamefont {M.~J.}\ \bibnamefont {Savage}},\ }\href@noop {} {\
  (\bibinfo {year} {2013}{\natexlab{b}})},\ \Eprint
  {http://arxiv.org/abs/1309.3556} {arXiv:1309.3556 [hep-lat]} \BibitemShut
  {NoStop}%
\bibitem [{\citenamefont {Briceno}\ \emph
  {et~al.}(2013{\natexlab{c}})\citenamefont {Briceno}, \citenamefont {Davoudi},
  \citenamefont {Luu},\ and\ \citenamefont {Savage}}]{Briceno:2013hya}%
  \BibitemOpen
  \bibfield  {author} {\bibinfo {author} {\bibfnamefont {R.~A.}\ \bibnamefont
  {Briceno}}, \bibinfo {author} {\bibfnamefont {Z.}~\bibnamefont {Davoudi}},
  \bibinfo {author} {\bibfnamefont {T.~C.}\ \bibnamefont {Luu}}, \ and\
  \bibinfo {author} {\bibfnamefont {M.~J.}\ \bibnamefont {Savage}},\
  }\href@noop {} {\  (\bibinfo {year} {2013}{\natexlab{c}})},\ \Eprint
  {http://arxiv.org/abs/1311.7686} {arXiv:1311.7686 [hep-lat]} \BibitemShut
  {NoStop}%
\bibitem [{\citenamefont {Liu}\ \emph {et~al.}(2006)\citenamefont {Liu},
  \citenamefont {Feng},\ and\ \citenamefont {He}}]{Liu:2005kr}%
  \BibitemOpen
  \bibfield  {author} {\bibinfo {author} {\bibfnamefont {C.}~\bibnamefont
  {Liu}}, \bibinfo {author} {\bibfnamefont {X.}~\bibnamefont {Feng}}, \ and\
  \bibinfo {author} {\bibfnamefont {S.}~\bibnamefont {He}},\ }\href {\doibase
  10.1142/S0217751X06032150} {\bibfield  {journal} {\bibinfo  {journal}
  {Int.J.Mod.Phys.}\ }\textbf {\bibinfo {volume} {A21}},\ \bibinfo {pages}
  {847} (\bibinfo {year} {2006})},\ \Eprint
  {http://arxiv.org/abs/hep-lat/0508022} {arXiv:hep-lat/0508022 [hep-lat]}
  \BibitemShut {NoStop}%
\bibitem [{\citenamefont {Li}\ and\ \citenamefont {Liu}(2013)}]{Li:2012bi}%
  \BibitemOpen
  \bibfield  {author} {\bibinfo {author} {\bibfnamefont {N.}~\bibnamefont
  {Li}}\ and\ \bibinfo {author} {\bibfnamefont {C.}~\bibnamefont {Liu}},\
  }\href {\doibase 10.1103/PhysRevD.87.014502} {\bibfield  {journal} {\bibinfo
  {journal} {Phys.Rev.}\ }\textbf {\bibinfo {volume} {D87}},\ \bibinfo {pages}
  {014502} (\bibinfo {year} {2013})},\ \Eprint {http://arxiv.org/abs/1209.2201}
  {arXiv:1209.2201 [hep-lat]} \BibitemShut {NoStop}%
\bibitem [{\citenamefont {Guo}\ \emph {et~al.}(2013)\citenamefont {Guo},
  \citenamefont {Dudek}, \citenamefont {Edwards},\ and\ \citenamefont
  {Szczepaniak}}]{Guo:2012hv}%
  \BibitemOpen
  \bibfield  {author} {\bibinfo {author} {\bibfnamefont {P.}~\bibnamefont
  {Guo}}, \bibinfo {author} {\bibfnamefont {J.}~\bibnamefont {Dudek}}, \bibinfo
  {author} {\bibfnamefont {R.}~\bibnamefont {Edwards}}, \ and\ \bibinfo
  {author} {\bibfnamefont {A.~P.}\ \bibnamefont {Szczepaniak}},\ }\href
  {\doibase 10.1103/PhysRevD.88.014501} {\bibfield  {journal} {\bibinfo
  {journal} {Phys.Rev.}\ }\textbf {\bibinfo {volume} {D88}},\ \bibinfo {pages}
  {014501} (\bibinfo {year} {2013})},\ \Eprint {http://arxiv.org/abs/1211.0929}
  {arXiv:1211.0929 [hep-lat]} \BibitemShut {NoStop}%
\bibitem [{\citenamefont {Bernard}\ \emph {et~al.}(2011)\citenamefont
  {Bernard}, \citenamefont {Lage}, \citenamefont {Meissner},\ and\
  \citenamefont {Rusetsky}}]{Bernard:2010fp}%
  \BibitemOpen
  \bibfield  {author} {\bibinfo {author} {\bibfnamefont {V.}~\bibnamefont
  {Bernard}}, \bibinfo {author} {\bibfnamefont {M.}~\bibnamefont {Lage}},
  \bibinfo {author} {\bibfnamefont {U.-G.}\ \bibnamefont {Meissner}}, \ and\
  \bibinfo {author} {\bibfnamefont {A.}~\bibnamefont {Rusetsky}},\ }\href
  {\doibase 10.1007/JHEP01(2011)019} {\bibfield  {journal} {\bibinfo  {journal}
  {JHEP}\ }\textbf {\bibinfo {volume} {1101}},\ \bibinfo {pages} {019}
  (\bibinfo {year} {2011})},\ \Eprint {http://arxiv.org/abs/1010.6018}
  {arXiv:1010.6018 [hep-lat]} \BibitemShut {NoStop}%
\bibitem [{\citenamefont {Li}\ \emph {et~al.}(2014)\citenamefont {Li},
  \citenamefont {Li},\ and\ \citenamefont {Liu}}]{Li:2014wga}%
  \BibitemOpen
  \bibfield  {author} {\bibinfo {author} {\bibfnamefont {N.}~\bibnamefont
  {Li}}, \bibinfo {author} {\bibfnamefont {S.-Y.}\ \bibnamefont {Li}}, \ and\
  \bibinfo {author} {\bibfnamefont {C.}~\bibnamefont {Liu}},\ }\href@noop {} {\
   (\bibinfo {year} {2014})},\ \Eprint {http://arxiv.org/abs/1401.5569}
  {arXiv:1401.5569 [hep-lat]} \BibitemShut {NoStop}%
\bibitem [{\citenamefont {Briceno}(2014)}]{Briceno:2014oea}%
  \BibitemOpen
  \bibfield  {author} {\bibinfo {author} {\bibfnamefont {R.~A.}\ \bibnamefont
  {Briceno}},\ }\href@noop {} {\  (\bibinfo {year} {2014})},\ \Eprint
  {http://arxiv.org/abs/1401.3312} {arXiv:1401.3312 [hep-lat]} \BibitemShut
  {NoStop}%
\bibitem [{\citenamefont {Roca}\ and\ \citenamefont
  {Oset}(2012)}]{Roca:2012rx}%
  \BibitemOpen
  \bibfield  {author} {\bibinfo {author} {\bibfnamefont {L.}~\bibnamefont
  {Roca}}\ and\ \bibinfo {author} {\bibfnamefont {E.}~\bibnamefont {Oset}},\
  }\href {\doibase 10.1103/PhysRevD.85.054507} {\bibfield  {journal} {\bibinfo
  {journal} {Phys.Rev.}\ }\textbf {\bibinfo {volume} {D85}},\ \bibinfo {pages}
  {054507} (\bibinfo {year} {2012})},\ \Eprint {http://arxiv.org/abs/1201.0438}
  {arXiv:1201.0438 [hep-lat]} \BibitemShut {NoStop}%
\bibitem [{\citenamefont {Polejaeva}\ and\ \citenamefont
  {Rusetsky}(2012)}]{Polejaeva:2012ut}%
  \BibitemOpen
  \bibfield  {author} {\bibinfo {author} {\bibfnamefont {K.}~\bibnamefont
  {Polejaeva}}\ and\ \bibinfo {author} {\bibfnamefont {A.}~\bibnamefont
  {Rusetsky}},\ }\href@noop {} {\bibfield  {journal} {\bibinfo  {journal}
  {Eur.Phys.J.}\ }\textbf {\bibinfo {volume} {A48}},\ \bibinfo {pages} {67}
  (\bibinfo {year} {2012})},\ \Eprint {http://arxiv.org/abs/1203.1241}
  {arXiv:1203.1241 [hep-lat]} \BibitemShut {NoStop}%
\bibitem [{\citenamefont {Briceno}\ and\ \citenamefont
  {Davoudi}(2012)}]{Briceno:2012rv}%
  \BibitemOpen
  \bibfield  {author} {\bibinfo {author} {\bibfnamefont {R.~A.}\ \bibnamefont
  {Briceno}}\ and\ \bibinfo {author} {\bibfnamefont {Z.}~\bibnamefont
  {Davoudi}},\ }\href {\doibase 10.1103/PhysRevD.87.094507} {\bibfield
  {journal} {\bibinfo  {journal} {Phys.Rev.}\ }\textbf {\bibinfo {volume}
  {D87}},\ \bibinfo {pages} {094507} (\bibinfo {year} {2012})},\ \Eprint
  {http://arxiv.org/abs/1212.3398} {arXiv:1212.3398 [hep-lat]} \BibitemShut
  {NoStop}%
\bibitem [{\citenamefont {Hansen}\ and\ \citenamefont
  {Sharpe}(2013)}]{Hansen:2013dla}%
  \BibitemOpen
  \bibfield  {author} {\bibinfo {author} {\bibfnamefont {M.~T.}\ \bibnamefont
  {Hansen}}\ and\ \bibinfo {author} {\bibfnamefont {S.~R.}\ \bibnamefont
  {Sharpe}},\ }\href@noop {} {\  (\bibinfo {year} {2013})},\ \Eprint
  {http://arxiv.org/abs/1311.4848} {arXiv:1311.4848 [hep-lat]} \BibitemShut
  {NoStop}%
\bibitem [{\citenamefont {Li}\ \emph {et~al.}(2007)\citenamefont {Li} \emph
  {et~al.}}]{Li:2007ey}%
  \BibitemOpen
  \bibfield  {author} {\bibinfo {author} {\bibfnamefont {X.}~\bibnamefont {Li}}
  \emph {et~al.} (\bibinfo {collaboration} {CLQCD Collaboration}),\ }\href
  {\doibase 10.1088/1126-6708/2007/06/053} {\bibfield  {journal} {\bibinfo
  {journal} {JHEP}\ }\textbf {\bibinfo {volume} {0706}},\ \bibinfo {pages}
  {053} (\bibinfo {year} {2007})},\ \Eprint
  {http://arxiv.org/abs/hep-lat/0703015} {arXiv:hep-lat/0703015 [hep-lat]}
  \BibitemShut {NoStop}%
\bibitem [{\citenamefont {Durr}\ \emph {et~al.}(2008)\citenamefont {Durr},
  \citenamefont {Fodor}, \citenamefont {Frison}, \citenamefont {Hoelbling},
  \citenamefont {Hoffmann} \emph {et~al.}}]{Durr:2008zz}%
  \BibitemOpen
  \bibfield  {author} {\bibinfo {author} {\bibfnamefont {S.}~\bibnamefont
  {Durr}}, \bibinfo {author} {\bibfnamefont {Z.}~\bibnamefont {Fodor}},
  \bibinfo {author} {\bibfnamefont {J.}~\bibnamefont {Frison}}, \bibinfo
  {author} {\bibfnamefont {C.}~\bibnamefont {Hoelbling}}, \bibinfo {author}
  {\bibfnamefont {R.}~\bibnamefont {Hoffmann}},  \emph {et~al.},\ }\href
  {\doibase 10.1126/science.1163233} {\bibfield  {journal} {\bibinfo  {journal}
  {Science}\ }\textbf {\bibinfo {volume} {322}},\ \bibinfo {pages} {1224}
  (\bibinfo {year} {2008})},\ \Eprint {http://arxiv.org/abs/0906.3599}
  {arXiv:0906.3599 [hep-lat]} \BibitemShut {NoStop}%
\bibitem [{\citenamefont {Beane}\ \emph
  {et~al.}(2011{\natexlab{a}})\citenamefont {Beane} \emph
  {et~al.}}]{Beane:2010hg}%
  \BibitemOpen
  \bibfield  {author} {\bibinfo {author} {\bibfnamefont {S.~R.}\ \bibnamefont
  {Beane}} \emph {et~al.} (\bibinfo {collaboration} {NPLQCD Collaboration}),\
  }\href {\doibase 10.1103/PhysRevLett.106.162001} {\bibfield  {journal}
  {\bibinfo  {journal} {Phys.Rev.Lett.}\ }\textbf {\bibinfo {volume} {106}},\
  \bibinfo {pages} {162001} (\bibinfo {year} {2011}{\natexlab{a}})},\ \Eprint
  {http://arxiv.org/abs/1012.3812} {arXiv:1012.3812 [hep-lat]} \BibitemShut
  {NoStop}%
\bibitem [{\citenamefont {Beane}\ \emph
  {et~al.}(2011{\natexlab{b}})\citenamefont {Beane}, \citenamefont {Chang},
  \citenamefont {Detmold}, \citenamefont {Joo}, \citenamefont {Lin} \emph
  {et~al.}}]{Beane:2011xf}%
  \BibitemOpen
  \bibfield  {author} {\bibinfo {author} {\bibfnamefont {S.~R.}\ \bibnamefont
  {Beane}}, \bibinfo {author} {\bibfnamefont {E.}~\bibnamefont {Chang}},
  \bibinfo {author} {\bibfnamefont {W.}~\bibnamefont {Detmold}}, \bibinfo
  {author} {\bibfnamefont {B.}~\bibnamefont {Joo}}, \bibinfo {author}
  {\bibfnamefont {H.}~\bibnamefont {Lin}},  \emph {et~al.},\ }\href {\doibase
  10.1142/S0217732311036978} {\bibfield  {journal} {\bibinfo  {journal}
  {Mod.Phys.Lett.}\ }\textbf {\bibinfo {volume} {A26}},\ \bibinfo {pages}
  {2587} (\bibinfo {year} {2011}{\natexlab{b}})},\ \Eprint
  {http://arxiv.org/abs/1103.2821} {arXiv:1103.2821 [hep-lat]} \BibitemShut
  {NoStop}%
\bibitem [{\citenamefont {Beane}\ \emph
  {et~al.}(2013{\natexlab{a}})\citenamefont {Beane}, \citenamefont {Chang},
  \citenamefont {Cohen}, \citenamefont {Detmold}, \citenamefont {Lin} \emph
  {et~al.}}]{Beane:2012vq}%
  \BibitemOpen
  \bibfield  {author} {\bibinfo {author} {\bibfnamefont {S.~R.}\ \bibnamefont
  {Beane}}, \bibinfo {author} {\bibfnamefont {E.}~\bibnamefont {Chang}},
  \bibinfo {author} {\bibfnamefont {S.~D.}\ \bibnamefont {Cohen}}, \bibinfo
  {author} {\bibfnamefont {W.}~\bibnamefont {Detmold}}, \bibinfo {author}
  {\bibfnamefont {H.}~\bibnamefont {Lin}},  \emph {et~al.},\ }\href {\doibase
  10.1103/PhysRevD.87.034506} {\bibfield  {journal} {\bibinfo  {journal}
  {Phys.Rev.}\ }\textbf {\bibinfo {volume} {D87}},\ \bibinfo {pages} {034506}
  (\bibinfo {year} {2013}{\natexlab{a}})},\ \Eprint
  {http://arxiv.org/abs/1206.5219} {arXiv:1206.5219 [hep-lat]} \BibitemShut
  {NoStop}%
\bibitem [{\citenamefont {Beane}\ \emph
  {et~al.}(2012{\natexlab{a}})\citenamefont {Beane}, \citenamefont {Chang},
  \citenamefont {Cohen}, \citenamefont {Detmold}, \citenamefont {Lin} \emph
  {et~al.}}]{Beane:2012ey}%
  \BibitemOpen
  \bibfield  {author} {\bibinfo {author} {\bibfnamefont {S.}~\bibnamefont
  {Beane}}, \bibinfo {author} {\bibfnamefont {E.}~\bibnamefont {Chang}},
  \bibinfo {author} {\bibfnamefont {S.}~\bibnamefont {Cohen}}, \bibinfo
  {author} {\bibfnamefont {W.}~\bibnamefont {Detmold}}, \bibinfo {author}
  {\bibfnamefont {H.-W.}\ \bibnamefont {Lin}},  \emph {et~al.},\ }\href
  {\doibase 10.1103/PhysRevLett.109.172001} {\bibfield  {journal} {\bibinfo
  {journal} {Phys.Rev.Lett.}\ }\textbf {\bibinfo {volume} {109}},\ \bibinfo
  {pages} {172001} (\bibinfo {year} {2012}{\natexlab{a}})},\ \Eprint
  {http://arxiv.org/abs/1204.3606} {arXiv:1204.3606 [hep-lat]} \BibitemShut
  {NoStop}%
\bibitem [{\citenamefont {Yamazaki}\ \emph {et~al.}(2012)\citenamefont
  {Yamazaki}, \citenamefont {Ishikawa}, \citenamefont {Kuramashi},\ and\
  \citenamefont {Ukawa}}]{Yamazaki:2012hi}%
  \BibitemOpen
  \bibfield  {author} {\bibinfo {author} {\bibfnamefont {T.}~\bibnamefont
  {Yamazaki}}, \bibinfo {author} {\bibfnamefont {K.-i.}\ \bibnamefont
  {Ishikawa}}, \bibinfo {author} {\bibfnamefont {Y.}~\bibnamefont {Kuramashi}},
  \ and\ \bibinfo {author} {\bibfnamefont {A.}~\bibnamefont {Ukawa}},\ }\href
  {\doibase 10.1103/PhysRevD.86.074514} {\bibfield  {journal} {\bibinfo
  {journal} {Phys.Rev.}\ }\textbf {\bibinfo {volume} {D86}},\ \bibinfo {pages}
  {074514} (\bibinfo {year} {2012})},\ \Eprint {http://arxiv.org/abs/1207.4277}
  {arXiv:1207.4277 [hep-lat]} \BibitemShut {NoStop}%
\bibitem [{\citenamefont {Beane}\ \emph
  {et~al.}(2012{\natexlab{b}})\citenamefont {Beane} \emph
  {et~al.}}]{Beane:2011iw}%
  \BibitemOpen
  \bibfield  {author} {\bibinfo {author} {\bibfnamefont {S.~R.}\ \bibnamefont
  {Beane}} \emph {et~al.} (\bibinfo {collaboration} {NPLQCD Collaboration}),\
  }\href {\doibase 10.1103/PhysRevD.85.054511} {\bibfield  {journal} {\bibinfo
  {journal} {Phys.Rev.}\ }\textbf {\bibinfo {volume} {D85}},\ \bibinfo {pages}
  {054511} (\bibinfo {year} {2012}{\natexlab{b}})},\ \Eprint
  {http://arxiv.org/abs/1109.2889} {arXiv:1109.2889 [hep-lat]} \BibitemShut
  {NoStop}%
\bibitem [{\citenamefont {Beane}\ \emph
  {et~al.}(2013{\natexlab{b}})\citenamefont {Beane}, \citenamefont {Chang},
  \citenamefont {Cohen}, \citenamefont {Detmold}, \citenamefont {Junnarkar}
  \emph {et~al.}}]{Beane:2013br}%
  \BibitemOpen
  \bibfield  {author} {\bibinfo {author} {\bibfnamefont {S.}~\bibnamefont
  {Beane}}, \bibinfo {author} {\bibfnamefont {E.}~\bibnamefont {Chang}},
  \bibinfo {author} {\bibfnamefont {S.}~\bibnamefont {Cohen}}, \bibinfo
  {author} {\bibfnamefont {W.}~\bibnamefont {Detmold}}, \bibinfo {author}
  {\bibfnamefont {P.}~\bibnamefont {Junnarkar}},  \emph {et~al.},\ }\href
  {\doibase 10.1103/PhysRevC.88.024003} {\bibfield  {journal} {\bibinfo
  {journal} {Phys.Rev.}\ }\textbf {\bibinfo {volume} {C88}},\ \bibinfo {pages}
  {024003} (\bibinfo {year} {2013}{\natexlab{b}})},\ \Eprint
  {http://arxiv.org/abs/1301.5790} {arXiv:1301.5790 [hep-lat]} \BibitemShut
  {NoStop}%
\bibitem [{\citenamefont {Beane}\ \emph
  {et~al.}(2012{\natexlab{c}})\citenamefont {Beane} \emph
  {et~al.}}]{Beane:2011sc}%
  \BibitemOpen
  \bibfield  {author} {\bibinfo {author} {\bibfnamefont {S.~R.}\ \bibnamefont
  {Beane}} \emph {et~al.} (\bibinfo {collaboration} {NPLQCD Collaboration}),\
  }\href {\doibase 10.1103/PhysRevD.85.034505} {\bibfield  {journal} {\bibinfo
  {journal} {Phys.Rev.}\ }\textbf {\bibinfo {volume} {D85}},\ \bibinfo {pages}
  {034505} (\bibinfo {year} {2012}{\natexlab{c}})},\ \Eprint
  {http://arxiv.org/abs/1107.5023} {arXiv:1107.5023 [hep-lat]} \BibitemShut
  {NoStop}%
\bibitem [{\citenamefont {Pelissier}\ \emph {et~al.}(2011)\citenamefont
  {Pelissier}, \citenamefont {Alexandru},\ and\ \citenamefont
  {Lee}}]{Pelissier:2011ib}%
  \BibitemOpen
  \bibfield  {author} {\bibinfo {author} {\bibfnamefont {C.~S.}\ \bibnamefont
  {Pelissier}}, \bibinfo {author} {\bibfnamefont {A.}~\bibnamefont
  {Alexandru}}, \ and\ \bibinfo {author} {\bibfnamefont {F.~X.}\ \bibnamefont
  {Lee}},\ }\href@noop {} {\bibfield  {journal} {\bibinfo  {journal} {PoS}\
  }\textbf {\bibinfo {volume} {LATTICE2011}},\ \bibinfo {pages} {134} (\bibinfo
  {year} {2011})},\ \Eprint {http://arxiv.org/abs/1111.2314} {arXiv:1111.2314
  [hep-lat]} \BibitemShut {NoStop}%
\bibitem [{\citenamefont {Aoki}\ \emph {et~al.}(2007)\citenamefont {Aoki} \emph
  {et~al.}}]{Aoki:2007rd}%
  \BibitemOpen
  \bibfield  {author} {\bibinfo {author} {\bibfnamefont {S.}~\bibnamefont
  {Aoki}} \emph {et~al.} (\bibinfo {collaboration} {CP-PACS Collaboration}),\
  }\href {\doibase 10.1103/PhysRevD.76.094506} {\bibfield  {journal} {\bibinfo
  {journal} {Phys.Rev.}\ }\textbf {\bibinfo {volume} {D76}},\ \bibinfo {pages}
  {094506} (\bibinfo {year} {2007})},\ \Eprint {http://arxiv.org/abs/0708.3705}
  {arXiv:0708.3705 [hep-lat]} \BibitemShut {NoStop}%
\bibitem [{\citenamefont {Lang}\ \emph {et~al.}(2011)\citenamefont {Lang},
  \citenamefont {Mohler}, \citenamefont {Prelovsek},\ and\ \citenamefont
  {Vidmar}}]{Lang:2011mn}%
  \BibitemOpen
  \bibfield  {author} {\bibinfo {author} {\bibfnamefont {C.}~\bibnamefont
  {Lang}}, \bibinfo {author} {\bibfnamefont {D.}~\bibnamefont {Mohler}},
  \bibinfo {author} {\bibfnamefont {S.}~\bibnamefont {Prelovsek}}, \ and\
  \bibinfo {author} {\bibfnamefont {M.}~\bibnamefont {Vidmar}},\ }\href
  {\doibase 10.1103/PhysRevD.84.054503} {\bibfield  {journal} {\bibinfo
  {journal} {Phys.Rev.}\ }\textbf {\bibinfo {volume} {D84}},\ \bibinfo {pages}
  {054503} (\bibinfo {year} {2011})},\ \Eprint {http://arxiv.org/abs/1105.5636}
  {arXiv:1105.5636 [hep-lat]} \BibitemShut {NoStop}%
\bibitem [{\citenamefont {Pelissier}\ and\ \citenamefont
  {Alexandru}(2013)}]{Pelissier:2012pi}%
  \BibitemOpen
  \bibfield  {author} {\bibinfo {author} {\bibfnamefont {C.}~\bibnamefont
  {Pelissier}}\ and\ \bibinfo {author} {\bibfnamefont {A.}~\bibnamefont
  {Alexandru}},\ }\href {\doibase 10.1103/PhysRevD.87.014503} {\bibfield
  {journal} {\bibinfo  {journal} {Phys.Rev.}\ }\textbf {\bibinfo {volume}
  {D87}},\ \bibinfo {pages} {014503} (\bibinfo {year} {2013})},\ \Eprint
  {http://arxiv.org/abs/1211.0092} {arXiv:1211.0092 [hep-lat]} \BibitemShut
  {NoStop}%
\bibitem [{\citenamefont {Ozaki}\ and\ \citenamefont
  {Sasaki}(2013)}]{Ozaki:2012ce}%
  \BibitemOpen
  \bibfield  {author} {\bibinfo {author} {\bibfnamefont {S.}~\bibnamefont
  {Ozaki}}\ and\ \bibinfo {author} {\bibfnamefont {S.}~\bibnamefont {Sasaki}},\
  }\href {\doibase 10.1103/PhysRevD.87.014506} {\bibfield  {journal} {\bibinfo
  {journal} {Phys.Rev.}\ }\textbf {\bibinfo {volume} {D87}},\ \bibinfo {pages}
  {014506} (\bibinfo {year} {2013})},\ \Eprint {http://arxiv.org/abs/1211.5512}
  {arXiv:1211.5512 [hep-lat]} \BibitemShut {NoStop}%
\bibitem [{\citenamefont {Buchoff}\ \emph {et~al.}(2012)\citenamefont
  {Buchoff}, \citenamefont {Luu},\ and\ \citenamefont
  {Wasem}}]{Buchoff:2012ja}%
  \BibitemOpen
  \bibfield  {author} {\bibinfo {author} {\bibfnamefont {M.~I.}\ \bibnamefont
  {Buchoff}}, \bibinfo {author} {\bibfnamefont {T.~C.}\ \bibnamefont {Luu}}, \
  and\ \bibinfo {author} {\bibfnamefont {J.}~\bibnamefont {Wasem}},\ }\href
  {\doibase 10.1103/PhysRevD.85.094511} {\bibfield  {journal} {\bibinfo
  {journal} {Phys.Rev.}\ }\textbf {\bibinfo {volume} {D85}},\ \bibinfo {pages}
  {094511} (\bibinfo {year} {2012})},\ \Eprint {http://arxiv.org/abs/1201.3596}
  {arXiv:1201.3596 [hep-lat]} \BibitemShut {NoStop}%
\bibitem [{\citenamefont {Dudek}\ \emph {et~al.}(2013)\citenamefont {Dudek},
  \citenamefont {Edwards},\ and\ \citenamefont {Thomas}}]{Dudek:2012xn}%
  \BibitemOpen
  \bibfield  {author} {\bibinfo {author} {\bibfnamefont {J.~J.}\ \bibnamefont
  {Dudek}}, \bibinfo {author} {\bibfnamefont {R.~G.}\ \bibnamefont {Edwards}},
  \ and\ \bibinfo {author} {\bibfnamefont {C.~E.}\ \bibnamefont {Thomas}},\
  }\href {\doibase 10.1103/PhysRevD.87.034505} {\bibfield  {journal} {\bibinfo
  {journal} {Phys.Rev.}\ }\textbf {\bibinfo {volume} {D87}},\ \bibinfo {pages}
  {034505} (\bibinfo {year} {2013})},\ \Eprint {http://arxiv.org/abs/1212.0830}
  {arXiv:1212.0830 [hep-ph]} \BibitemShut {NoStop}%
\bibitem [{\citenamefont {Dudek}\ \emph {et~al.}(2012)\citenamefont {Dudek},
  \citenamefont {Edwards},\ and\ \citenamefont {Thomas}}]{Dudek:2012gj}%
  \BibitemOpen
  \bibfield  {author} {\bibinfo {author} {\bibfnamefont {J.~J.}\ \bibnamefont
  {Dudek}}, \bibinfo {author} {\bibfnamefont {R.~G.}\ \bibnamefont {Edwards}},
  \ and\ \bibinfo {author} {\bibfnamefont {C.~E.}\ \bibnamefont {Thomas}},\
  }\href {\doibase 10.1103/PhysRevD.86.034031} {\bibfield  {journal} {\bibinfo
  {journal} {Phys.Rev.}\ }\textbf {\bibinfo {volume} {D86}},\ \bibinfo {pages}
  {034031} (\bibinfo {year} {2012})},\ \Eprint {http://arxiv.org/abs/1203.6041}
  {arXiv:1203.6041 [hep-ph]} \BibitemShut {NoStop}%
\bibitem [{\citenamefont {Mohler}\ \emph {et~al.}(2013)\citenamefont {Mohler},
  \citenamefont {Lang}, \citenamefont {Leskovec}, \citenamefont {Prelovsek},\
  and\ \citenamefont {Woloshyn}}]{Mohler:2013rwa}%
  \BibitemOpen
  \bibfield  {author} {\bibinfo {author} {\bibfnamefont {D.}~\bibnamefont
  {Mohler}}, \bibinfo {author} {\bibfnamefont {C.}~\bibnamefont {Lang}},
  \bibinfo {author} {\bibfnamefont {L.}~\bibnamefont {Leskovec}}, \bibinfo
  {author} {\bibfnamefont {S.}~\bibnamefont {Prelovsek}}, \ and\ \bibinfo
  {author} {\bibfnamefont {R.}~\bibnamefont {Woloshyn}},\ }\href {\doibase
  10.1103/PhysRevLett.111.222001} {\bibfield  {journal} {\bibinfo  {journal}
  {Phys.Rev.Lett.}\ }\textbf {\bibinfo {volume} {111}},\ \bibinfo {pages}
  {222001} (\bibinfo {year} {2013})},\ \Eprint {http://arxiv.org/abs/1308.3175}
  {arXiv:1308.3175 [hep-lat]} \BibitemShut {NoStop}%
\bibitem [{\citenamefont {Lang}\ \emph {et~al.}(2014)\citenamefont {Lang},
  \citenamefont {Leskovec}, \citenamefont {Mohler},\ and\ \citenamefont
  {Prelovsek}}]{Lang:2014tia}%
  \BibitemOpen
  \bibfield  {author} {\bibinfo {author} {\bibfnamefont {C.}~\bibnamefont
  {Lang}}, \bibinfo {author} {\bibfnamefont {L.}~\bibnamefont {Leskovec}},
  \bibinfo {author} {\bibfnamefont {D.}~\bibnamefont {Mohler}}, \ and\ \bibinfo
  {author} {\bibfnamefont {S.}~\bibnamefont {Prelovsek}},\ }\href@noop {} {\
  (\bibinfo {year} {2014})},\ \Eprint {http://arxiv.org/abs/1401.2088}
  {arXiv:1401.2088 [hep-lat]} \BibitemShut {NoStop}%
\bibitem [{\citenamefont {Guo}(2013)}]{Guo:2013vsa}%
  \BibitemOpen
  \bibfield  {author} {\bibinfo {author} {\bibfnamefont {P.}~\bibnamefont
  {Guo}},\ }\href {\doibase 10.1103/PhysRevD.88.014507} {\bibfield  {journal}
  {\bibinfo  {journal} {Phys.Rev.}\ }\textbf {\bibinfo {volume} {D88}},\
  \bibinfo {pages} {014507} (\bibinfo {year} {2013})},\ \Eprint
  {http://arxiv.org/abs/1304.7812} {arXiv:1304.7812 [hep-lat]} \BibitemShut
  {NoStop}%
\bibitem [{\citenamefont {Dudek}\ \emph {et~al.}(2014)\citenamefont {Dudek},
  \citenamefont {Edwards}, \citenamefont {Thomas},\ and\ \citenamefont
  {Wilson}}]{Dudek:2014qha}%
  \BibitemOpen
  \bibfield  {author} {\bibinfo {author} {\bibfnamefont {J.~J.}\ \bibnamefont
  {Dudek}}, \bibinfo {author} {\bibfnamefont {R.~G.}\ \bibnamefont {Edwards}},
  \bibinfo {author} {\bibfnamefont {C.~E.}\ \bibnamefont {Thomas}}, \ and\
  \bibinfo {author} {\bibfnamefont {D.~J.}\ \bibnamefont {Wilson}},\
  }\href@noop {} {\  (\bibinfo {year} {2014})},\ \Eprint
  {http://arxiv.org/abs/1406.4158} {arXiv:1406.4158 [hep-ph]} \BibitemShut
  {NoStop}%
\bibitem [{\citenamefont {Byers}\ and\ \citenamefont
  {Yang}(1961)}]{PhysRevLett.7.46}%
  \BibitemOpen
  \bibfield  {author} {\bibinfo {author} {\bibfnamefont {N.}~\bibnamefont
  {Byers}}\ and\ \bibinfo {author} {\bibfnamefont {C.~N.}\ \bibnamefont
  {Yang}},\ }\href {\doibase 10.1103/PhysRevLett.7.46} {\bibfield  {journal}
  {\bibinfo  {journal} {Phys. Rev. Lett.}\ }\textbf {\bibinfo {volume} {7}},\
  \bibinfo {pages} {46} (\bibinfo {year} {1961})}\BibitemShut {NoStop}%
\bibitem [{\citenamefont {Johnson}(1982)}]{Johnson:1982yq}%
  \BibitemOpen
  \bibfield  {author} {\bibinfo {author} {\bibfnamefont {R.}~\bibnamefont
  {Johnson}},\ }\href {\doibase 10.1016/0370-2693(82)90134-4} {\bibfield
  {journal} {\bibinfo  {journal} {Phys.Lett.}\ }\textbf {\bibinfo {volume}
  {B114}},\ \bibinfo {pages} {147} (\bibinfo {year} {1982})}\BibitemShut
  {NoStop}%
\bibitem [{\citenamefont {Moore}\ and\ \citenamefont
  {Fleming}(2006{\natexlab{a}})}]{Moore:2005dw}%
  \BibitemOpen
  \bibfield  {author} {\bibinfo {author} {\bibfnamefont {D.~C.}\ \bibnamefont
  {Moore}}\ and\ \bibinfo {author} {\bibfnamefont {G.~T.}\ \bibnamefont
  {Fleming}},\ }\href {\doibase 10.1103/PhysRevD.73.014504,
  10.1103/PhysRevD.74.079905} {\bibfield  {journal} {\bibinfo  {journal}
  {Phys.Rev.}\ }\textbf {\bibinfo {volume} {D73}},\ \bibinfo {pages} {014504}
  (\bibinfo {year} {2006}{\natexlab{a}})},\ \Eprint
  {http://arxiv.org/abs/hep-lat/0507018} {arXiv:hep-lat/0507018 [hep-lat]}
  \BibitemShut {NoStop}%
\bibitem [{\citenamefont {Basak}\ \emph {et~al.}(2005)\citenamefont {Basak}
  \emph {et~al.}}]{Basak:2005ir}%
  \BibitemOpen
  \bibfield  {author} {\bibinfo {author} {\bibfnamefont {S.}~\bibnamefont
  {Basak}} \emph {et~al.} (\bibinfo {collaboration} {Lattice Hadron Physics
  Collaboration (LHPC)}),\ }\href {\doibase 10.1103/PhysRevD.72.074501}
  {\bibfield  {journal} {\bibinfo  {journal} {Phys.Rev.}\ }\textbf {\bibinfo
  {volume} {D72}},\ \bibinfo {pages} {074501} (\bibinfo {year} {2005})},\
  \Eprint {http://arxiv.org/abs/hep-lat/0508018} {arXiv:hep-lat/0508018
  [hep-lat]} \BibitemShut {NoStop}%
\bibitem [{\citenamefont {Moore}\ and\ \citenamefont
  {Fleming}(2006{\natexlab{b}})}]{Moore:2006ng}%
  \BibitemOpen
  \bibfield  {author} {\bibinfo {author} {\bibfnamefont {D.~C.}\ \bibnamefont
  {Moore}}\ and\ \bibinfo {author} {\bibfnamefont {G.~T.}\ \bibnamefont
  {Fleming}},\ }\href {\doibase 10.1103/PhysRevD.74.054504} {\bibfield
  {journal} {\bibinfo  {journal} {Phys.Rev.}\ }\textbf {\bibinfo {volume}
  {D74}},\ \bibinfo {pages} {054504} (\bibinfo {year} {2006}{\natexlab{b}})},\
  \Eprint {http://arxiv.org/abs/hep-lat/0607004} {arXiv:hep-lat/0607004
  [hep-lat]} \BibitemShut {NoStop}%
\bibitem [{\citenamefont {Dudek}\ \emph {et~al.}(2009)\citenamefont {Dudek},
  \citenamefont {Edwards}, \citenamefont {Peardon}, \citenamefont {Richards},\
  and\ \citenamefont {Thomas}}]{Dudek:2009qf}%
  \BibitemOpen
  \bibfield  {author} {\bibinfo {author} {\bibfnamefont {J.~J.}\ \bibnamefont
  {Dudek}}, \bibinfo {author} {\bibfnamefont {R.~G.}\ \bibnamefont {Edwards}},
  \bibinfo {author} {\bibfnamefont {M.~J.}\ \bibnamefont {Peardon}}, \bibinfo
  {author} {\bibfnamefont {D.~G.}\ \bibnamefont {Richards}}, \ and\ \bibinfo
  {author} {\bibfnamefont {C.~E.}\ \bibnamefont {Thomas}},\ }\href {\doibase
  10.1103/PhysRevLett.103.262001} {\bibfield  {journal} {\bibinfo  {journal}
  {Phys.Rev.Lett.}\ }\textbf {\bibinfo {volume} {103}},\ \bibinfo {pages}
  {262001} (\bibinfo {year} {2009})},\ \Eprint {http://arxiv.org/abs/0909.0200}
  {arXiv:0909.0200 [hep-ph]} \BibitemShut {NoStop}%
\bibitem [{\citenamefont {Dudek}\ \emph {et~al.}(2010)\citenamefont {Dudek},
  \citenamefont {Edwards}, \citenamefont {Peardon}, \citenamefont {Richards},\
  and\ \citenamefont {Thomas}}]{Dudek:2010wm}%
  \BibitemOpen
  \bibfield  {author} {\bibinfo {author} {\bibfnamefont {J.~J.}\ \bibnamefont
  {Dudek}}, \bibinfo {author} {\bibfnamefont {R.~G.}\ \bibnamefont {Edwards}},
  \bibinfo {author} {\bibfnamefont {M.~J.}\ \bibnamefont {Peardon}}, \bibinfo
  {author} {\bibfnamefont {D.~G.}\ \bibnamefont {Richards}}, \ and\ \bibinfo
  {author} {\bibfnamefont {C.~E.}\ \bibnamefont {Thomas}},\ }\href {\doibase
  10.1103/PhysRevD.82.034508} {\bibfield  {journal} {\bibinfo  {journal}
  {Phys.Rev.}\ }\textbf {\bibinfo {volume} {D82}},\ \bibinfo {pages} {034508}
  (\bibinfo {year} {2010})},\ \Eprint {http://arxiv.org/abs/1004.4930}
  {arXiv:1004.4930 [hep-ph]} \BibitemShut {NoStop}%
\bibitem [{\citenamefont {Luu}\ and\ \citenamefont
  {Savage}(2011)}]{Luu:2011ep}%
  \BibitemOpen
  \bibfield  {author} {\bibinfo {author} {\bibfnamefont {T.}~\bibnamefont
  {Luu}}\ and\ \bibinfo {author} {\bibfnamefont {M.~J.}\ \bibnamefont
  {Savage}},\ }\href {\doibase 10.1103/PhysRevD.83.114508} {\bibfield
  {journal} {\bibinfo  {journal} {Phys.Rev.}\ }\textbf {\bibinfo {volume}
  {D83}},\ \bibinfo {pages} {114508} (\bibinfo {year} {2011})},\ \Eprint
  {http://arxiv.org/abs/1101.3347} {arXiv:1101.3347 [hep-lat]} \BibitemShut
  {NoStop}%
\bibitem [{\citenamefont {Edwards}\ \emph {et~al.}(2011)\citenamefont
  {Edwards}, \citenamefont {Dudek}, \citenamefont {Richards},\ and\
  \citenamefont {Wallace}}]{Edwards:2011jj}%
  \BibitemOpen
  \bibfield  {author} {\bibinfo {author} {\bibfnamefont {R.~G.}\ \bibnamefont
  {Edwards}}, \bibinfo {author} {\bibfnamefont {J.~J.}\ \bibnamefont {Dudek}},
  \bibinfo {author} {\bibfnamefont {D.~G.}\ \bibnamefont {Richards}}, \ and\
  \bibinfo {author} {\bibfnamefont {S.~J.}\ \bibnamefont {Wallace}},\ }\href
  {\doibase 10.1103/PhysRevD.84.074508} {\bibfield  {journal} {\bibinfo
  {journal} {Phys.Rev.}\ }\textbf {\bibinfo {volume} {D84}},\ \bibinfo {pages}
  {074508} (\bibinfo {year} {2011})},\ \Eprint {http://arxiv.org/abs/1104.5152}
  {arXiv:1104.5152 [hep-ph]} \BibitemShut {NoStop}%
\bibitem [{\citenamefont {Thomas}\ \emph {et~al.}(2012)\citenamefont {Thomas},
  \citenamefont {Edwards},\ and\ \citenamefont {Dudek}}]{Thomas:2011rh}%
  \BibitemOpen
  \bibfield  {author} {\bibinfo {author} {\bibfnamefont {C.~E.}\ \bibnamefont
  {Thomas}}, \bibinfo {author} {\bibfnamefont {R.~G.}\ \bibnamefont {Edwards}},
  \ and\ \bibinfo {author} {\bibfnamefont {J.~J.}\ \bibnamefont {Dudek}},\
  }\href {\doibase 10.1103/PhysRevD.85.014507, 10.1103/PhysRevD.85.039901}
  {\bibfield  {journal} {\bibinfo  {journal} {Phys.Rev.}\ }\textbf {\bibinfo
  {volume} {D85}},\ \bibinfo {pages} {014507} (\bibinfo {year} {2012})},\
  \Eprint {http://arxiv.org/abs/1107.1930} {arXiv:1107.1930 [hep-lat]}
  \BibitemShut {NoStop}%
\bibitem [{\citenamefont {Drut}\ and\ \citenamefont
  {Nicholson}(2013)}]{Drut:2012md}%
  \BibitemOpen
  \bibfield  {author} {\bibinfo {author} {\bibfnamefont {J.~E.}\ \bibnamefont
  {Drut}}\ and\ \bibinfo {author} {\bibfnamefont {A.~N.}\ \bibnamefont
  {Nicholson}},\ }\href {\doibase 10.1088/0954-3899/40/4/043101} {\bibfield
  {journal} {\bibinfo  {journal} {J.Phys.}\ }\textbf {\bibinfo {volume}
  {G40}},\ \bibinfo {pages} {043101} (\bibinfo {year} {2013})},\ \Eprint
  {http://arxiv.org/abs/1208.6556} {arXiv:1208.6556 [cond-mat.stat-mech]}
  \BibitemShut {NoStop}%
\bibitem [{\citenamefont {Endres}\ \emph {et~al.}(2013)\citenamefont {Endres},
  \citenamefont {Kaplan}, \citenamefont {Lee},\ and\ \citenamefont
  {Nicholson}}]{Endres:2012cw}%
  \BibitemOpen
  \bibfield  {author} {\bibinfo {author} {\bibfnamefont {M.~G.}\ \bibnamefont
  {Endres}}, \bibinfo {author} {\bibfnamefont {D.~B.}\ \bibnamefont {Kaplan}},
  \bibinfo {author} {\bibfnamefont {J.-W.}\ \bibnamefont {Lee}}, \ and\
  \bibinfo {author} {\bibfnamefont {A.~N.}\ \bibnamefont {Nicholson}},\ }\href
  {\doibase 10.1103/PhysRevA.87.023615} {\bibfield  {journal} {\bibinfo
  {journal} {Phys.Rev.}\ }\textbf {\bibinfo {volume} {A87}},\ \bibinfo {pages}
  {023615} (\bibinfo {year} {2013})},\ \Eprint {http://arxiv.org/abs/1203.3169}
  {arXiv:1203.3169 [hep-lat]} \BibitemShut {NoStop}%
\bibitem [{\citenamefont {Bedaque}\ \emph {et~al.}(2006)\citenamefont
  {Bedaque}, \citenamefont {Sato},\ and\ \citenamefont
  {Walker-Loud}}]{Bedaque:2006yi}%
  \BibitemOpen
  \bibfield  {author} {\bibinfo {author} {\bibfnamefont {P.~F.}\ \bibnamefont
  {Bedaque}}, \bibinfo {author} {\bibfnamefont {I.}~\bibnamefont {Sato}}, \
  and\ \bibinfo {author} {\bibfnamefont {A.}~\bibnamefont {Walker-Loud}},\
  }\href {\doibase 10.1103/PhysRevD.73.074501} {\bibfield  {journal} {\bibinfo
  {journal} {Phys.Rev.}\ }\textbf {\bibinfo {volume} {D73}},\ \bibinfo {pages}
  {074501} (\bibinfo {year} {2006})},\ \Eprint
  {http://arxiv.org/abs/hep-lat/0601033} {arXiv:hep-lat/0601033 [hep-lat]}
  \BibitemShut {NoStop}%
\bibitem [{\citenamefont {Sato}\ and\ \citenamefont
  {Bedaque}(2007)}]{Sato:2007ms}%
  \BibitemOpen
  \bibfield  {author} {\bibinfo {author} {\bibfnamefont {I.}~\bibnamefont
  {Sato}}\ and\ \bibinfo {author} {\bibfnamefont {P.~F.}\ \bibnamefont
  {Bedaque}},\ }\href {\doibase 10.1103/PhysRevD.76.034502} {\bibfield
  {journal} {\bibinfo  {journal} {Phys.Rev.}\ }\textbf {\bibinfo {volume}
  {D76}},\ \bibinfo {pages} {034502} (\bibinfo {year} {2007})},\ \Eprint
  {http://arxiv.org/abs/hep-lat/0702021} {arXiv:hep-lat/0702021 [HEP-LAT]}
  \BibitemShut {NoStop}%
\bibitem [{\citenamefont {Chen}\ and\ \citenamefont
  {Oset}(2013)}]{Chen:2012rp}%
  \BibitemOpen
  \bibfield  {author} {\bibinfo {author} {\bibfnamefont {H.-X.}\ \bibnamefont
  {Chen}}\ and\ \bibinfo {author} {\bibfnamefont {E.}~\bibnamefont {Oset}},\
  }\href {\doibase 10.1103/PhysRevD.87.016014} {\bibfield  {journal} {\bibinfo
  {journal} {Phys.Rev.}\ }\textbf {\bibinfo {volume} {D87}},\ \bibinfo {pages}
  {016014} (\bibinfo {year} {2013})},\ \Eprint {http://arxiv.org/abs/1202.2787}
  {arXiv:1202.2787 [hep-lat]} \BibitemShut {NoStop}%
\bibitem [{\citenamefont {Albaladejo}\ \emph {et~al.}(2013)\citenamefont
  {Albaladejo}, \citenamefont {Rios}, \citenamefont {Oller},\ and\
  \citenamefont {Roca}}]{Albaladejo:2013bra}%
  \BibitemOpen
  \bibfield  {author} {\bibinfo {author} {\bibfnamefont {M.}~\bibnamefont
  {Albaladejo}}, \bibinfo {author} {\bibfnamefont {G.}~\bibnamefont {Rios}},
  \bibinfo {author} {\bibfnamefont {J.}~\bibnamefont {Oller}}, \ and\ \bibinfo
  {author} {\bibfnamefont {L.}~\bibnamefont {Roca}},\ }\href@noop {} {\
  (\bibinfo {year} {2013})},\ \Eprint {http://arxiv.org/abs/1307.5169}
  {arXiv:1307.5169 [hep-lat]} \BibitemShut {NoStop}%
\bibitem [{\citenamefont {Fu}(2012)}]{Fu:2011xz}%
  \BibitemOpen
  \bibfield  {author} {\bibinfo {author} {\bibfnamefont {Z.}~\bibnamefont
  {Fu}},\ }\href {\doibase 10.1103/PhysRevD.85.014506} {\bibfield  {journal}
  {\bibinfo  {journal} {Phys.Rev.}\ }\textbf {\bibinfo {volume} {D85}},\
  \bibinfo {pages} {014506} (\bibinfo {year} {2012})},\ \Eprint
  {http://arxiv.org/abs/1110.0319} {arXiv:1110.0319 [hep-lat]} \BibitemShut
  {NoStop}%
\bibitem [{\citenamefont {Beringer}\ \emph {et~al.}(2012)\citenamefont
  {Beringer} \emph {et~al.}}]{Beringer:1900zz}%
  \BibitemOpen
  \bibfield  {author} {\bibinfo {author} {\bibfnamefont {J.}~\bibnamefont
  {Beringer}} \emph {et~al.} (\bibinfo {collaboration} {Particle Data Group}),\
  }\href {\doibase 10.1103/PhysRevD.86.010001} {\bibfield  {journal} {\bibinfo
  {journal} {Phys.Rev.}\ }\textbf {\bibinfo {volume} {D86}},\ \bibinfo {pages}
  {010001} (\bibinfo {year} {2012})}\BibitemShut {NoStop}%
\bibitem [{\citenamefont {Bedaque}\ and\ \citenamefont
  {Chen}(2005)}]{Bedaque:2004ax}%
  \BibitemOpen
  \bibfield  {author} {\bibinfo {author} {\bibfnamefont {P.~F.}\ \bibnamefont
  {Bedaque}}\ and\ \bibinfo {author} {\bibfnamefont {J.-W.}\ \bibnamefont
  {Chen}},\ }\href {\doibase 10.1016/j.physletb.2005.04.045} {\bibfield
  {journal} {\bibinfo  {journal} {Phys.Lett.}\ }\textbf {\bibinfo {volume}
  {B616}},\ \bibinfo {pages} {208} (\bibinfo {year} {2005})},\ \Eprint
  {http://arxiv.org/abs/hep-lat/0412023} {arXiv:hep-lat/0412023 [hep-lat]}
  \BibitemShut {NoStop}%
\bibitem [{\citenamefont {Doring}\ \emph {et~al.}(2011)\citenamefont {Doring},
  \citenamefont {Meissner}, \citenamefont {Oset},\ and\ \citenamefont
  {Rusetsky}}]{Doring:2011vk}%
  \BibitemOpen
  \bibfield  {author} {\bibinfo {author} {\bibfnamefont {M.}~\bibnamefont
  {Doring}}, \bibinfo {author} {\bibfnamefont {U.-G.}\ \bibnamefont
  {Meissner}}, \bibinfo {author} {\bibfnamefont {E.}~\bibnamefont {Oset}}, \
  and\ \bibinfo {author} {\bibfnamefont {A.}~\bibnamefont {Rusetsky}},\ }\href
  {\doibase 10.1140/epja/i2011-11139-7} {\bibfield  {journal} {\bibinfo
  {journal} {Eur.Phys.J.}\ }\textbf {\bibinfo {volume} {A47}},\ \bibinfo
  {pages} {139} (\bibinfo {year} {2011})},\ \Eprint
  {http://arxiv.org/abs/1107.3988} {arXiv:1107.3988 [hep-lat]} \BibitemShut
  {NoStop}%
\bibitem [{\citenamefont {Agadjanov}\ \emph {et~al.}(2013)\citenamefont
  {Agadjanov}, \citenamefont {Meissner},\ and\ \citenamefont
  {Rusetsky}}]{Agadjanov:2013wqa}%
  \BibitemOpen
  \bibfield  {author} {\bibinfo {author} {\bibfnamefont {D.}~\bibnamefont
  {Agadjanov}}, \bibinfo {author} {\bibfnamefont {U.-G.}\ \bibnamefont
  {Meissner}}, \ and\ \bibinfo {author} {\bibfnamefont {A.}~\bibnamefont
  {Rusetsky}},\ }\href@noop {} {\  (\bibinfo {year} {2013})},\ \Eprint
  {http://arxiv.org/abs/1310.7875} {arXiv:1310.7875 [hep-lat]} \BibitemShut
  {NoStop}%
\end{thebibliography}%

\end{document}